\documentclass[preprint]{aastex}
\shortauthors{Logsdon et al.}

\usepackage{color,amsmath, longtable}
\usepackage{natbib}
\usepackage{graphicx}
\usepackage{hyperref}

\newcommand{\teff}{\textit{T}$_{\rm eff}$}
\newcommand{\logg}{$\log{g}$}
\newcommand{\fsed}{$f_{\rm sed}$}

\newcommand{\meth}{CH$_4$}
\newcommand{\water}{H$_2$O}

\newcommand{\hh}{H$_2$}

\newcommand{\microm}{$\mu$m}
\newcommand{\midtilde}{$\sim$}
\newcommand{\asec}{$^{\prime\prime}$}

\newcommand{\rotated}{\@pt@rottrue}
\defcitealias{Mace2013b}{M13b}

\begin{document}

\title{Probing Late-type T dwarf \textit{$J-H$} Color Outliers for Signs of Age\textsuperscript{*}}

\author{
Sarah E.\ Logsdon\altaffilmark{1,3},
Gregory N.\ Mace\altaffilmark{2}, 
Ian S.\ McLean\altaffilmark{1}, and
Emily C.\ Martin\altaffilmark{1}
}

\altaffiltext{1}{Department of Physics and Astronomy, University of California Los Angeles, 430 Portola Plaza, Box 951547, Los Angeles, CA 90095-1547, USA}
\altaffiltext{2}{McDonald Observatory \& Department of Astronomy, University of Texas at Austin, 2515 Speedway, Stop C1400, Austin, TX 78712-1205, USA}
\altaffiltext{3}{Current address: NASA Goddard Space Flight Center, 8800 Greenbelt Road, Greenbelt, MD 20771, USA; sarah.e.logsdon@nasa.gov}

\begin{abstract}

We present the results of a Keck/NIRSPEC follow-up survey of thirteen late-type T dwarfs (T6-T9), twelve of which have unusually red or blue \textit{$J-H$} colors. Previous work suggests that \textit{$J-H$} color outliers may represent the high-gravity, low-metallicity (old) and low-gravity, solar-metallicity (young) extremes of the late-type T dwarf population. We use medium-resolution \textit{Y}- and \textit{H}-band spectroscopy to probe regions of T dwarf atmospheres that are more sensitive to gravity and metallicity variations than the \textit{J} band. We find that the spectral morphologies of our sample are largely homogeneous, with peak-normalized, \textit{Y}- and \textit{H}-band morphologies consistent with spectral standards. However, three objects stand out as potentially old, with overluminous \textit{Y}-band spectra compared to their respective spectral standards, and a fourth object stands out as potentially young, with an underluminous \textit{Y} band. Of these four objects, three have been previously identified as potential metallicity/gravity outliers, including the one object in our sample with a normal \textit{J-H} color. We fit publicly available atmospheric model grids to our spectra and find that the best-fit physical parameters vary depending on the model used. As we continue to probe the characteristics of the late-T population, differences in synthetic spectra of \midtilde10-20\% in the blue wing of the \textit{Y} band and \midtilde45\% at 1.65 \microm, for the same physical parameters, must be reconciled. Further development and public availability of nonsolar metallicity models is also recommended. Future progress toward deciphering  the impacts of gravity, metallicity, and variability in the late-type T dwarf population will also require high signal-to-noise, multiwavelength and multi-epoch photometry and spectroscopy. 
\end{abstract}

\renewcommand{\thefootnote}{\fnsymbol{footnote}}
\footnotetext[1]{The data presented herein were obtained at the W.M. Keck Observatory, which is operated as a scientific partnership among the California Institute of Technology, the University of California and the National Aeronautics and Space Administration. The Observatory was made possible by the generous financial support of the W.M. Keck Foundation.}


\section{Introduction}
\renewcommand{\thefootnote}{\arabic{footnote}}
Like stars, brown dwarfs are classified into spectral types (M, L, T, and Y) based on changes in their observed spectral morphologies. These changes are predominantly, but not exclusively, driven by changes in temperature, with the M dwarf class comprising both low-mass stars and the warmest brown dwarfs (\teff\ $\gtrsim$ 2500 K), the L dwarf class comprising low-mass stars and brown dwarfs with 2500 K $\gtrsim$ \teff\ $\gtrsim$ 1400 K, and the T and Y dwarf classes comprising the coldest brown dwarfs (\teff\ $\lesssim$ 1400 K; e.g. \citealt{Kirkpatrick2005, Cushing2011}). Unlike main-sequence stars, brown dwarfs are not massive enough to maintain stable hydrogen fusion in their cores. This lack of a sustainable energy source means that brown dwarfs continuously cool as they age, creating a degeneracy between luminosity (temperature and radius), mass, and age.  Thus, independently determining the mass and/or age of individual field brown dwarfs has long presented a challenge to observers.

Brown dwarfs whose mass and/or age can be independently determined are so-called ``benchmark'' brown dwarfs (e.g. \citealt{Pinfield2006, Liu2008}; see also \citealt{Marocco2017} and the references therein for a recent discussion in the context of \textit{Gaia}). Brown dwarf companions in assumed coeval systems where the age and metallicity of the primary star are known and the mass of the brown dwarf can be dynamically measured are ideal benchmarks, but such systems are rare (e.g. \citealt{Dupuy2009, Crepp2012}). For single field objects, inferred properties from kinematics (e.g. \citealt{Dahn2002, Vrba2004, Schmidt2007, Faherty2009, DupuyLiu2012}), and spectral signatures of gravity (e.g. \citealt{McGovern2004, AllersLiu2013, Martin2017}) and metallicity (e.g. \citealt{Burgasser2003, Kirkpatrick2014, Zhang2017}) can help differentiate ages. A young brown dwarf will have a smaller mass (M) and larger radius (R), and thus a lower surface gravity (\textit{g} = GM/R$^2$), than an older brown dwarf at that same temperature. The metallicity of a brown dwarf is dependent on the chemical enrichment of its natal environment. Like stars, brown dwarfs with subsolar metallicities are likely old, having formed in more pristine environments than their younger counterparts, and brown dwarfs with supersolar metallicity are predominantly young. Thus, young brown dwarfs tend to have low gravity and solar or supersolar metallicity, while old brown dwarfs mostly have higher gravity and solar or subsolar metallicity. A well-characterized sample of brown dwarfs at a large range of temperatures, gravities, and metallicities serves as a powerful probe of atmospheric evolution at the lowest masses and temperatures and informs our broader understanding of Galactic evolution. 

Within the last decade, spectroscopic follow-up of T dwarf candidates identified from near-to-mid infrared imaging surveys such as the \textit{Wide-Field Infrared Survey Explorer} (\textit{WISE}; \citealt{Wright2010}) and the UKIRT Infrared Deep Sky Survey (UKIDSS; \citealt{Lawrence2007}), has more than doubled the number of known T dwarfs, with the largest increases in the population of T dwarfs with spectral types T5 and later (see e.g. DwarfArchives.org, and the compilation in \citealt{MacePHD2014}). With these additions, we now have enough confirmed late-T dwarfs to identify photometric and spectroscopic trends (e.g. \citealt{Kirkpatrick2011, Kirkpatrick2012, Burningham2013, Mace2013}) and to investigate outliers. 

T dwarfs are typically classified by their near-infrared spectra \citep{Burgasser2006b}, which are broadly shaped by methane (\meth) and water (\water) absorption features. The strengths of \meth\ and \water\ absorption are largely influenced by changes in temperature in the atmospheres of these T dwarfs. However, the overall impact of secondary parameters such as gravity, metallicity, and clouds on T dwarf atmospheres can be significant (e.g. \citealt{Burrows2002, Burgasser2006c}). For example, collision-induced absorption (CIA) of \hh\ is  strongly dependent on gravity and is also metallicity dependent. Increased \hh\ CIA opacity in high-density, high-gravity atmospheres leads to a suppression of the near-infrared flux in T dwarfs and is dominant in the \textit{K} band (e.g. \citealt{Saumon2012}). At shorter wavelengths, the shape of the blue wing of the \textit{Z/Y} band is impacted by pressure-broadening of the Na D (\midtilde5890 $\AA$) and K I (\midtilde7700 $\AA$) doublets in the red-optical part of the spectrum. In low-metallicity, high-pressure photospheres, these strong alkali lines are expected to enhance the blue wing of the \textit{Z/Y} band (e.g. \citealt{Burrows2002, Burrows2006, Burgasser2006c}). 

The past five years have seen the discovery of the first bonafide late-T subdwarf age benchmarks, which serve to inform the roles subsolar metallicity and high gravity play in shaping the emergent spectral morphology of the late-T dwarf population. \citet{Mace2013b} (henceforth \citetalias{Mace2013b}) presented the first unambiguous late-T subdwarf discovery, Wolf 1130C. Wolf 1130C is a sdT8 companion to a sdM and ultramassive white dwarf binary system. Inferring metallicity from the M dwarf, Wolf 1130C has the lowest known metallicity of a T dwarf ([Fe/H]$=-$0.7 $\pm$ 0.12 dex, \citealt{Mace2018}). This low metallicity is most evident in the peak-normalized \textit{Y}-band spectrum of Wolf 1130C, which is unusually blue compared to the T8 spectral standard (see Figure 5 in \citetalias{Mace2013b} and Figure \ref{fig:allvsWolf} in this text). Shortly after the discovery of Wolf 1130C, \citet{Pinfield2014} presented the discovery of WISE J001354.39+063448.2 (WISE J0013+0634; T8) and WISE J083337.83+005214.2 (WISE J0833+0052; T9). Both T dwarfs have thick disk/halo kinematics derived using spectrophotometric distance estimates and their \textit{K}-band spectra are suppressed; they are likely subdwarfs. \citet{Burningham2014} presented the discovery of the sdT6.5 dwarf ULAS J131610.28+075553.0 (ULAS J1316+0755), which displays an unusually blue \textit{Y}-band spectrum and a suppressed \textit{K} band, again indicative of low metallicity and/or high gravity. Other late-T age benchmarks with inferred subsolar metallicities include HIP 73786B \citep{Scholz2010, Murray2011}, a T6p dwarf companion to a K8V star \citep{Gray2003} with a metallicity of [Fe/H] = -0.3 $\pm$ 0.1 \citep{Cenarro2007}, and BD +01$^\circ$ 2920B (T8p), a companion to a G1 dwarf with a metallicity of  [Fe/H] = -0.38 $\pm$ 0.06 (\citealt{Pinfield2012}; references therein). HIP 73786B is discussed in more detail below and in Section \ref{sec:characteristics}.

\begin{figure}[!h]
\centering
\includegraphics[width=4.5in]{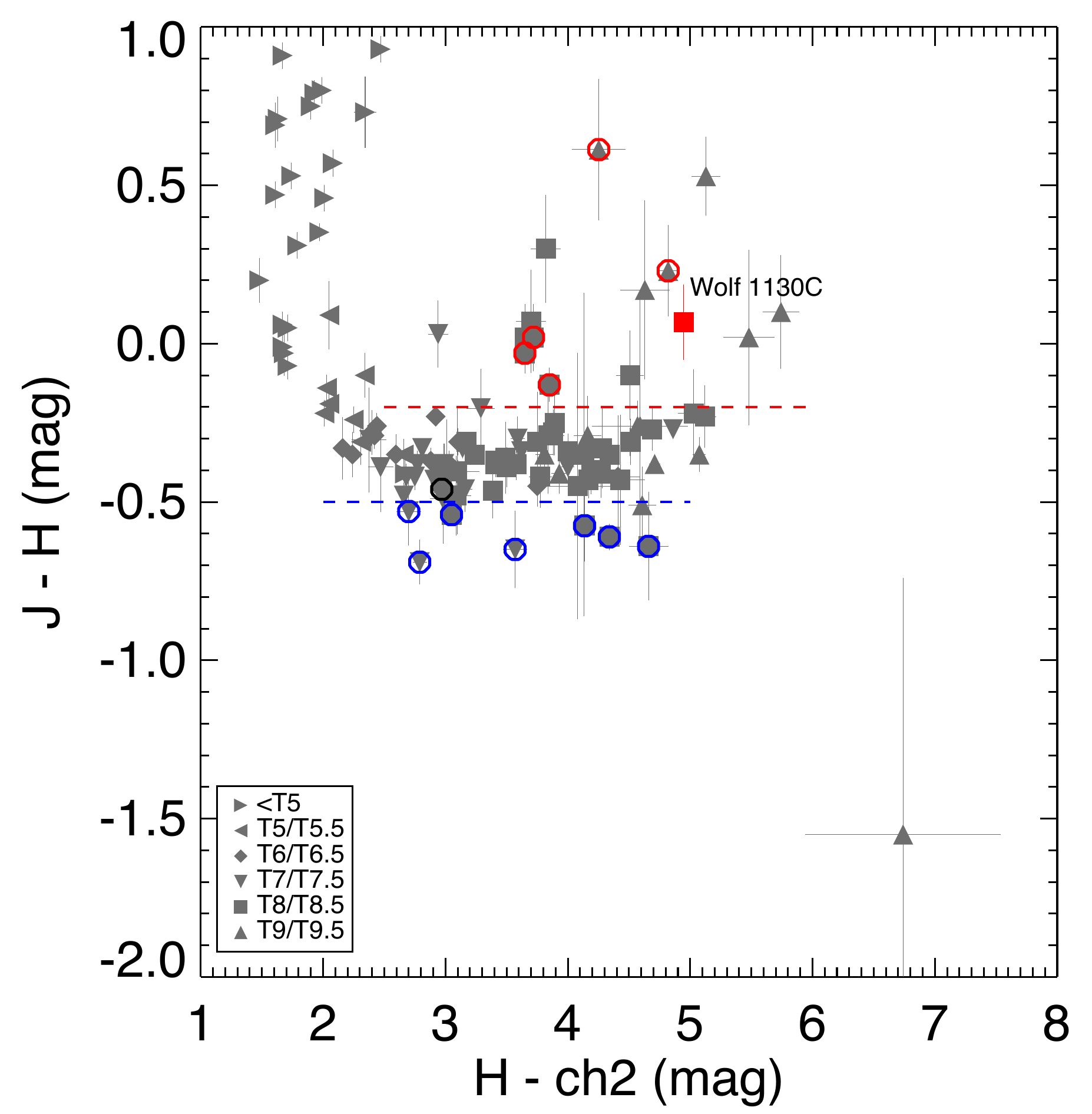}
\caption{\small A modified version of Figure 8 in \citetalias{Mace2013b}: \textit{$J-H$} vs. \textit{$H-ch2$} of T dwarfs from the literature (photometry from \citealt{Albert2011}, \citealt{Burningham2010, Burningham2013}, Dupuy \& Liu 2012, \citealt{Kirkpatrick2011, Kirkpatrick2012}, \citealt{Leggett2010, Leggett2013, Leggett2017}, \citealt{Mace2013, Mace2013b}, \citealt{Murray2011}, \citealt{Scholz2010}, \citealt{Thompson2013}). For T dwarfs with spectral types later than T5, \textit{$H-ch2$} serves as a proxy for spectral type, with redder \textit{$H-ch2$} colors implying later spectral types \citep{Kirkpatrick2011}. The red and blue dashed lines denote the color criteria of \citetalias{Mace2013b}; Wolf 1130C is denoted with a red square and labeled for emphasis. Circled objects denote the targets in our sample. The color of the circle denotes a target's \citetalias{Mace2013b} color classification. Late-T dwarfs largely follow a well-defined sequence in \textit{$J-H$} vs. \textit{$H-ch2$}, though outliers do exist. Investigating the cause(s) of the unusual outlier photometry is the aim of this study.}
\label{fig:JH_vs_HCH2}
\end{figure}

Wolf 1130C not only stands out as the most metal-poor, late-T subdwarf discovered to date, it also stands out in \textit{$J-H$} color space with \textit{$J-H$} = 0.068 $\pm$ 0.119 \citep{Mace2013b}. While \textit{$J-H$} colors become increasingly bluer with spectral type from late-L to mid-T, T dwarfs with spectral types later than T5 tend to plateau in \textit{$J-H$} color (e.g. \citealt{Burningham2013, Mace2013}; see Figure \ref{fig:JH_vs_HCH2}). As discussed in \citetalias{Mace2013b}, there are several late-T dwarfs with unusual colors that stand out from the plateau in  \textit{$J-H$} color space.  Based on MKO photometry from the literature, \citetalias{Mace2013b} defined the width of the late-T,  \textit{$J-H$}$_{MKO}$ plateau to be $-0.5 \leq$ \textit{$J-H$}$_{MKO}$ $\leq -0.2$ (see Figure \ref{fig:JH_vs_HCH2}). Objects with a \textit{$J-H$}$_{MKO}$ color $> -0.2$ were defined as``red'' and objects with a  \textit{$J-H$}$_{MKO}$ color $< -0.5$ were defined as ``blue.'' Here we refer to objects that lie between the \citetalias{Mace2013b} ``red'' and ``blue'' color cuts (-0.5 $\leq$ \textit{$J-H$}$_{MKO}$ $\leq$ -0.2) as ``normal.'' \citetalias{Mace2013b} hypothesized that red \textit{$J-H$} color outliers (like Wolf 1130C) may represent the metal-poor/high-gravity objects in the late-T dwarf population, and that the blue \textit{$J-H$} color outliers may represent the converse, namely the metal-rich/low-gravity objects. As discussed in \citetalias{Mace2013b}, several of the known red \textit{$J-H$} outliers, such as Wolf 1130C, show signs of old age, and in contrast, blue outliers show evidence of youth. However, there are exceptions to the red and blue designations. For example, HIP 73786B has a normal \textit{$J-H$} color (\textit{$J-H$} = -0.46 $\pm$ 0.04; \citealt{Murray2011}), only 1$\sigma$ from the blue color cut of \citetalias{Mace2013b}, which would suggest that it is a potentially metal-rich/low-gravity object, however, as discussed above, its inferred metallicity is subsolar. 

The goal of this work is to spectroscopically investigate the nature(s) of the late-T dwarf \textit{$J-H$} color outlier population, and thus to test the \citetalias{Mace2013b} hypothesis. To identify metallicity/gravity trends in the late-T dwarf population, we compare spectral standards and atmospheric model grids from BT-Settl \citep{Allard2011,Allard2012}, \citet{Burrows2006}, \citet{Morley2012}, and \citet{Saumon2012} to medium-resolution, Keck/NIRSPEC \textit{Y}- and \textit{H}-band spectra of thirteen late-T dwarfs. The \textit{Y} band was chosen because of its ability to separate temperature, which is correlated with \water\ absorption at the long-wavelengths of the \textit{Y} band, from gravity and metallicity, which,  as discussed above, modulate the flux at bluer wavelengths. As our objects have all been previously spectral-typed and the spectral typing of these late-type objects is predominantly done in the \textit{J} band, we expect the least variation in \textit{J}-band morphology within a given spectral type bin. Thus, to spectroscopically test for the impact of gravity, metallicity, and additional atmospheric parameters like clouds (e.g. \citealt{Marley2010, Burgasser2010}) that may lead to unusual \textit{$J-H$} colors, we also observed our targets in the \textit{H} band. In Section 2 we describe the sample selection, observations, and data reduction technique. Spectroscopic results are presented in Section 3. Section 4 presents an analysis of the individual spectra, including atmospheric model fitting and a detailed comparison to spectral standards. Section 5 discusses the implications of these results and a summary is provided in Section 6.

\section{Sample, Observations, and Data Reduction}
\subsection{Sample Motivation and Selection}
Here we present a sample of thirteen T6-T9 brown dwarfs, twelve of which are identified as \textit{$J-H$}$_{MKO}$ color outliers following the criteria outlined in \citetalias{Mace2013b}. The thirteenth object is HIP 73786B, which provides a potentially interesting counter-example to the \citetalias{Mace2013b} color hypothesis as discussed above. Beyond imposing the \citetalias{Mace2013b} color criteria, we further require that all of the objects in the sample are observable from the Northern Hemisphere and are bright enough for medium-resolution spectroscopic follow-up with the NIRSPEC instrument \citep{McLean1998} on Keck II (\textit{J} \textless\ 19.5 mag). The twelve color outliers presented here, combined with Wolf 1130C, encompass over half (13/23) of the late-T dwarf \textit{$J-H$} color outlier population as presented in Table 3 of \citetalias{Mace2013b}.  A summary of our observations is presented in Table \ref{table:observations}. Of the thirteen dwarfs in the sample, five are classified as ``red,'' seven are classified as ``blue,'' and one, HIP 73786B, is classified as ``normal.'' 

\subsection{NIRSPEC Observations and Data Reduction}\label{sec:data_reduction}
All data were obtained between 2012 June and 2014 December using the medium-resolution (\textit{R}\midtilde2000) mode of the NIRSPEC instrument \citep{McLean1998} on the Keck II telescope. In most cases, the targets were observed in both NIRSPEC's N1 filter (equivalent to \textit{Y} band; \midtilde0.95-1.12 \microm) and N5 filter (equivalent to \textit{H} band; \midtilde1.5-1.78 \microm) configurations. The exceptions are WISE J075946.98-490454.0 (henceforth WISE J0759-4904; \citealt{Kirkpatrick2011}, \citealt{Mace2013}), ULAS J101721.40+011817.9 (ULAS 1017+0118; \citealt{Burningham2011}), and WISE J161441.46+173935.5 (WISE J1614+1739; \citealt{Kirkpatrick2011}). Only \textit{Y-}band spectra were obtained for WISE J0759-4904 and WISE J1614+1739. Both \textit{Y-} and \textit{H-}band data were obtained for ULAS J1017+0118, but the signal-to-noise ratio (S/N) for the \textit{Y-}band data was \midtilde1-2 per resolution element and was deemed too low for this analysis.  For two objects, WISE J000517.48+373720.5 (WISE J0005+3732) and WISE J054047.00+483232.4 (WISE J0540+4832), existing NIRSPEC N3 (equivalent to \textit{J} band; \midtilde1.15-1.35 \microm) spectra are presented in \citet{Mace2013}. Table \ref{table:observations} also lists previously unpublished NIRSPEC \textit{Y-} and \textit{H}-band spectra for T dwarf spectral standards used in our analysis. 

Unless otherwise noted in Table \ref{table:observations}, targets and standards were observed using NIRSPEC's 0.57\asec\ (3 pixel) slit in the typical AB (or ABBA) nod pattern in order to enable sky background subtraction. Single nod exposure times were either 300 s or 600 s depending on target brightness. For telluric corrections, an A0 standard star was observed at a similar airmass to the target star either proceeding or following target observations. Flat field, dark, Ne and Ar arc lamp frames (for wavelength calibration) were also observed with each target in each observing mode. All data were reduced in IDL using the publicly available REDSPEC package\footnote{\footnotesize See \url{https://www2.keck.hawaii.edu/inst/nirspec/redspec.html}}. REDSPEC performs the standard wavelength calibration, background subtraction, flat fielding, telluric correction, and source extraction. Absorption features in the A0 spectra are removed by interpolation before the calibrator is used for telluric correction. The REDSPEC software is described in more detail in \citet{McLean2003}. After reduction, individual target nod pairs are averaged together to improve S/N and a barycentric velocity correction is applied.

\begin{deluxetable}{cccccccccc}
\rotate
\label{table:observations}
\tablecolumns{10}
\tablecaption{NIRSPEC Observations}
\tabletypesize{\tiny}
\tablewidth{0in}
\tablehead{
\colhead{Object Name} &
\colhead{Short Name} &
\colhead{Discovery Ref.\tablenotemark{a}}&
\colhead{SpT}&
\colhead{SpT Ref.\tablenotemark{a}}&
\colhead{Band \tablenotemark{b}}&
\colhead{UT Date Observed}&
\colhead{Exp. Time (s)}&
\colhead{A0}&
\colhead{Slit Width (\asec)}
}

\startdata
WISE J000517.48+373720.5	&	WISE J0005+3737	&	1	&	T9	&	1	&	N1	&	2014 Nov 11	&	3600	&	HD 222749    	&	0.57	\\
	&		&	1	&	T9	&	1	&	N5	&	2014 Nov 11	&	3000	&	HD 222749    	&	0.57	\\
ULAS J013939.77+004813.8	&	ULAS J0139+0048	&	2	&	T7.5	&	2	&	N1	&	2014 Dec 2	&	4800	&	HD 18571 	&	0.57	\\
	&		&	2	&	T7.5	&	2	&	N5	&	2014 Dec 2	&	3600	&	HD 18571 	&	0.57	\\
CFBDS J030135.11-161418.0	&	CFBDS J0301-1614	&	3	&	T7p	&	3	&	N1	&	2014 Nov 11&	3600	&	HD 23683       	&	0.57	\\	
	&		&	3	&	T7p	&	3	&	N5	&	2014 Nov 12	&	4200	&	HD 23683       	&	0.57	\\
WISE J054047.00+483232.4	&	WISE J0540+4832	&	1	&	T8.5	&	1	&	N1	&	2014 Nov 11&	3000	&	HD 45105 	&	0.57	\\	
	&		&	1	&	T8.5	&	1	&	N5	&	2014 Nov 11	&	2400	&	HD 45105 	&	0.57	\\
WISE J075946.98-490454.0	&	WISE J0759-4904	&	4	&	T8	&	4	&	N1	&	2014 Dec 2	&	4200	&	HD 74042	&	0.57	\\
CFBDS J092250.12+152741.4	&	CFBDS J0922+1527	&	3	&	T7	&	3	&	N1	&	2014 Apr 14	&	3600	&	HD 111744 	&	0.57	\\
	&		&	3	&	T7	&	3	&	N5	&	2014 Dec 3	&	3000	&	HD 79108	&	0.57	\\
ULAS J095047.28+011734.3	&	ULAS J0950+0117	&	5	&	T8p	&	5	&	N1	&	2014 Apr 13	&	3600	&	HD 95126	&	0.57	\\
	&		&	5	&	T8p	&	5	&	N5	&	2014 Dec 2	&	3000	&	HD 79108	&	0.57	\\
ULAS J101721.40+011817.9	&	ULAS J1017+0118	&	6	&	T8p	&	6	&	N5	&	2014 Dec 3	&	3000	&	HD 79108	&	0.57	\\
ULAS J150457.65+053800.8	&	HIP 73786B	&	7	&	T6p	&	8	&	N1	&	2014 Apr 14	&	1800	&	7 Ser	&	0.57	\\
	&		&	7	&	T6p	&	8	&	N5	&	2014 Jun 20	&	2400	&	 HD 123233	&	0.57	\\
WISE J161441.46+173935.5	&	WISE J1614+1739	&	4	&	T9	&	4	&	N1	&	2014 Apr 13	&	2400	&	26 Ser	&	0.57	\\
	&		&	4	&	T9	&	4	&	N1	&	2014 Apr 14	&	2400	&	26 Ser	&	0.57	\\
WISE J161705.74+180714.1	&	WISE J1617+1807	&	9	&	T8	&	9	&	N1	&	2014 Apr 12	&	2400	&	q Her	&	0.57	\\
	&		&	9	&	T8	&	9	&	N5	&	2014 Jun 20 &	2400	&	q Her	&	0.57	\\	
WISE J181210.85+272144.3	&	WISE J1812+2721	&	9	&	T8.5	&	9	&	N1	&	2014 Jun 21	&	1800	&	HD 199217	&	0.57	\\
	&		&	9	&	T8.5	&	9	&	N5	&	2014 Jun 20	&	2400	&	HD 192538 	&	0.57	\\
ULAS J214638.83-001038.7	&	Wolf 940B	&	10	&	T8.5	&	10	&	N1	&	2014 Jun 21	&	2400	&	HD 210501	&	0.57	\\
	&		&	10	&	T8.5	&	10	&	N5	&	2014 Nov 11	&	2400	&	HD 210501	&	0.57	\\
\cutinhead{Previously Unpublished NIRSPEC Spectral Standard Observations}																			
2MASSI J0415195-093506	&	2MASS J0415-0935	&	11	&	T8	&	12	&	N1	&	2002 Dec 23	&	1200	&	HD 34481	&	0.38	\\
UGPS J072227.51-054031.2	&	UGPS J0722-0540	&	13	&	T9	&	14	&	N1	&	2014 Dec 2	&	1800	&	HD 64653	&	0.57	\\
	&		&	13	&	T9	&	14	&	N5	&	2014 Dec 3	&	1800	&	HD 64653	&	0.57	\\
2MASSI J0727182+171001	&	2MASS J0727+1710	&	11	&	T7	&	12	&	N1	&	2014 Nov 12 &	1800	&	HD 57208	&	0.57	\\	
	&		&	11	&	T7	&	12	&	N5	&	2014 Nov 12	&	1800	&	HD 57208	&	0.57	\\
\cutinhead{Previously Unpublished Brown Dwarf Spectroscopic Survey (BDSS; \citealt{McLean2003}) Comparison Object Observation}																			
2MASSI J0937347+293142	&	2MASS J0937+2931	&	11	&	T6p	&	12	&	N1	&	2001 Mar 7 &	1200 &	AG +27 1006 &	  0.38\\				
\enddata

\tablenotetext{a}{Discovery and Spectral Type references: (1) \citet{Mace2013}, (2) \citet{Chiu2008}, (3) \citet{Albert2011}, (4) \citet{Kirkpatrick2011}, (5) \citet{Burningham2013},  (6) \citet{Burningham2008}, (7) \citet{Scholz2010}, (8) \citet{Murray2011}, (9) \citet{Burgasser2011}, (10) \citet{Burningham2009}, (11) \citet{Burgasser2002}, (12) \citet{Burgasser2006b}, (13) \citet{Lucas2010}, (14) \citet{Cushing2011}}
\tablenotetext{b}{Keck/NIRSPEC \textit{J-}band spectra of WISE J0540+4832 and WISE J0005+3737 were presented in \citet{Mace2013}. A Keck/NIRSPEC \textit{J-}band spectrum of 2MASS J0727+1710 and both \textit{J-} and \textit{H-} band spectra of 2MASS J0415-0935 were presented in \citet{McLean2003}.}
\label{table:observations}
\end{deluxetable}

\section{Results}\label{sec:results}
\begin{figure}
\centering
\includegraphics[width=5.5in]{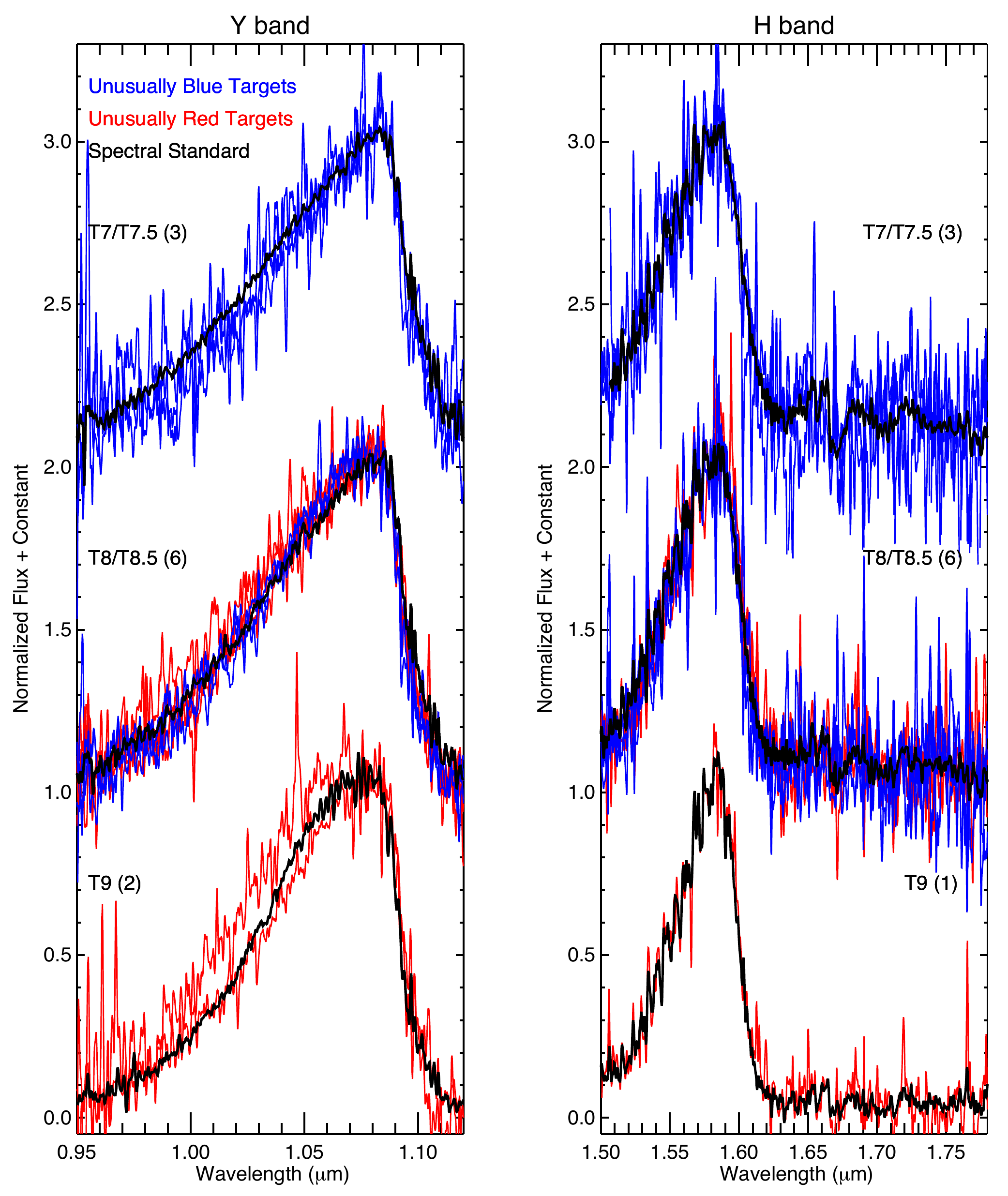}
\caption{\footnotesize NIRSPEC \textit{Y}-band (left) and \textit{H}-band (right) spectra of the color outliers in the sample plotted by spectral type from T7 (top) to T9 (bottom). The spectra are normalized at \midtilde1.08 $\mu$m and \midtilde1.58 $\mu$m, respectively. Objects with unusually red and blue \textit{$J-H$}$_{MKO}$ colors as defined by \citetalias{Mace2013b} are plotted by color accordingly. The number given in parenthesis next to each spectral type indicates the number of sample spectra over-plotted for that spectral type and band. Spectral standards from the Brown Dwarf Spectroscopic Survey (BDSS; \citealt{McLean2003}), are also over-plotted in black for comparison, but are not included in the number counts in parentheses. The spectral standards are as follows: 2MASSI J0727182+171001 (T7; \citealt{Burgasser2002, Burgasser2006b}), 2MASSI J0415195-093506 (T8; \citealt{Burgasser2002, Burgasser2006b}), and UGPS J072227.51-054031.2 (T9; \citealt{Lucas2010, Cushing2011}).}
\label{fig:allresults}
\end{figure}

\begin{figure}[!ht]
\centering
\includegraphics[width=3in]{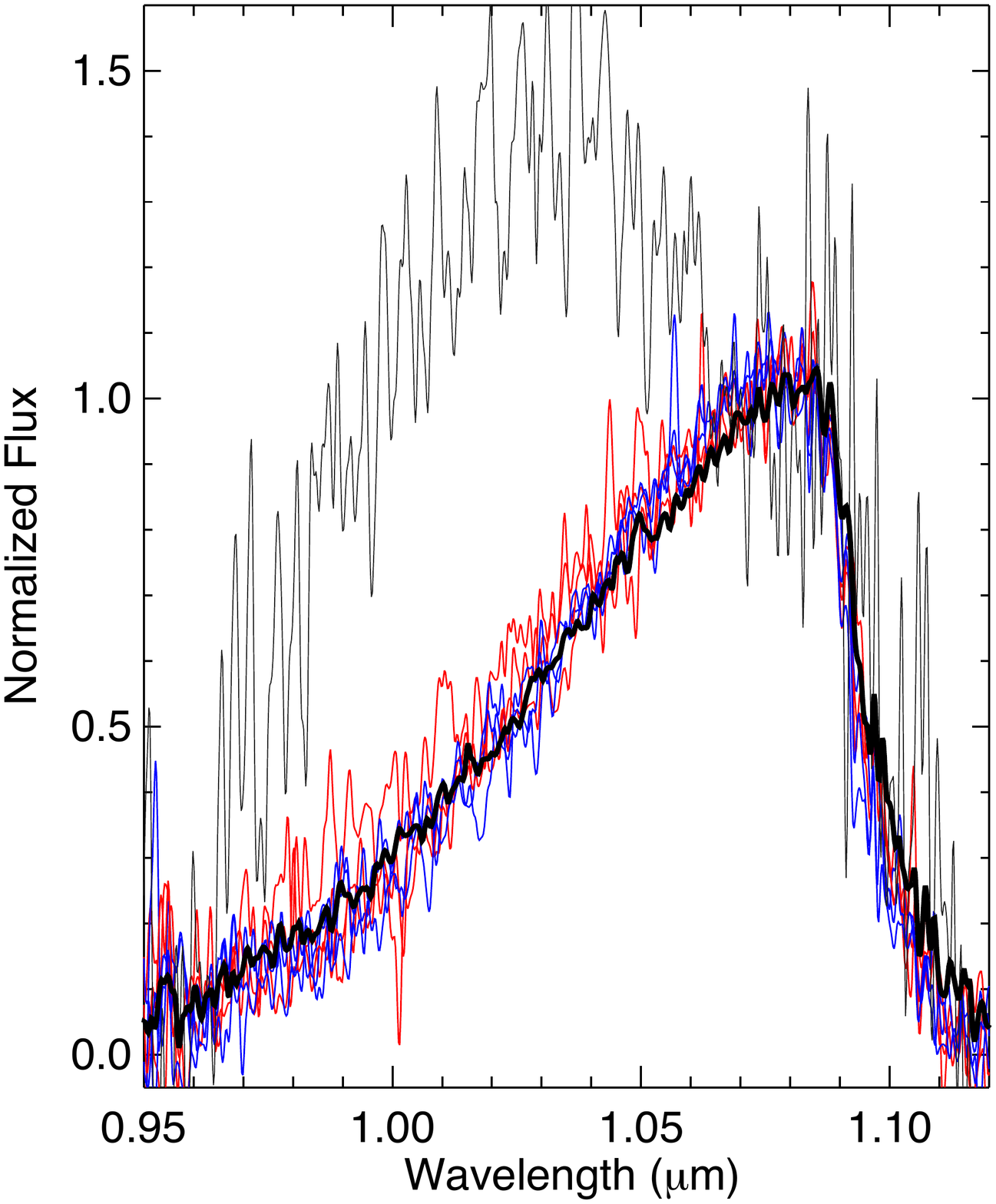}
\includegraphics[width=3in]{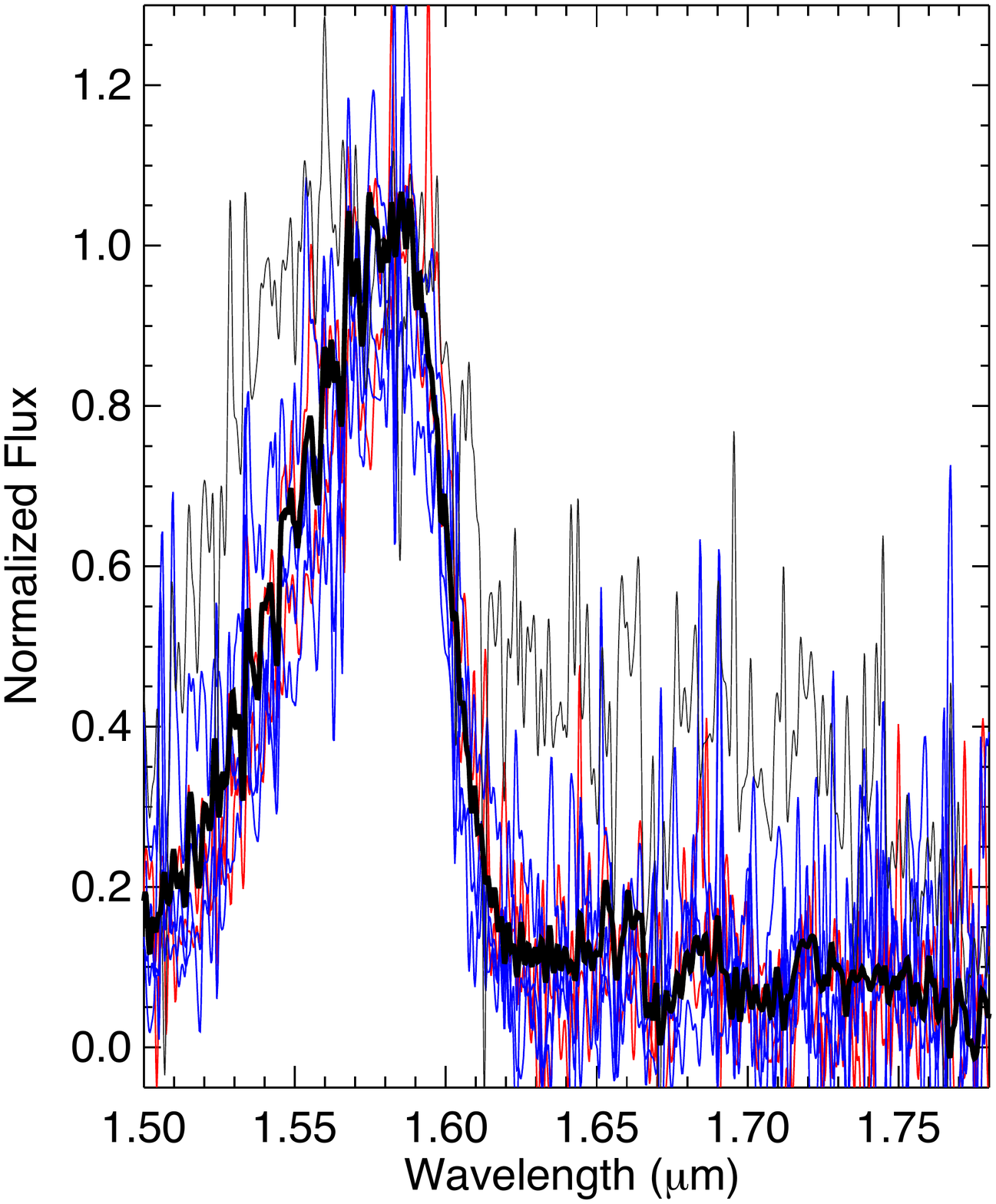}
\caption[NIRSPEC \textit{Y}- and \textit{H}-band Comparison: T8 /T8.5 dwarfs to Wolf 1130C]{NIRSPEC \textit{Y}- and \textit{H}-band spectra of the T8 and T8.5 dwarfs in this sample (see Figure \ref{fig:allresults} for details) compared to the\textit{Y}- and \textit{H}-band  spectra of the metal-poor, sdT8 Wolf 1130C (gray; \citetalias{Mace2013b}). The \textit{Y} band of Wolf 1130C is significantly broadened toward the blue end of the band and Wolf 1130C's \textit{H} band features both a broader peak and decreased methane absorption relative to the flux-normalized T8/T8.5 spectra in our sample.}
\label{fig:allvsWolf}
\end{figure}

The sample is comprised of one T6 dwarf, three T7/T7.5 dwarfs, seven T8/T8.5 dwarfs, and two T9 dwarfs. In Figure \ref{fig:allresults} we present the \textit{Y}- and \textit{H}-band NIRSPEC spectra of the color outliers in our sample, plotted together by spectral type. In order to detect morphological variations on a per band basis, each \textit{Y}- and \textit{H}-band spectrum is normalized to the flux peak in that band, as detailed in Section \ref{sec:data_v_model}. Spectra are colored according to their \textit{$J-H$} color (i.e. targets with a ``red" \textit{$J-H$} color are plotted in red, and targets with a ``blue" \textit{$J-H$} color are plotted in blue).  Medium-resolution spectral standards observed as part of the NIRSPEC Brown Dwarf Spectroscopic Survey\footnote{\footnotesize BDSS spectra are available at \url{http://bdssarchive.org} or by request.} (BDSS; e.g. \citealt{McLean2003}) are also plotted for comparison.  Except for the \textit{H}-band spectrum of the T8 standard, presented in \citet{McLean2003}, the \textit{Y}- and \textit{H}-band NIRSPEC  BDSS spectra of the standards are presented here for the first time (see Table \ref{table:observations}). All spectral standards were reduced in the same manner as described in Section \ref{sec:data_reduction}. Before plotting, all data (targets and standards) were smoothed using a Gaussian profile with a width of 3 pixels. Figure \ref{fig:allvsWolf} compares the peak-normalized T8/T8.5 spectra with the spectrum of the extremely metal-poor, sdT8 Wolf 1130C presented in \citetalias{Mace2013b}.

Somewhat surprisingly, visual comparison of the targets with each other and with Wolf 1130C suggests that the spectral morphologies of the dwarfs in this sample are more consistent with spectral standards than Wolf 1130C. For the T8/T8.5 objects, which comprise seven of the thirteen objects in the sample, there is \textit{no clear delineation} between the unusually red and blue objects in the sample. Instead, the targets tend to cluster around the spectral standard. The broad similarities among the T8/T8.5 targets are particularly evident in the \textit{Y} band, where the flux of the peak-normalized Wolf 1130C is significantly broadened toward the blue end of the band, indicative of subsolar metallicity/high gravity, whereas the targets in this sample all have much narrower and redder flux peaks. Why is the sample so homogeneous, and why do none of the color outliers share the same morphology as Wolf 1130C? We examine potential explanations for this homogeneity in Section \ref{sec:discussion}.

Though fairly homogeneous overall, the individual targets in the sample are not identical. For example, one of the two T9 objects in Figure \ref{fig:allresults} is visually separated from the other color outlier and the spectral standard in the \textit{Y} band. In Section \ref{sec:analysis} we take a closer look at the individual objects in the sample and use model and spectral standard comparison to investigate the variations among the targets.

\section{Analysis}\label{sec:analysis}
In the following section, we investigate the range of model parameter space probed by the targets in the sample. We also highlight and discuss the model fits and known properties of four objects that visually deviate from their respective spectral standard. This section begins with a brief discussion of the four publicly available atmospheric models employed in this analysis.

\subsection{A Comparison of the Models}\label{sec:model_comparison}
With temperatures $<$ 1000 K, the latest-type T dwarfs pose a challenge for atmospheric modeling due to the complex molecular chemistries, clouds, and disequilibrium processes present in their atmospheres (for a detailed review of brown dwarf atmospheric model development and construction see \citealt{MarleyRobinson2015}). Nevertheless, several teams have made significant strides toward modeling the atmospheres of these cool objects. We consider four sets of model atmosphere grids in this analysis: the cloudy CIFIST 2011 version of the BT-Settl models (henceforth BT-Settl models; \citealt{Allard2011, Allard2012}), the cloud-free models of \citet{Burrows2006} (henceforth Burrows models), the sulfide cloud models of \citet{Morley2012} (henceforth Morley models), and the cloud-free models of \citet{Saumon2012} (henceforth Saumon models). While the publicly available BT-Settl models provide nonsolar metallicity grids, the complete nonsolar set only extends down to \teff\ =1000 K. subsolar metallicity spectra for limited \logg\ values down to \teff\ = 800 K are also available, but the limited nonsolar metallicity spectra were not considered in this analysis. Though the Burrows model grid is lower resolution than the NIRSPEC data and only reaches a 
\clearpage
\begin{deluxetable}{cccccc}
\tablecolumns{6}
\tablecaption{Model Comparison}
\tabletypesize{\small}
\tablewidth{0pt}

\tablehead{
\colhead{Model Short Name} &
\colhead{Reference} &
\colhead{\teff\ Range} &
\colhead{\logg\ (cgs) Range}&
\colhead{Clouds?}&
\colhead{[Fe/H]}}

\startdata
BT-Settl&	Allard et al. 2011, 2012 & 400-7000 K	&	3.5-5	&	yes&	0.0\tablenotemark{a}\\
Burrows	&	Burrows et al. 2006 & 700-2300 K	&	4.5-5.5	&	no\tablenotemark{b} &    -0.5, 0.0, +0.5\tablenotemark{c}\\
Morley	&   Morley et al. 2012	&   400-1300 K	&	4-5.5	&	yes&	0.0\\
Saumon	& Saumon et al. 2012	& 300-1500 K	&	3.75-5.5&	no &	0.0\\
\enddata
\tablenotetext{a}{\small Nonsolar [Fe/H] CIFIST 2011 BT-Settl models are available, but do not cover the the entire range of temperatures considered in this analysis and were thus not included in this work.}
\tablenotetext{b}{\small Cloudy Burrows models are available, but we only consider the clear models in this analysis.}
\tablenotetext{c}{\small At \teff\ = 700 K, [Fe/H] = +0.5 dex is only available for the \logg\ = 5.0 dex case.} 
\label{tab:model_comparison}
\end{deluxetable}
\clearpage

\noindent
minimum temperature of \teff\ = 700 K, it is the only publicly available model set that includes both subsolar ([Fe/H] = -0.5 dex) and supersolar ([Fe/H] = +0.5 dex) metallicities at late-T dwarf temperatures. The Morley and Saumon models only consider solar metallicities, but the two models have the advantage that their main difference is in the treatment of clouds, allowing for a direct test of the impact of clouds on the data. Table \ref{tab:model_comparison} summarizes the parameter space covered by each model used in this analysis.

\begin{figure}[]
\centering
\includegraphics[width=2.5in]{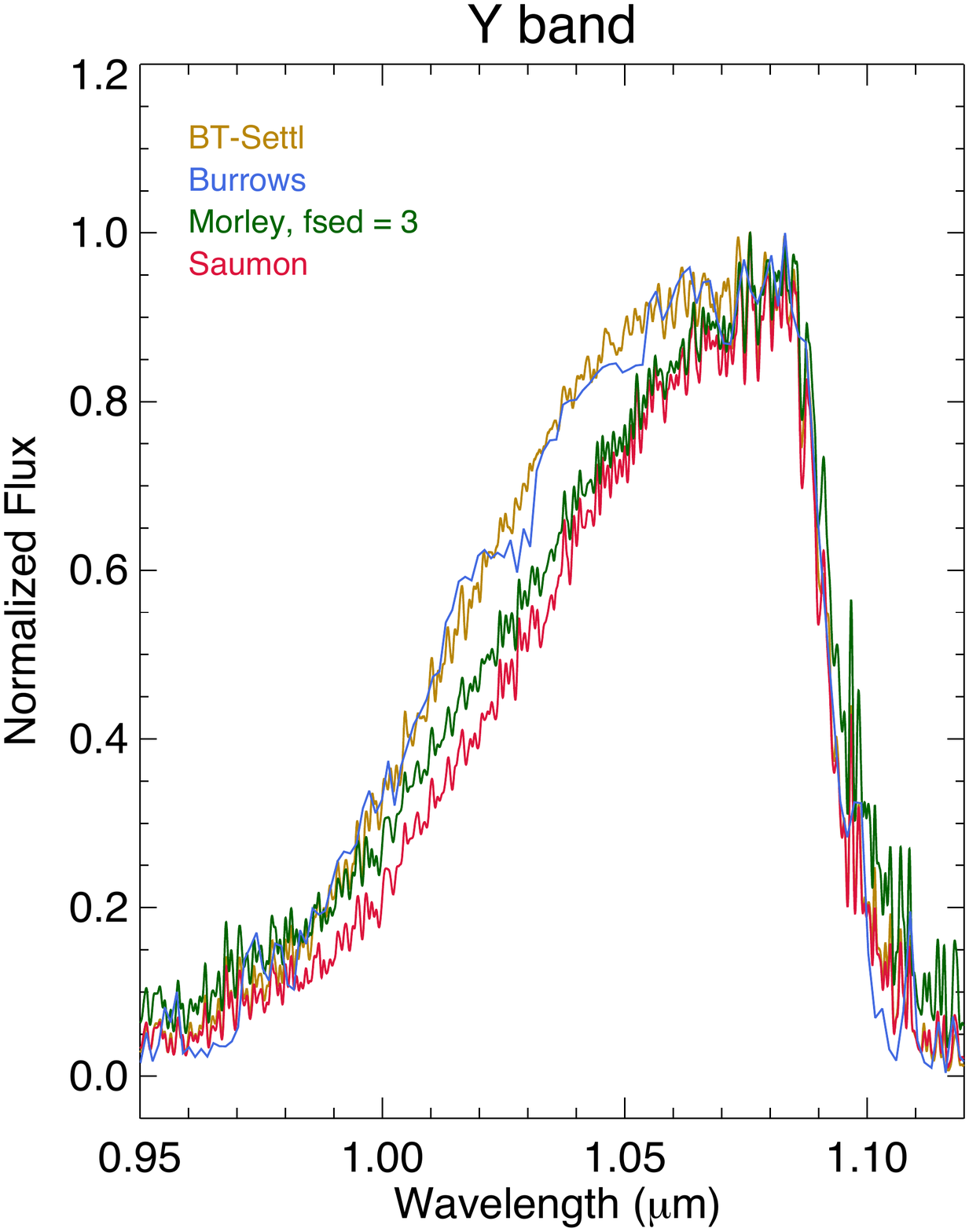}
\includegraphics[width=2.5in]{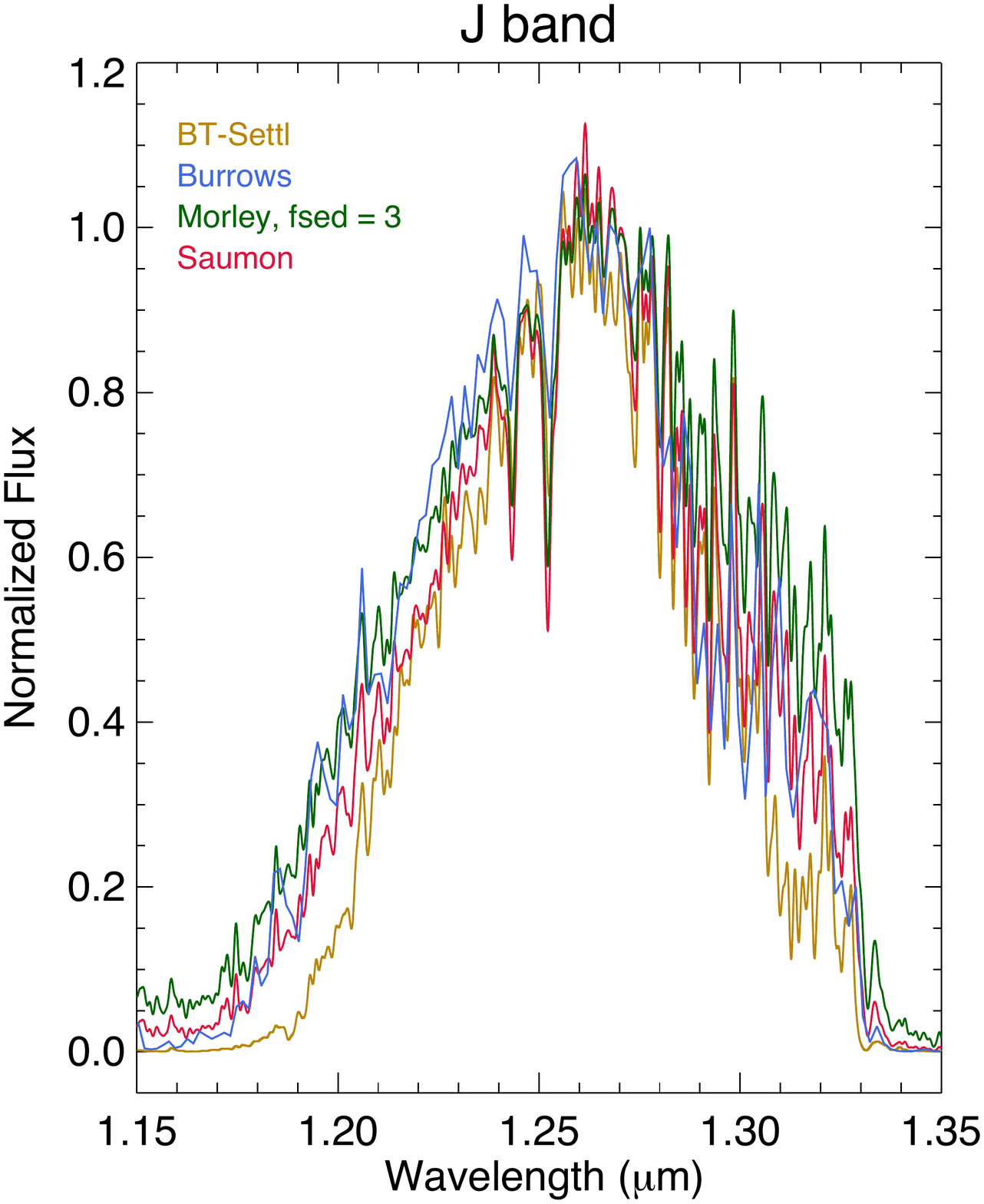}
\includegraphics[width=2.5in]{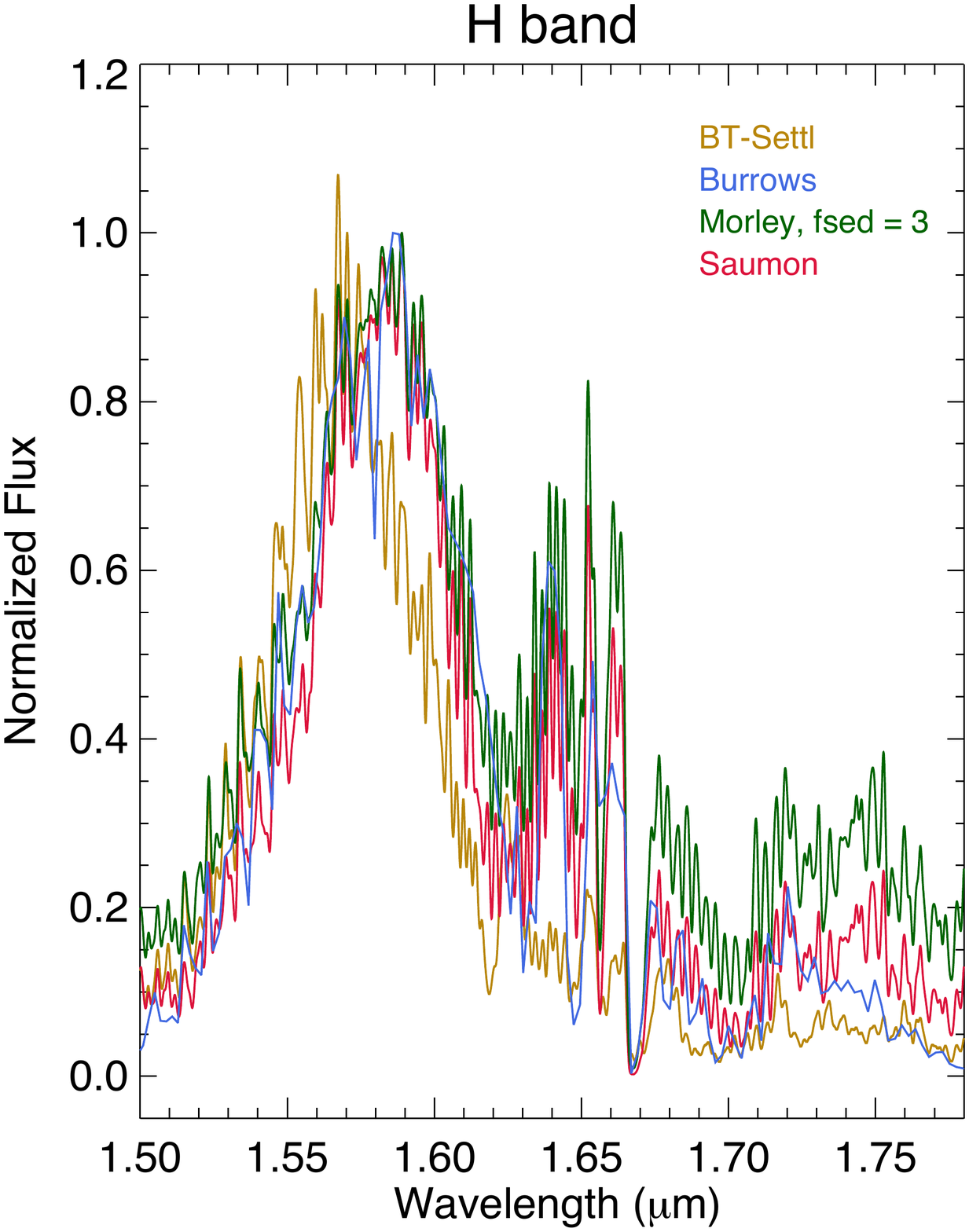}
\caption{\footnotesize Direct comparison of the BT-Settl (gold, \citealt{Allard2011, Allard2012}), Burrows (blue; \citealt{Burrows2006}), Morley (green; \citealt{Morley2012}), and Saumon (red; \citealt{Saumon2012}) atmospheric models in \textit{Y}, \textit{J}, and \textit{H} band. All models have been peak-normalized in each spectral band and assume \teff = 700 K, \logg = 4.5 dex, and solar metallicity. The BT-Settl, Saumon, and Morley models have been binned down to match the resolution of NIRSPEC. The Burrows models are lower resolution than the NIRSPEC data, and thus are plotted at their native resolution. The Morley model assumes \fsed = 3. There are clear differences in the peak-normalized spectral morphologies predicted by the models at these low temperatures, particularly in the blue-wing shape of the \textit{Y} band and in the strength of the methane dropout in the \textit{H} band. These differences are discussed further in Section \ref{sec:data_v_model} of the text.}
\label{fig:modelcomparison}
\end{figure}

Figure \ref{fig:modelcomparison} is a direct comparison of the peak-normalized, \textit{Y}-, \textit{J}-, and \textit{H}-band spectra output from the four models given identical input parameters (\teff\ = 700 K, \logg\ = 4.5 dex (cgs), [Fe/H] = 0 dex; \fsed\footnote{\footnotesize As described in \citet{Morley2012}, \fsed\ describes the sedimentation efficiency of the atmosphere. In the Morley models, \fsed\ ranges from 2 to 5, with 2 indicating small particle sizes/optically thick clouds and 5 indicating large particle sizes/optically thinner clouds.} = 3 for the Morley model spectrum). While all four models are in relatively good agreement in the \textit{J} band, the BT-Settl models predict deeper methane absorption in the \textit{H} band, and both the BT-Settl and Burrows model spectra show enhanced flux in the blue wing of the \textit{Y} band when compared to the Morley and Saumon model spectra.  While these variations are not entirely unexpected given the differences in molecular line lists, particularly for methane, and varied treatment of alkali metals and clouds among the models, these differences must be kept in mind when comparing the models with each other and with observed data. The physical implications of these differences are discussed in more detail in Section \ref{sec:data_v_model}.

\subsection{Model Fitting}\label{sec:data_v_model}

Using chi-squared minimization, we compare the target spectra to the four model grids discussed above. We normalize both data and model spectra to the flux peak in each band ($\sim$1.08 \microm\ in \textit{Y} and $\sim$1.58 \microm\ in \textit{H} band) prior to fitting by calculating the robust mean of the data within $\pm$ 0.01 \microm\ of the flux peak and dividing each data point by that value. We fit the \textit{Y}- and \textit{H}-band spectra individually for each target. Each model spectrum is smoothed to the resolution of the target data by convolving the model data with a Gaussian profile and interpolated onto the target's wavelength solution before fitting. In the case of the lower-resolution Burrows models, the spectra are only interpolated onto the NIRSPEC target's wavelength solution before fitting.

\begin{figure}
\center
\includegraphics[angle=90, width=4in]{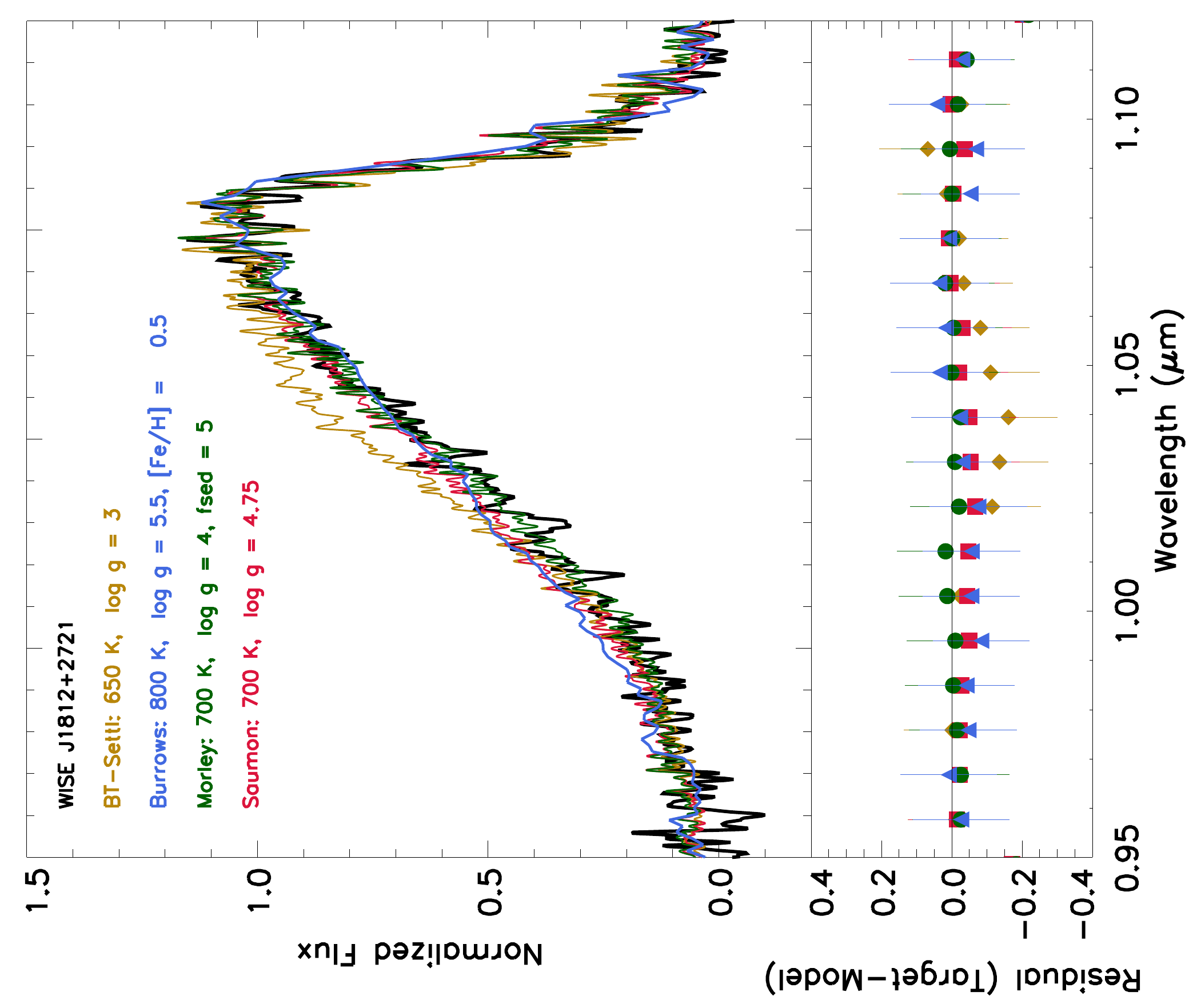}
\includegraphics[angle=90, width=4in]{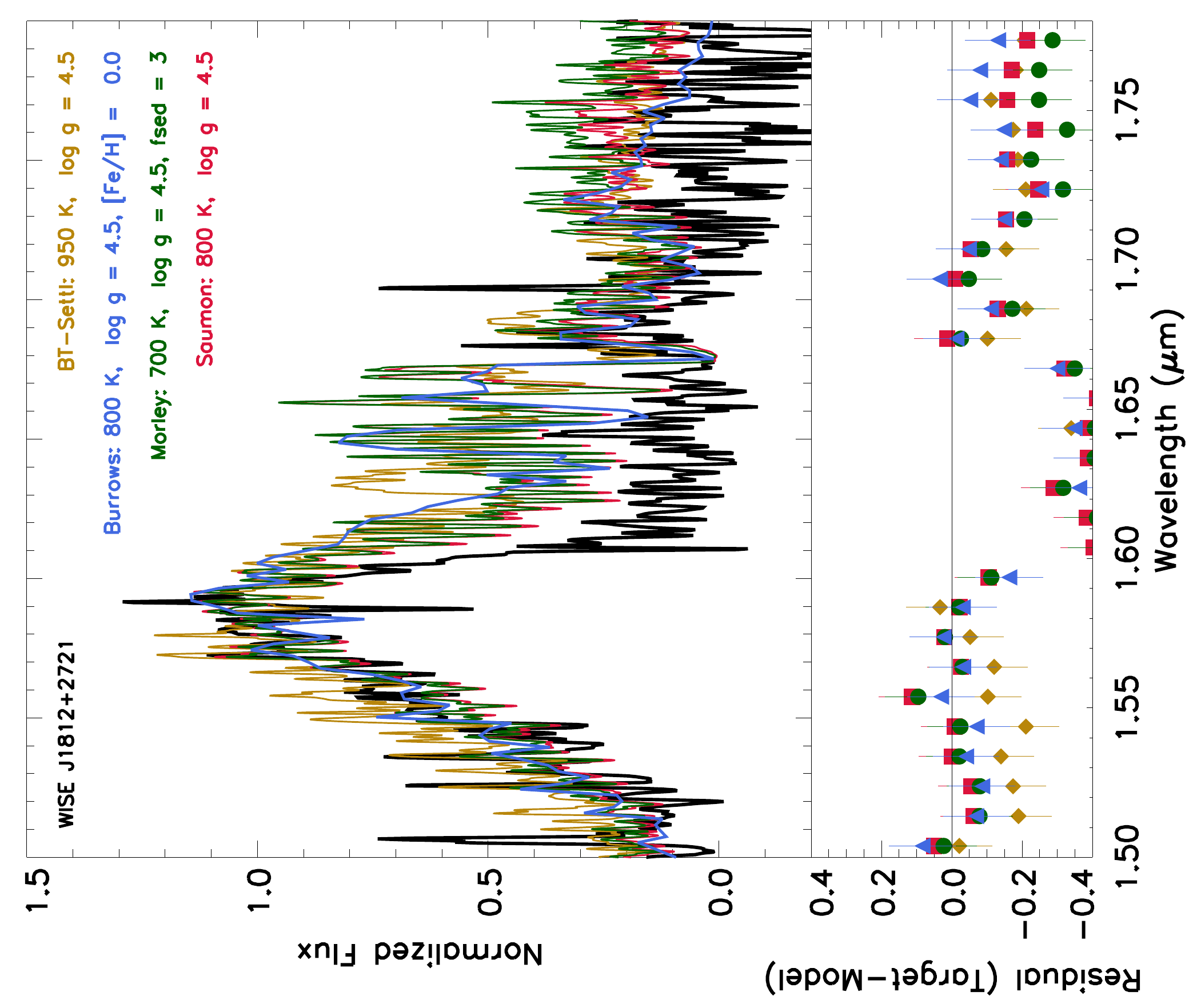}
\caption{\footnotesize Comparison of the \textit{Y}- (left) and \textit{H}-band (right) spectra for the bluest T8.5 dwarf in the sample, WISE J1812+2721, to the best-fit atmospheric models for the BT-Settl (gold), Burrows (blue), Morley (green), and Saumon (red) grids obtained while holding \teff\ fixed to fall within the 1$\sigma$ range of \teff\ as defined in \citealt{Filipazzo2015} and described in Section \ref{sec:data_v_model}. The target spectrum is plotted in black. The best-fit model spectrum and the resulting residuals, with 1$\sigma$ target error bars, for the BT-Settl (gold diamonds), Burrows (blue triangles), Morley (green circles), and Saumon (red squares) model atmosphere grids are also plotted. For the \textit{H}-band spectrum, the models are only fit to the 1.5-1.59 \microm\ region. Note that the BT-Settl model overestimates the flux of the target in the 1.05 \microm\ region of the \textit{Y} band.}
\label{fig:datamodel_filipazzo}
\end{figure}

\begin{figure}
\center
\includegraphics[angle=90, width=4.25in]{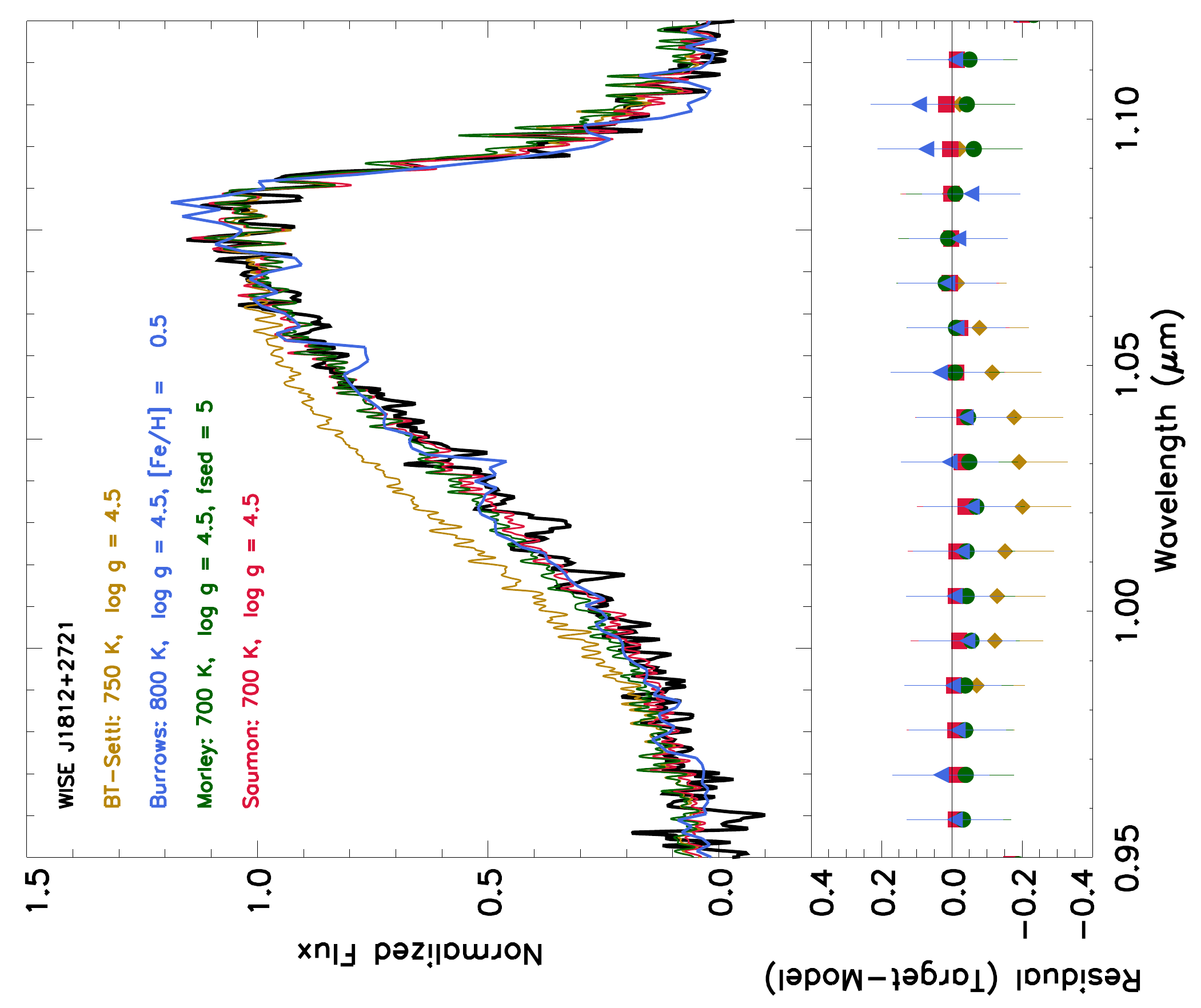}
\includegraphics[angle=90, width=4.25in]{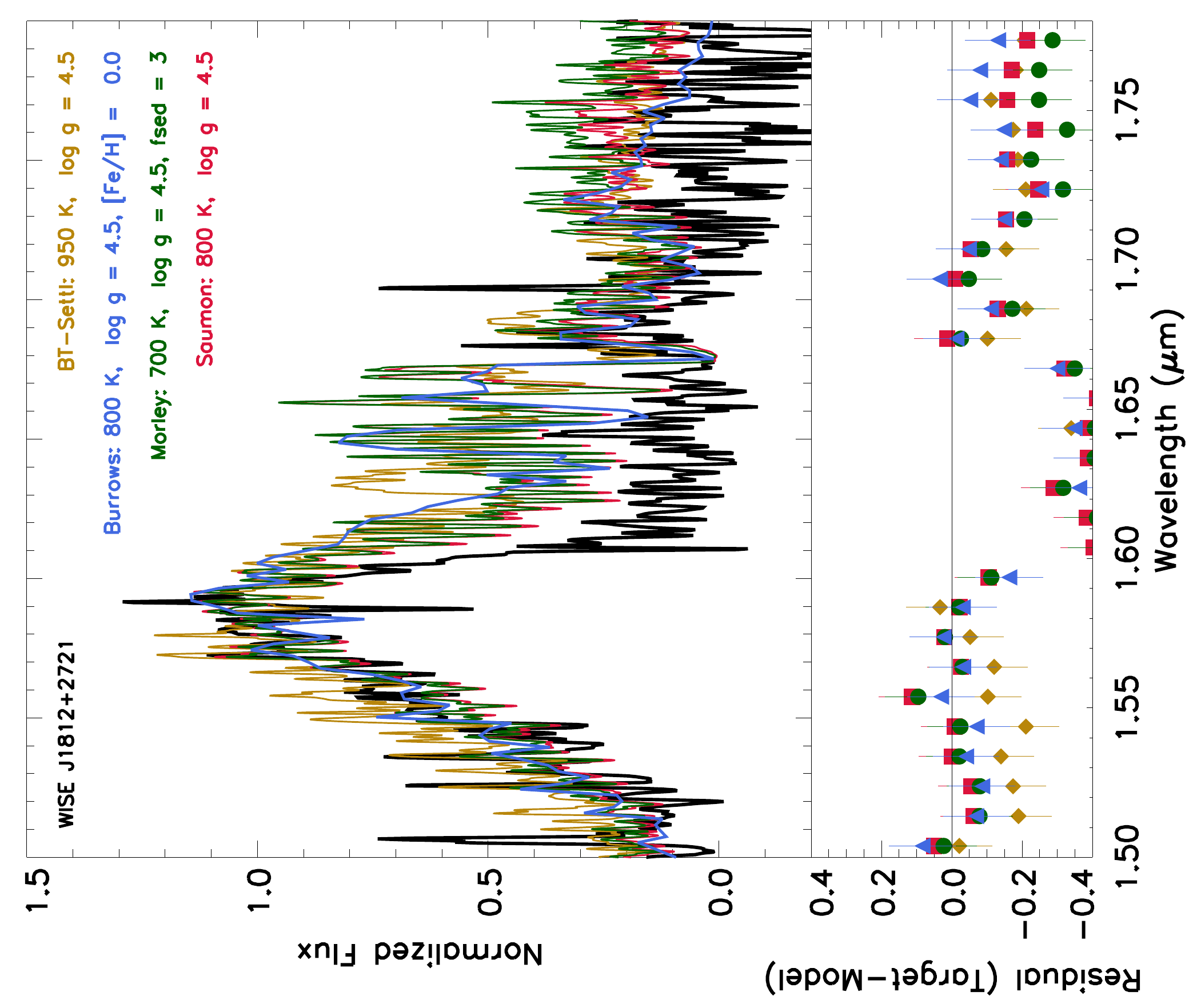}
\caption{\small Comparison of the \textit{Y}- (left) and \textit{H}-band (right) spectra for the bluest T8.5 dwarf in the sample, WISE J1812+2721, to the best-fit atmospheric models for the BT-Settl, Burrows, Morley, and Saumon grids while holding \logg\ = 4.5 dex.  See Figure \ref{fig:datamodel_filipazzo} for more details.}
\label{fig:datamodel_fixedg}
\end{figure}

To look for trends in our data set, we run a series of model fit tests holding one quantity fixed (either \teff\ or \logg) and allowing all other quantities to vary (see Tables  \ref{table:model_fits_yh_teff} and \ref{table:model_fits_yh_logg}). Holding one quantity fixed (e.g. \teff) allows us to reduce the number of free parameters probed in a given fit and to more fully explore trends in the data driven by the other parameters (e.g. \logg, [Fe/H], clouds). In the fixed effective temperature case, we hold \teff\ to fall within the 1$\sigma$ range of \teff\ for that spectral type as derived in \citet{Filipazzo2015}\footnote{\footnotesize \citet{Filipazzo2015} used parallaxes (or, in a few cases, kinematic distances) and a combination of optical to mid-infrared photometry and spectroscopy to determine bolometric luminosities (L$_{bol}$) for a sample of late-M, L, and T dwarfs. They then combined their measured L$_{bol}$ with radii determined from evolutionary models to derive a semi-empirical \teff\ versus spectral type relation for the young and field brown dwarf populations. The rms uncertainty on the \teff\ fit is 113 K.}, rounded to the nearest 50 K or 100 K to match the temperature grid spacing of the models,  and allow gravity, \fsed\ (for the Morley models), and metallicity (for the Burrows models) to vary. However, the Burrows models do not reach cold enough temperatures to be in the \teff\ range for T9 dwarfs or in the full \teff\ range for T8 dwarfs as defined in \citet{Filipazzo2015}. Thus, we fit our T8 and T9 spectra to the two coldest Burrows model grids (\teff\ $=$ 700 K, 800K). In the fixed gravity-case, we hold gravity fixed at \logg\ = 4.5 dex and allow \teff, \fsed, and metallicity (where applicable) to vary. A \logg\ of 4.5-5.0 is typical for late-type T dwarfs in the field (e.g. \citealt{Burrows1997, Knapp2004}). We then visually inspected the best-fit results in each scenario by plotting the best-fit model solution for each of the four atmospheric model grids against each target. We also calculated and examined the Target $-$ Model residuals for each model. Example best-fit results for WISE J1812+2721, a T8.5 ``blue'' target in our sample, at fixed \teff\ and gravity are shown in Figures \ref{fig:datamodel_filipazzo} and \ref{fig:datamodel_fixedg} respectively. The residuals have been binned down for clarity.

\subsubsection{\textit{Y-}band Fits for Fixed \teff\ Range}
With \teff\ fixed, we are able to investigate the impact of gravity, metallicity, and \fsed\, where applicable, on the data set (see Table \ref{table:model_fits_yh_teff}). The intra-model variation between the best-fit values for each object of a given spectral type is typically small. For example, when compared to the Burrows model grid, five of the six T8/T8.5 targets in our sample with \textit{Y}-band spectra are best-fit by the exact same model (\teff\ = 800 K, \logg\ = 5.5 dex, and [Fe/H] = +0.5 dex). The sixth target, ULAS J0950+0117, is best-fit by the \teff\ = 800 K, \logg\ = 4.5 dex, and [Fe/H] = +0.5 dex Burrows model spectrum and is discussed in more detail in  Section \ref{sec:characteristics}. The standard deviation in best-fit model parameters for the BT-Settl, Morley, and Saumon models are typically within one step\footnote{\footnotesize The typical parameter step sizes for the model grids considered in this analysis are $\Delta$\teff\ = 100 K, $\Delta$\logg\ = 0.5 dex, $\Delta$\fsed\ = 1, though some model grids have finer sampling for smaller ranges of \teff\ and \logg.} in the model grid ($\Delta$\teff\ $<$ 100 K, $\Delta$\logg\ $\le$ 0.5 dex, $\Delta$\fsed\ $<$ 1). The lack of intra-model dispersion between individual target fits of the same spectral type implies that there is no distinct trend in model fit with \textit{$J-H$} color (see Section \ref{sec:discussion} for a discussion of the homogeneity of the sample).

While the variation between the best-fit results for each spectrum given a specific model grid is typically less than a step size, there are more significant variations in best-fit results across the four model grids. As discussed in Section \ref{sec:model_comparison}, the BT-Settl models and the Burrows models predict enhanced flux in the blue wing of the \textit{Y} band when compared to the Morley and Saumon models at the same \teff\ and \logg\ values. When all four models are independently fit to the same target with \teff\ fixed, the best-fit BT-Settl models tend to have a lower gravity (for most spectra, \logg\ = 3-3.5 dex) compared to the best-fit model spectra from the other three model grids. The lower gravity serves to narrow the overall \textit{Y}-band flux peak, notably decreasing the flux in the blue wing of the \textit{Y} band as discussed in the Introduction. Even with lower best-fit gravities than the other model grids, the BT-Settl models tend to overestimate the flux for most of our targets and the NIRSPEC spectral standards in the $\sim$1.05 \microm\ region (see e.g. Figure \ref{fig:datamodel_filipazzo}). The Burrows model grid also predicts enhanced flux in the blue wing of the \textit{Y} band compared to the equivalent \teff\ and \logg\ Morley and Saumon model grids. However, the Burrows model grid does not span the full range of gravity space that the BT-Settl grid probes, only allowing 4.5 dex $<$ \logg\ $<$ 5.5 dex. Instead, to account for an excess blue-wing flux, the Burrows best-fit model spectra for nine of the twelve targets in our sample with \textit{Y}-band spectra feature a supersolar [Fe/H], which similarly decreases the blue-wing flux. It should be noted that these nine objects include both red and blue outliers, so again there is no color trend associated with this result. The three remaining targets (HIP 73786B, CFBDS J0301-1614, and WISE J1614+1739) are fit by solar-metallicity Burrows models and are discussed individually in Section \ref{sec:characteristics}. In most cases the Morley and Saumon best-fit model spectra for a given object agree to within $\Delta$\teff~$\leq$ 100 K and \logg\ $\leq$ 0.5 dex, and most targets are fit with Morley \fsed\ = 4 or 5 (see Section \ref{sec:clouds} for more discussion on clouds and variability).

\subsubsection{\textit{Y}-band Fits for Fixed \logg}\label{sec:Y_fixedlogg}
With gravity fixed to \logg\ = 4.5 dex, we are able to probe the impact of \teff, metallicity, and \fsed\ on the data set (see Table \ref{table:model_fits_yh_logg}). The fixed gravity results are very similar to those determined in the fixed \teff\ scenario. For every target, the Burrows models reproduce the same best-fit [Fe/H] value in both scenarios. For most targets, the best-fit \teff\ at fixed gravity for each model grid agrees within 100 K of the best-fit value derived for that model grid in the fixed \teff\ scenario. There are only three targets that have at least one model fit \teff\ value more than 150 K from the value obtained for that model in the fixed \teff\ scenario. These exceptions are CFBDS J0301-1614, HIP 73786B, and WISE J1617+1807. For CFBDS J0301-1614 and HIP 73786B, the Morley \teff\ values vary by 300 K and 350 K, respectively, and the \fsed\ values also vary by 3 and 2 steps, respectively. Both of these objects stand out in the fixed \teff\ case as well and are discussed in more detail below. For WISE J1617+1807 the Burrows temperatures differ by 200 K and the Morley temperatures differ by 250 K. WISE J1617+1807 is well matched by the T8 spectral standard (see Appendix \ref{sec:appendix}), showing no strong evidence for spectral peculiarity. WISE 1617+1807 highlights the need for caution when assigning physical properties to individual targets based on a single model fit over a narrow wavelength region. We measure and provide individual fit values in Tables  \ref{table:model_fits_yh_teff} and \ref{table:model_fits_yh_logg} to identify trends and outliers, but do not assign specific \teff, \logg, and metallicity values to individual targets.  

\begin{deluxetable}{ccllccccccc}
\tablecolumns{11}
\tabletypesize{\scriptsize}
\tablewidth{0pt}
\tablecaption{\textit{Y}- and \textit{H}-band Model Fits: Fixed \teff}

\tablehead{
\colhead{} &
\colhead{} &
\colhead{} &
\multicolumn{4}{c}{\textit{Y} band} &
\multicolumn{4}{c}{\textit{H} band}
\\
\colhead{Short Name} &
\colhead{SpT}&
\colhead{Model \tablenotemark{a, b}}&
\colhead{\teff} &
\colhead{\logg}&
\colhead{[Fe/H]}&
\colhead{\fsed}&
\colhead{\teff} &
\colhead{\logg}&
\colhead{[Fe/H]}&
\colhead{\fsed}
}

\startdata
WISE J0005+3737	&	T9	&		&		&		&		&		&		&		&		&		\\
	&		&	BT-Settl (fixed \teff\ range)	&	550	&	3	&	-	&	-	&	600	&	5	&	-	&	-	\\
	&		&	Burrows (fixed \teff\ range)	&	800\tablenotemark{c}	&	4.5	&	0.5	&	-	&	700\tablenotemark{c}	&	4.5	&	0	&	-	\\
	&		&	Morley (fixed \teff\ range)	&	450	&	4	&	-	&	4	&	600	&	5.5	&	-	&	4	\\
	&		&	Saumon (fixed \teff\ range)	&	450	&	4.25	&	-	&	-	&	600	&	3	&	-	&	-	\\
ULAS J0139+0048	&	T7.5	&		&		&		&		&		&		&		&		&		\\
	&		&	BT-Settl (fixed \teff\ range)	&	800	&	3	&	-	&	-	&	950	&	5.5	&	-	&	-	\\
	&		&	Burrows (fixed \teff\ range)	&	900	&	4.5	&	0.5	&	-	&	900	&	4.5	&	-0.5	&	-	\\
	&		&	Morley (fixed \teff\ range)	&	800	&	4	&	-	&	5	&	900	&	5.5	&	-	&	4	\\
	&		&	Saumon (fixed \teff\ range)	&	800	&	3	&	-	&	-	&	950	&	3.75	&	-	&	-	\\
CFBDS J0301-1614	&	T7p	&		&		&		&		&		&		&		&		&		\\
	&		&	BT-Settl (fixed \teff\ range)	&	700	&	3.5	&	-	&	-	&	700	&	5	&	-	&	-	\\
	&		&	Burrows (fixed \teff\ range)	&	700	&	5	&	0	&	-	&	700	&	4.5	&	0	&	-	\\
	&		&	Morley (fixed \teff\ range)	&	700	&	4	&	-	&	5	&	700	&	4.5	&	-	&	5	\\
	&		&	Saumon (fixed \teff\ range)	&	700	&	4.5	&	-	&	-	&	700	&	5.5	&	-	&	-	\\
WISE J0540+4832	&	T8.5	&		&		&		&		&		&		&		&		&		\\
	&		&	BT-Settl (fixed \teff\ range)	&	800	&	3.5	&	-	&	-	&	800	&	3	&	-	&	-	\\
	&		&	Burrows (fixed \teff\ range)	&	800\tablenotemark{c}	&	5.5	&	0.5	&	-	&	800\tablenotemark{c}	&	5.5	&	0.5	&	-	\\
	&		&	Morley (fixed \teff\ range)	&	700	&	4.5	&	-	&	5	&	700	&	5	&	-	&	5	\\
	&		&	Saumon (fixed \teff\ range)	&	750	&	4.75	&	-	&	-	&	800	&	5	&	-	&	-	\\
WISE J0759-4904	&	T8	&		&		&		&		&		&		&		&		&		\\
	&		&	BT-Settl (fixed \teff\ range)	&	800	&	3.5	&	-	&	-	&	N/A	&	N/A	&	N/A	&	N/A	\\
	&		&	Burrows (fixed \teff\ range)	&	800\tablenotemark{c}	&	5.5	&	0.5	&	-	&	N/A	&	N/A	&	N/A	&	N/A	\\
	&		&	Morley (fixed \teff\ range)	&	700	&	4	&	-	&	3	&	N/A	&	N/A	&	N/A	&	N/A	\\
	&		&	Saumon (fixed \teff\ range)	&	800	&	4.75	&	-	&	-	&	N/A	&	N/A	&	N/A	&	N/A	\\
CFBDS J0922+1527	&	T7	&		&		&		&		&		&		&		&		&		\\
	&		&	BT-Settl (fixed \teff\ range)	&	850	&	3.5	&	-	&	-	&	950	&	3.5	&	-	&	-	\\
	&		&	Burrows (fixed \teff\ range)	&	900	&	5.5	&	0.5	&	-	&	900	&	5.5	&	0	&	-	\\
	&		&	Morley (fixed \teff\ range)	&	800	&	4.5	&	-	&	3	&	900	&	5	&	-	&	4	\\
	&		&	Saumon (fixed \teff\ range)	&	900	&	5	&	-	&	-	&	950	&	5	&	-	&	-	\\
ULAS J0950+0117	&	T8p	&		&		&		&		&		&		&		&		&		\\
	&		&	BT-Settl (fixed \teff\ range)	&	550	&	4	&	-	&	-	&	800	&	3	&	-	&	-	\\
	&		&	Burrows (fixed \teff\ range)	&	800\tablenotemark{c}	&	4.5	&	0.5	&	-	&	800\tablenotemark{c}	&	5.5	&	0	&	-	\\
	&		&	Morley (fixed \teff\ range)	&	550	&	4	&	-	&	5	&	800	&	5	&	-	&	5	\\
	&		&	Saumon (fixed \teff\ range)	&	550	&	4	&	-	&	-	&	800	&	5	&	-	&	-	\\
ULAS J1017+0118	&	T8p	&		&		&		&		&		&		&		&		&		\\
	&		&	BT-Settl (fixed \teff\ range)	&	N/A	&	N/A	&	N/A	&	N/A	&	800	&	3.5	&	-	&	-	\\
	&		&	Burrows (fixed \teff\ range)	&	N/A	&	N/A	&	N/A	&	N/A	&	800\tablenotemark{c}	&	4.5	&	-0.5	&	-	\\
	&		&	Morley (fixed \teff\ range)	&	N/A	&	N/A	&	N/A	&	N/A	&	800	&	5	&	-	&	4	\\
	&		&	Saumon (fixed \teff\ range)	&	N/A	&	N/A	&	N/A	&	N/A	&	800	&	5.5	&	-	&	-	\\
HIP 73786B	&	T6p	&		&		&		&		&		&		&		&		&		\\
	&		&	BT-Settl (fixed \teff\ range)	&	950	&	5.5	&	-	&	-	&	1100	&	5	&	-	&	-	\\
	&		&	Burrows (fixed \teff\ range)	&	1000	&	5.5	&	0	&	-	&	1100	&	5.5	&	0	&	-	\\
	&		&	Morley (fixed \teff\ range)	&	900	&	5.5	&	-	&	4	&	1000	&	5.5	&	-	&	4	\\
	&		&	Saumon (fixed \teff\ range)	&	900	&	5.5	&	-	&	-	&	1100	&	5.5	&	-	&	-	\\
WISE J1614+1739	&	T9	&		&		&		&		&		&		&		&		&		\\
	&		&	BT-Settl (fixed \teff\ range)	&	550	&	3	&	-	&	-	&	N/A	&	N/A	&	N/A	&	N/A	\\
	&		&	Burrows (fixed \teff\ range)	&	700\tablenotemark{c}	&	4.5	&	0	&	-	&	N/A	&	N/A	&	N/A	&	N/A	\\
	&		&	Morley (fixed \teff\ range)	&	600	&	5	&	-	&	5	&	N/A	&	N/A	&	N/A	&	N/A	\\
	&		&	Saumon (fixed \teff\ range)	&	550	&	3	&	-	&	-	&	N/A	&	N/A	&	N/A	&	N/A	\\
WISE J1617+1807	&	T8	&		&		&		&		&		&		&		&		&		\\
	&		&	BT-Settl (fixed \teff\ range)	&	800	&	4.5	&	-	&	-	&	800	&	3	&	-	&	-	\\
	&		&	Burrows (fixed \teff\ range)	&	800\tablenotemark{c}	&	5.5	&	0.5	&	-	&	800\tablenotemark{c}	&	5	&	0	&	-	\\
	&		&	Morley (fixed \teff\ range)	&	550	&	5	&	-	&	5	&	800	&	5	&	-	&	5	\\
	&		&	Saumon (fixed \teff\ range)	&	800	&	5.5	&	-	&	-	&	800	&	5.5	&	-	&	-	\\
WISE J1812+2721	&	T8.5	&		&		&		&		&		&		&		&		&		\\
	&		&	BT-Settl (fixed \teff\ range)	&	650	&	3	&	-	&	-	&	800	&	3	&	-	&	-	\\
	&		&	Burrows (fixed \teff\ range)	&	800\tablenotemark{c}	&	5.5	&	0.5	&	-	&	800\tablenotemark{c}	&	5.5	&	0.5	&	-	\\
	&		&	Morley (fixed \teff\ range)	&	700	&	4	&	-	&	5	&	700	&	5	&	-	&	5	\\
	&		&	Saumon (fixed \teff\ range)	&	700	&	4.75	&	-	&	-	&	800	&	5	&	-	&	-	\\
Wolf 940B	&	T8.5	&		&		&		&		&		&		&		&		&		\\
	&		&	BT-Settl (fixed \teff\ range)	&	800	&	3.5	&	-	&	-	&	800	&	3	&	-	&	-	\\
	&		&	Burrows (fixed \teff\ range)	&	800\tablenotemark{c}	&	5.5	&	0.5	&	-	&	800\tablenotemark{c}	&	5.5	&	0.5	&	-	\\
	&		&	Morley (fixed \teff\ range)	&	700	&	4.5	&	-	&	5	&	700	&	5	&	-	&	4	\\
	&		&	Saumon (fixed \teff\ range)	&	600	&	4.75	&	-	&	-	&	800	&	5.5	&	-	&	-	\\
\enddata
\tablenotetext{a}{Model Citations: BT-Settl -- \citet{Allard2011,Allard2012}, Burrows -- \citet{Burrows2006}, Morley -- \citet{Morley2012}, Saumon -- \citet{Saumon2012} }	
\tablenotetext{b}{``Fixed \teff\ range'' implies the best fit is obtained for each model while holding \teff\ fixed to fall within the 1$\sigma$ range of \teff\ for a given spectral type (rounded to the nearest 50 K or 100 K to match the temperature grid spacing of the models) as defined in Filippazzo et al. (2015)\nocite{Filipazzo2015}. All other parameters (\logg, [Fe/H], \fsed) are allowed to vary.}
\tablenotetext{c}{The Burrows models used in this analysis do not reach cold enough temperatures to be in the \teff\ range for T9 dwarfs or in the full \teff\ range for T8 dwarfs as defined in \citet{Filipazzo2015}. Thus, we fit our T8 and T9 spectra to the two coldest Burrows model grids (\teff\ = 700 K, 800 K).}
\label{table:model_fits_yh_teff}
\end{deluxetable}	

\begin{deluxetable}{ccllccccccc}
\tablecolumns{11}
\tabletypesize{\scriptsize}
\tablewidth{0pt}
\tablecaption{\textit{Y}- and \textit{H}-band Model Fits: Fixed \logg}

\tablehead{
\colhead{} &
\colhead{} &
\colhead{} &
\multicolumn{4}{c}{\textit{Y} band} &
\multicolumn{4}{c}{\textit{H} band}
\\
\colhead{Short Name} &
\colhead{SpT}&
\colhead{Model \tablenotemark{a, b}}&
\colhead{\teff} &
\colhead{\logg}&
\colhead{[Fe/H]}&
\colhead{\fsed}&
\colhead{\teff} &
\colhead{\logg}&
\colhead{[Fe/H]}&
\colhead{\fsed}
}

\startdata
WISE J0005+3737	&	T9	&		&		&		&		&		&		&		&		&		\\
	&		&	BT-Settl (fixed \logg)	&	700	&	4.5	&	-	&	-	&	500	&	4.5	&	-	&	-	\\
	&		&	Burrows (fixed \logg)	&	800	&	4.5	&	0.5	&	-	&	700	&	4.5	&	0	&	-	\\
	&		&	Morley (fixed \logg)	&	450	&	4.5	&	-	&	5	&	600	&	4.5	&	-	&	3	\\
	&		&	Saumon (fixed \logg)	&	500	&	4.5	&	-	&	-	&	700	&	4.5	&	-	&	-	\\
ULAS J0139+0048	&	T7.5	&		&		&		&		&		&		&		&		&		\\
	&		&	BT-Settl (fixed \logg)	&	750	&	4.5	&	-	&	-	&	1000	&	4.5	&	-	&	-	\\
	&		&	Burrows (fixed \logg)	&	900	&	4.5	&	0.5	&	-	&	900	&	4.5	&	-0.5	&	-	\\
	&		&	Morley (fixed \logg)	&	800	&	4.5	&	-	&	5	&	1000	&	4.5	&	-	&	2	\\
	&		&	Saumon (fixed \logg)	&	850	&	4.5	&	-	&	-	&	1150	&	4.5	&	-	&	-	\\
CFBDS J0301-1614	&	T7p	&		&		&		&		&		&		&		&		&		\\
	&		&	BT-Settl (fixed \logg)	&	750	&	4.5	&	-	&	-	&	550	&	4.5	&	-	&	-	\\
	&		&	Burrows (fixed \logg)	&	800	&	4.5	&	0	&	-	&	700	&	4.5	&	0	&	-	\\
	&		&	Morley (fixed \logg)	&	400	&	4.5	&	-	&	2	&	600	&	4.5	&	-	&	3	\\
	&		&	Saumon (fixed \logg)	&	700	&	4.5	&	-	&	-	&	700	&	4.5	&	-	&	-	\\
WISE J0540+4832	&	T8.5	&		&		&		&		&		&		&		&		&		\\
	&		&	BT-Settl (fixed \logg)	&	800	&	4.5	&	-	&	-	&	950	&	4.5	&	-	&	-	\\
	&		&	Burrows (fixed \logg)	&	900	&	4.5	&	0.5	&	-	&	800	&	4.5	&	0	&	-	\\
	&		&	Morley (fixed \logg)	&	700	&	4.5	&	-	&	5	&	700	&	4.5	&	-	&	4	\\
	&		&	Saumon (fixed \logg)	&	750	&	4.5	&	-	&	-	&	800	&	4.5	&	-	&	-	\\
WISE J0759-4904	&	T8	&		&		&		&		&		&		&		&		&		\\
	&		&	BT-Settl (fixed \logg)	&	800	&	4.5	&	-	&	-	&	N/A	&	N/A	&	N/A	&	N/A	\\
	&		&	Burrows (fixed \logg)	&	900	&	4.5	&	0.5	&	-	&	N/A	&	N/A	&	N/A	&	N/A	\\
	&		&	Morley (fixed \logg)	&	800	&	4.5	&	-	&	5	&	N/A	&	N/A	&	N/A	&	N/A	\\
	&		&	Saumon (fixed \logg)	&	800	&	4.5	&	-	&	-	&	N/A	&	N/A	&	N/A	&	N/A	\\
CFBDS J0922+1527	&	T7	&		&		&		&		&		&		&		&		&		\\
	&		&	BT-Settl (fixed \logg)	&	850	&	4.5	&	-	&	-	&	950	&	4.5	&	-	&	-	\\
	&		&	Burrows (fixed \logg)	&	1000	&	4.5	&	0.5	&	-	&	1100	&	4.5	&	0	&	-	\\
	&		&	Morley (fixed \logg)	&	800	&	4.5	&	-	&	3	&	1000	&	4.5	&	-	&	5	\\
	&		&	Saumon (fixed \logg)	&	900	&	4.5	&	-	&	-	&	1050	&	4.5	&	-	&	-	\\
ULAS J0950+0117	&	T8p	&		&		&		&		&		&		&		&		&		\\
	&		&	BT-Settl (fixed \logg)	&	550	&	4.5	&	-	&	-	&	950	&	4.5	&	-	&	-	\\
	&		&	Burrows (fixed \logg)	&	800	&	4.5	&	0.5	&	-	&	700	&	4.5	&	-0.5	&	-	\\
	&		&	Morley (fixed \logg)	&	400	&	4.5	&	-	&	5	&	800	&	4.5	&	-	&	3	\\
	&		&	Saumon (fixed \logg)	&	400	&	4.5	&	-	&	-	&	900	&	4.5	&	-	&	-	\\
ULAS J1017+0118	&	T8p	&		&		&		&		&		&		&		&		&		\\
	&		&	BT-Settl (fixed \logg)	&	N/A	&	N/A	&	N/A	&	N/A	&	950	&	4.5	&	-	&	-	\\
	&		&	Burrows (fixed \logg)	&	N/A	&	N/A	&	N/A	&	N/A	&	1000	&	4.5	&	0	&	-	\\
	&		&	Morley (fixed \logg)	&	N/A	&	N/A	&	N/A	&	N/A	&	900	&	4.5	&	-	&	5	\\
	&		&	Saumon (fixed \logg)	&	N/A	&	N/A	&	N/A	&	N/A	&	950	&	4.5	&	-	&	-	\\
HIP 73786B	&	T6p	&		&		&		&		&		&		&		&		&		\\
	&		&	BT-Settl (fixed \logg)	&	950	&	4.5	&	-	&	-	&	1050	&	4.5	&	-	&	-	\\
	&		&	Burrows (fixed \logg)	&	1100	&	4.5	&	0	&	-	&	1000	&	4.5	&	-0.5	&	-	\\
	&		&	Morley (fixed \logg)	&	550	&	4.5	&	-	&	2	&	1100	&	4.5	&	-	&	2	\\
	&		&	Saumon (fixed \logg)	&	1000	&	4.5	&	-	&	-	&	1300	&	4.5	&	-	&	-	\\
WISE J1614+1739	&	T9	&		&		&		&		&		&		&		&		&		\\
	&		&	BT-Settl (fixed \logg)	&	700	&	4.5	&	-	&	-	&	N/A	&	N/A	&	N/A	&	N/A	\\
	&		&	Burrows (fixed \logg)	&	700	&	4.5	&	0	&	-	&	N/A	&	N/A	&	N/A	&	N/A	\\
	&		&	Morley (fixed \logg)	&	550	&	4.5	&	-	&	5	&	N/A	&	N/A	&	N/A	&	N/A	\\
	&		&	Saumon (fixed \logg)	&	550	&	4.5	&	-	&	-	&	N/A	&	N/A	&	N/A	&	N/A	\\
WISE J1617+1807	&	T8	&		&		&		&		&		&		&		&		&		\\
	&		&	BT-Settl (fixed \logg)	&	800	&	4.5	&	-	&	-	&	950	&	4.5	&	-	&	-	\\
	&		&	Burrows (fixed \logg)	&	1000	&	4.5	&	0.5	&	-	&	900	&	4.5	&	0	&	-	\\
	&		&	Morley (fixed \logg)	&	800	&	4.5	&	-	&	4	&	800	&	4.5	&	-	&	4	\\
	&		&	Saumon (fixed \logg)	&	900	&	4.5	&	-	&	-	&	900	&	4.5	&	-	&	-	\\
WISE J1812+2721	&	T8.5	&		&		&		&		&		&		&		&		&		\\
	&		&	BT-Settl (fixed \logg)	&	750	&	4.5	&	-	&	-	&	950	&	4.5	&	-	&	-	\\
	&		&	Burrows (fixed \logg)	&	800	&	4.5	&	0.5	&	-	&	800	&	4.5	&	0	&	-	\\
	&		&	Morley (fixed \logg)	&	700	&	4.5	&	-	&	5	&	700	&	4.5	&	-	&	3	\\
	&		&	Saumon (fixed \logg)	&	700	&	4.5	&	-	&	-	&	800	&	4.5	&	-	&	-	\\
Wolf 940B	&	T8.5	&		&		&		&		&		&		&		&		&		\\
	&		&	BT-Settl (fixed \logg)	&	800	&	4.5	&	-	&	-	&	950	&	4.5	&	-	&	-	\\
	&		&	Burrows (fixed \logg)	&	900	&	4.5	&	0.5	&	-	&	800	&	4.5	&	0	&	-	\\
	&		&	Morley (fixed \logg)	&	700	&	4.5	&	-	&	5	&	700	&	4.5	&	-	&	3	\\
	&		&	Saumon (fixed \logg)	&	750	&	4.5	&	-	&	-	&	800	&	4.5	&	-	&	-	\\
\enddata
\tablenotetext{a}{Model Citations: BT-Settl -- \citet{Allard2011,Allard2012}, Burrows -- \citet{Burrows2006}, Morley -- \citet{Morley2012}, Saumon -- \citet{Saumon2012} }	
\tablenotetext{b}{`Fixed \logg' reports the best fit obtained for each model while holding \logg\ = 4.5 dex and allowing all other parameters to vary.}
\label{table:model_fits_yh_logg}
\end{deluxetable}	
\clearpage

\subsubsection{\textit{H}-band Fits}\label{sec:Hband}

The 1.6 \microm\ methane absorption in the \textit{H} band of the target spectra is poorly matched by all four atmospheric model grids considered in this analysis. Incomplete methane line lists in the models are likely to be the dominant cause of the lack of agreement (e.g. \citealt{Saumon2012}). To minimize the impact of a poor methane fit on the overall model fit results, we only fit the 1.5-1.59 \microm\ region of the \textit{H}-band spectrum. As illustrated in Figures \ref{fig:datamodel_filipazzo} and \ref{fig:datamodel_fixedg}, even in the 1.5-1.59 \microm\ region the best-fit atmospheric models tend to provide a poorer fit to the data than the \textit{Y}-band fits. This is reflected in a larger scatter in the best-fit intra-model \textit{H}-band results for both the fixed \teff\ and fixed \logg\ cases compared to the \textit{Y-}band results. However, as is the case for the \textit{Y} band, neither the intra-model or the inter-model scatter is associated with a trend in \textit{$J-H$} color. 

Even with the larger scatter, there are some general trends among the model fits. In the fixed \teff\ case, the best-fit BT-Settl models tend to predict a lower gravity (for most spectra, \logg\ = 3-3.5 dex) than the best-fit model spectra from the other three model grids, as they did in the \textit{Y} band. Also, the Morley and Saumon \logg\ results are on average about 0.5 dex higher than in the \textit{Y}-band fits. In the fixed \logg\ case, the BT-Settl, Morley, and Saumon best-fit results for each object are either warmer or the same temperature than the best-fit result for that object in the \textit{Y} band. The only exception to this is CFBDS J0301-1614, which is poorly fit by all model grids in the \textit{H} band and is discussed in more detail below. Constrained to a fixed \logg\ = 4.5 dex, seven of the eleven targets with \textit{H}-band spectra are fit by significantly cloudy (\fsed\ = 2-3) Morley models. This last result is likely unphysical, driven by forcing \logg\ = 4.5 dex. A lower gravity decreases the width of the flux peak in a normalized \textit{H} band, and a cloudy atmosphere compensates by increasing the width of the blue wing of the \textit{H} band (see Figure \ref{fig:morley_fsed}). When \logg\ is allowed to vary in the fixed \teff\ scenario, the \textit{H}-band Morley best-fits on average have \logg\ = 5.0 dex and \fsed\ = 4-5. 

Ultimately, comparing the \textit{Y}- and \textit{H}-band spectra of the target sample to the BT-Settl, Burrows, Morley, and Saumon models serves to illustrate the differences among the models and highlights the difficulty of modeling the atmospheres of the coldest products of the star formation process. However, the overall homogeneity of the sample is supported by the similarity of the best-fit results for each model grid. The \textit{Y}-band model fits in particular also identify four potentially unusual targets worth further discussion. These targets are addressed in the next section.

\subsection{Characteristics of Unusual Objects in the Sample}\label{sec:characteristics}
\begin{figure}
\centering
\includegraphics[width=7in]{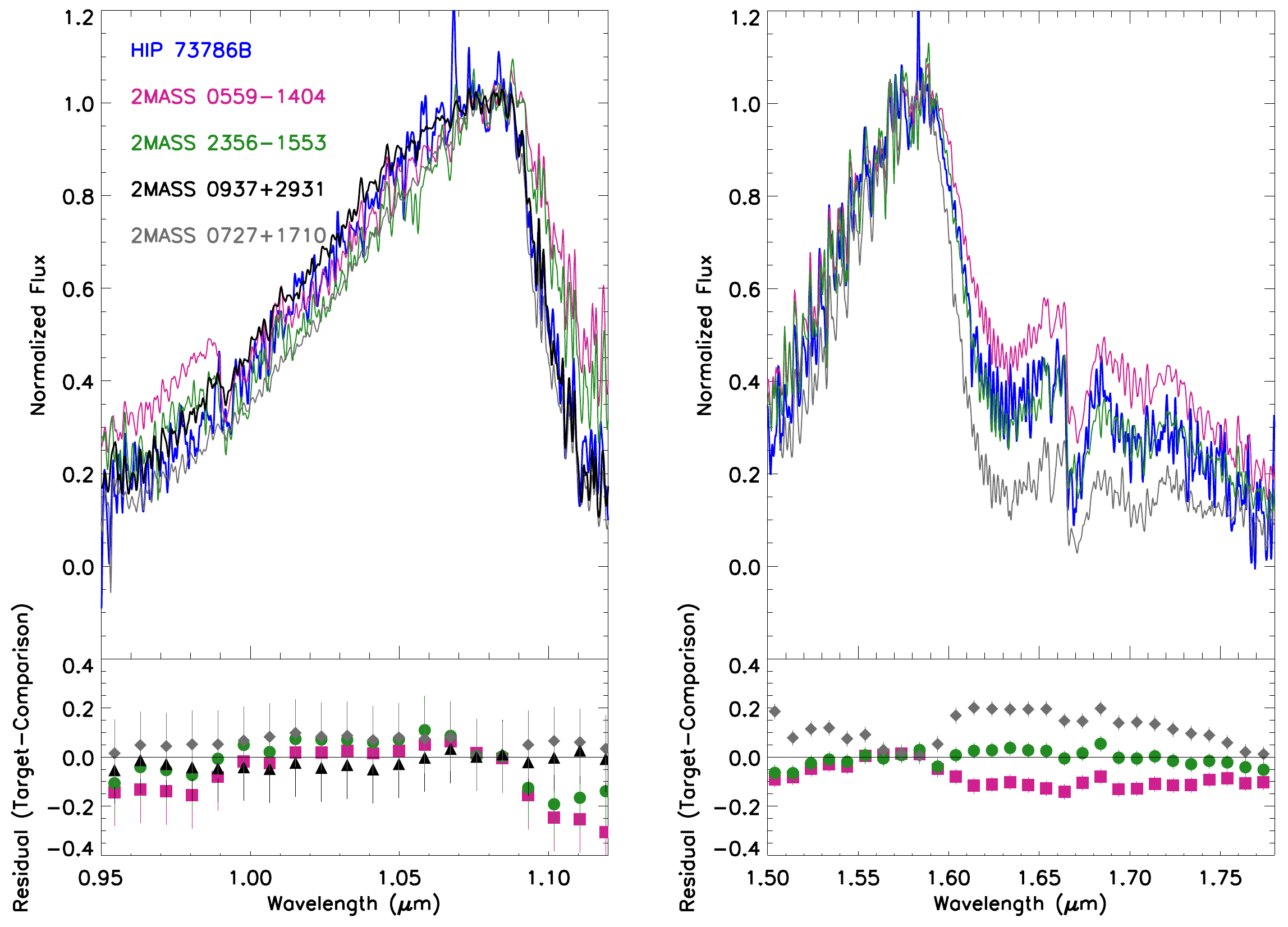}
\caption[Individual Outliers: HIP 73786B vs BDSS Brown Dwarfs]{\footnotesize Comparison of the \textit{Y}- (left) and \textit{H}-band (right) spectra of HIP 73786B (T6p; blue) to BDSS T4.5-T7 brown dwarfs with \textit{Y}- and \textit{H}-band spectra including: the T4.5 dwarf, 2MASS J05591914-1404488 (2MASS J0559-1404; \citealt{Burgasser2000b, Burgasser2006b}) plotted in pink, the T5.5 dwarf 2MASSI J2356547-155310 (2MASS J2356-1553; \citealt{Burgasser2002, Burgasser2006b}) plotted in green, the likely subsolar metallicity/high-gravity T6p dwarf,  2MASS J09373487+2931409 (2MASS J0937+2931; \citealt{Burgasser2002, Burgasser2006b}) plotted in black, and the T7 spectral standard, 2MASSI J0727182+171001 (2MASS J0727+1710; \citealt{Burgasser2002, Burgasser2006b}) plotted in gray. The target-comparison residuals show that 2MASS J0937+2931 (black triangles) is the best match for HIP 73786B in the \textit{Y} band, supporting the \citet{Burningham2014} result and further suggesting that HIP 73786B is metal-poor and/or has a high gravity, consistent with the slightly metal-poor primary. The BDSS library does not include a NIRSPEC N5 (\textit{H}-band) spectrum of 2MASS J0937+2931, but the \textit{H} band is well matched by 2MASS J2356-1553 (green circles), a typical field dwarf of a similar spectral type to HIP 73786B. The agreement between 2MASS J2356-1553 and HIP 73786B in the \textit{H} band implies that this object, while likely metal-poor/high gravity, has less extreme parameters than Wolf 1130C. The 2MASS J0559-1404 residuals are shown in pink triangles and the 2MASS J0727+1710 residuals are shown in gray diamonds.}
\label{fig:HIP73786B_bdss}
\end{figure}

\noindent
\textit{ULAS J150457.65+053800.8 (HIP 73786B):}\\ HIP 73786B (T6p) is a known wide-separation, proper motion companion to HIP 73786A (\citealt{Scholz2010, Murray2011}), which has a K8V spectral type \citep{Gray2003}. Based on WFCAM photometry from \citet{Murray2011}, HIP 73768B's \textit{$J-H$} color is classified as ``normal.'' However there is strong evidence that this object is unusual. HIP 73786B has an inferred distance of 18.6 $\pm$ 0.97 pc (\citealt{vanLeeuwen2007}) and a metallicity of [Fe/H] = -0.3 $\pm$ 0.1 dex \citep{Cenarro2007}, inferred from the properties of the primary. \citet{Burningham2014} presented a \textit{R}\midtilde120 IRTF/SpeX \citep{Rayner2003} spectrum of this object and showed that the \textit{Y}-band flux peak is enhanced and the \textit{K}-band is depressed, indicative of low metallicity/high gravity. In Figure \ref{fig:HIP73786B_bdss}, we compare the peak-normalized NIRSPEC \textit{Y}- and \textit{H}-band spectra of HIP 73786B to BDSS T dwarfs from the literature \citep{McLean2003} and to a previously unpublished (see Table \ref{table:observations}) BDSS \textit{Y}-band spectrum of the T6p dwarf 2MASS J09373487+2931409 (2MASS J0937+2931; \citealt{Burgasser2002, Burgasser2006b}). Supporting the \citet{Burningham2014} result, we find that the \textit{Y}-band NIRSPEC data is best matched by the \textit{Y}-band spectrum of 2MASS J0937+2931, which is likely metal-poor/high gravity (e.g. \citealt{Burgasser2002, Burgasser2003b, Burgasser2006b}). All four best-fit model spectra at a fixed \teff\ range are also consistent with high gravity in the \textit{Y} band and the truncated \textit{H} band, including the BT-Settl model (see Table \ref{table:model_fits_yh_teff}). Unlike Wolf 1130C, which has an enhanced \textit{H} band compared to other T8 brown dwarfs, the \textit{H}-band spectrum of HIP 73786B is reasonably well matched by the T5.5 dwarf, 2MASSI J2356547-155310 (2MASS I2356-1553, \citealt{Burgasser2002, Burgasser2006b}). The comparatively ``normal'' \textit{H}-band spectrum may explain why HIP 73786B does not have an unusually red \textit{$J-H$} color for its spectral type.

\begin{figure}
\centering
\includegraphics[width=7in]{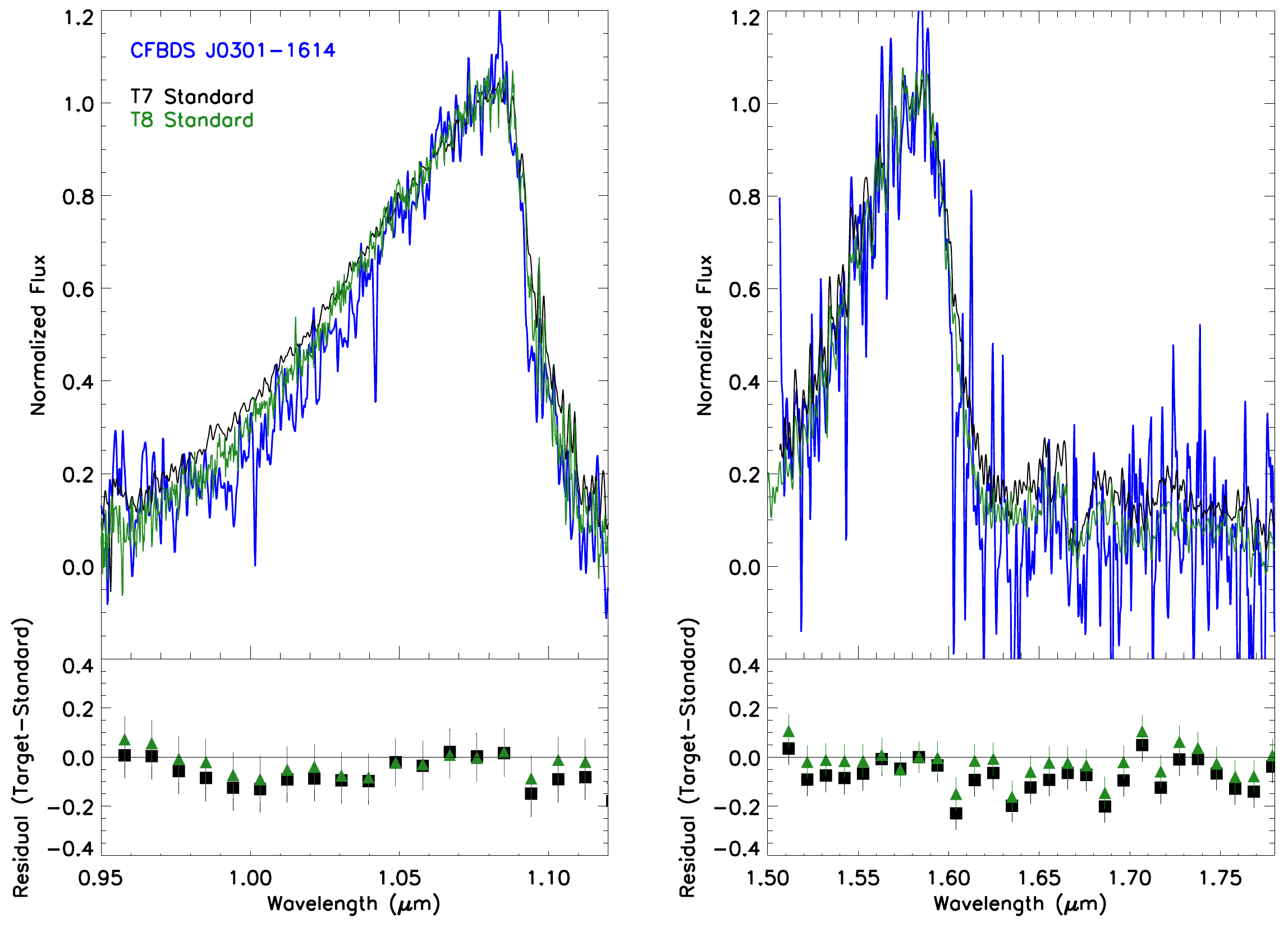}
\caption[Individual Outliers: CFBDS J0301-1614]{CFBDS J0301-1614 (T7; \citealt{Albert2011}) compared to the T7 (black) and T8 (green) spectral standards. CFBDS J0301-1614 is plotted in blue to reflect its blue \textit{$J-H$} color. Target $-$ standard residuals for the T7 standard are plotted in black squares and the Target $-$ standard residuals for the T8 standard are plotted in green triangles. In the \textit{Y} band (left), CFBDS J0301-1614 is underluminous in the blue-wing when compared to both spectral standards. In the \textit{H} band (right), the target is marginally suppressed compared to both spectral standards. Both the \textit{Y}- and \textit{H}-band results are suggestive of high metallicity/low gravity.}
\label{fig:CFBDS0301-1614}
\end{figure}

\noindent
\textit{CFBDS J030135.11-161418.0:}\\ \citet{Albert2011} present an  \textit{H}-band spectrum of CFBDS J0301-1614 and use \textit{H}-band spectral indices from \citet{Burgasser2000b} to classify this object as a T7 dwarf. Comparing CFBDS J0301-1614's red \textit{$J-K_{s}$} and \textit{$H-K_{s}$} colors to BT-Settl models, Albert et al. 2011 find that CFBDS J0301-1614 may be low gravity and/or high metallicity. In Figure \ref{fig:CFBDS0301-1614}, we compare our peak-normalized CFBDS J0301-1614 \textit{Y}- and \textit{H}-band spectra to both the T7 and T8 spectral standards. We find that CFBDS J0301-1614 is underluminous in both wings of the \textit{Y} band when compared to the T7 spectral standard. The T8 spectral standard provides a better fit to the red wing of the \textit{Y} band, but the blue wing of CFBDS J0301-1614 is still depressed relative to the T8 standard, consistent with the low-gravity/high-metallicity hypothesis of \citet{Albert2011}. In the \textit{H} band, the target is slightly better matched by the T8 standard than the T7 standard, which is potential further evidence of low gravity/high metallicity as high gravity/low metallicity enhances the \textit{H} band as seen in Wolf 1130C (e.g. Figure 3). Like the other three targets in this section, CFBDS J0301-1614's best-fit model results in the \textit{Y} band stand out from the bulk of the sample, though the model results vary. CFBDS J0301-1614 is classified as blue based on the \citetalias{Mace2013b} color criteria and does indeed show spectral signatures low gravity/high metallicity, indicative of youth.  Additional spectral coverage in \textit{J} band, to confirm its \textit{H}-band determined spectral type, and \textit{K} band, to further test for the impact of low gravity/high metallicity, would be valuable for further investigation into this peculiar object. 

\begin{figure}
\centering
\includegraphics[width=7in]{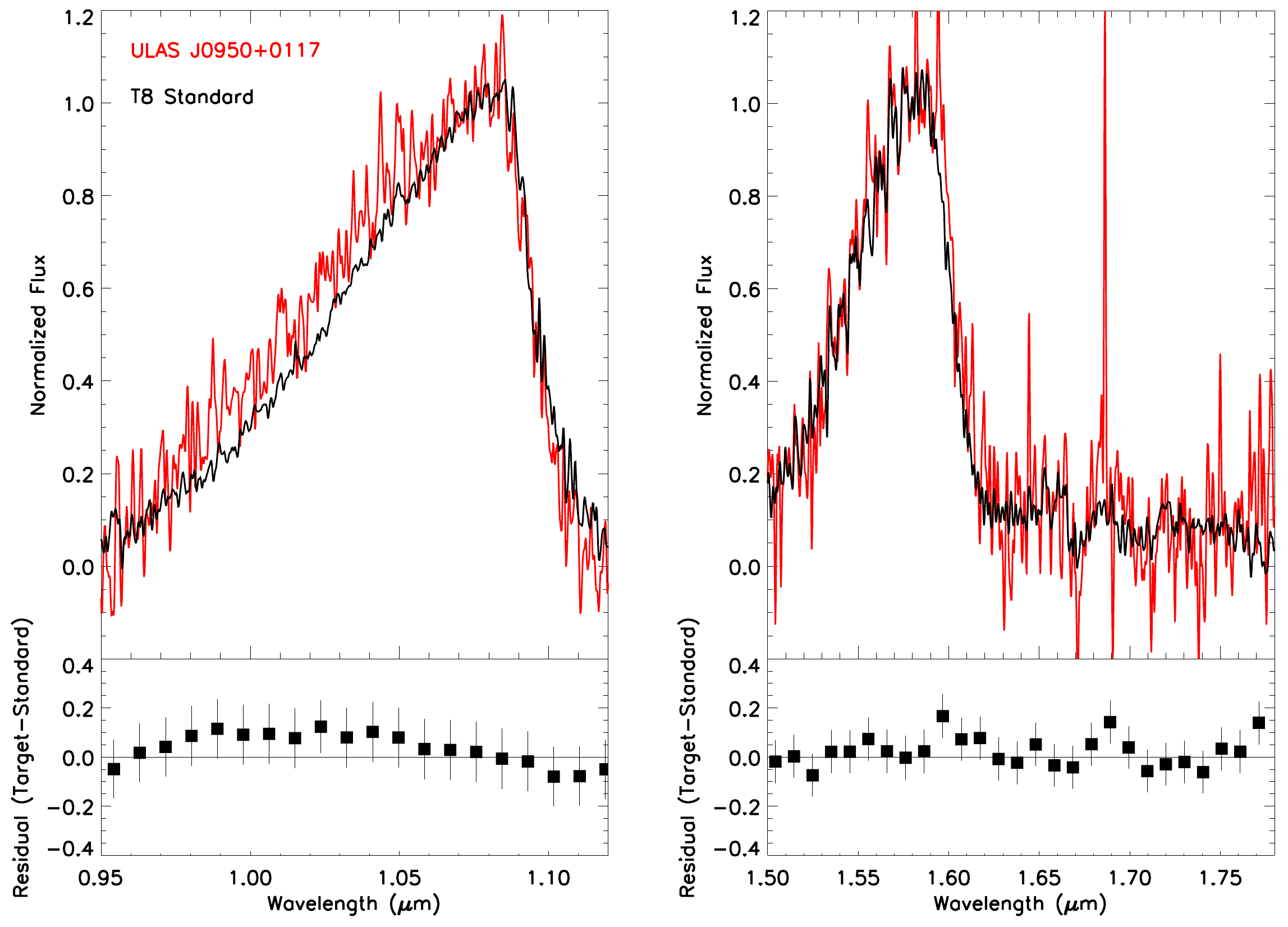}
\caption[Individual Outliers: ULAS J0950+0227]{ULAS J0950+0117 (T8p; \citealt{Luhman2012, Burningham2013}) compared to the T8 spectral standard (black). Target $-$ standard residuals are plotted in black squares. In the \textit{Y} band, ULAS J0950+0117 is overluminous in the blue wing when compared to the standard indicative of low metallicity/high gravity, though the \textit{H} band is well matched by the T8 standard.}
\label{fig:ulas0950+0117}
\end{figure}

\noindent
\textit{ULAS J095047.28+011734.3:} \\Like Wolf 1130C, ULAS J0950+0117 (T8p) was identified as a wide-separation, binary companion to a low-metallicity, M dwarf primary (LHS 6176, [Fe/H] = -0.3 $\pm$ 0.1 dex; \citealt{Luhman2012, Burningham2013}). We find that the \textit{H}-band NIRSPEC spectrum of ULAS J0950+0117 is well-fit by the T8 spectral standard, but that the blue wing of the \textit{Y} band is marginally enhanced relative to the spectral standard (see Figure \ref{fig:ulas0950+0117}), potentially consistent with low-metallicity/high-gravity. \citet{Mace2013} present a Magellan/FIRE spectrum of ULAS J0950+0117 and also find that the \textit{Y} band is slightly enhanced. They also detect a slightly enhanced \textit{H} band and suppressed \textit{K} band, though they do not classify the object as peculiar due to a low S/N. Our NIRSPEC \textit{Y}-band spectrum is poorly fit by the atmospheric models, though the truncated \textit{H}-band model fits are consistent with the other T8/T8.5 brown dwarfs in this sample. The lack of peculiarity in the \textit{H} band is unexpected given the unusually red \textit{$J-H$} color presented in \citet{Mace2013}. However, ULAS J0950+0117 has another \textit{H}-band photometric measurement from \citet{Burningham2013} that implies a normal \textit{$J-H$} color (see Table \ref{table:colors}). As discussed in Section \ref{sec:discussion}, further photometric follow-up to resolve the discrepant photometry is recommended.

\begin{figure}
\centering
\includegraphics[width=6in]{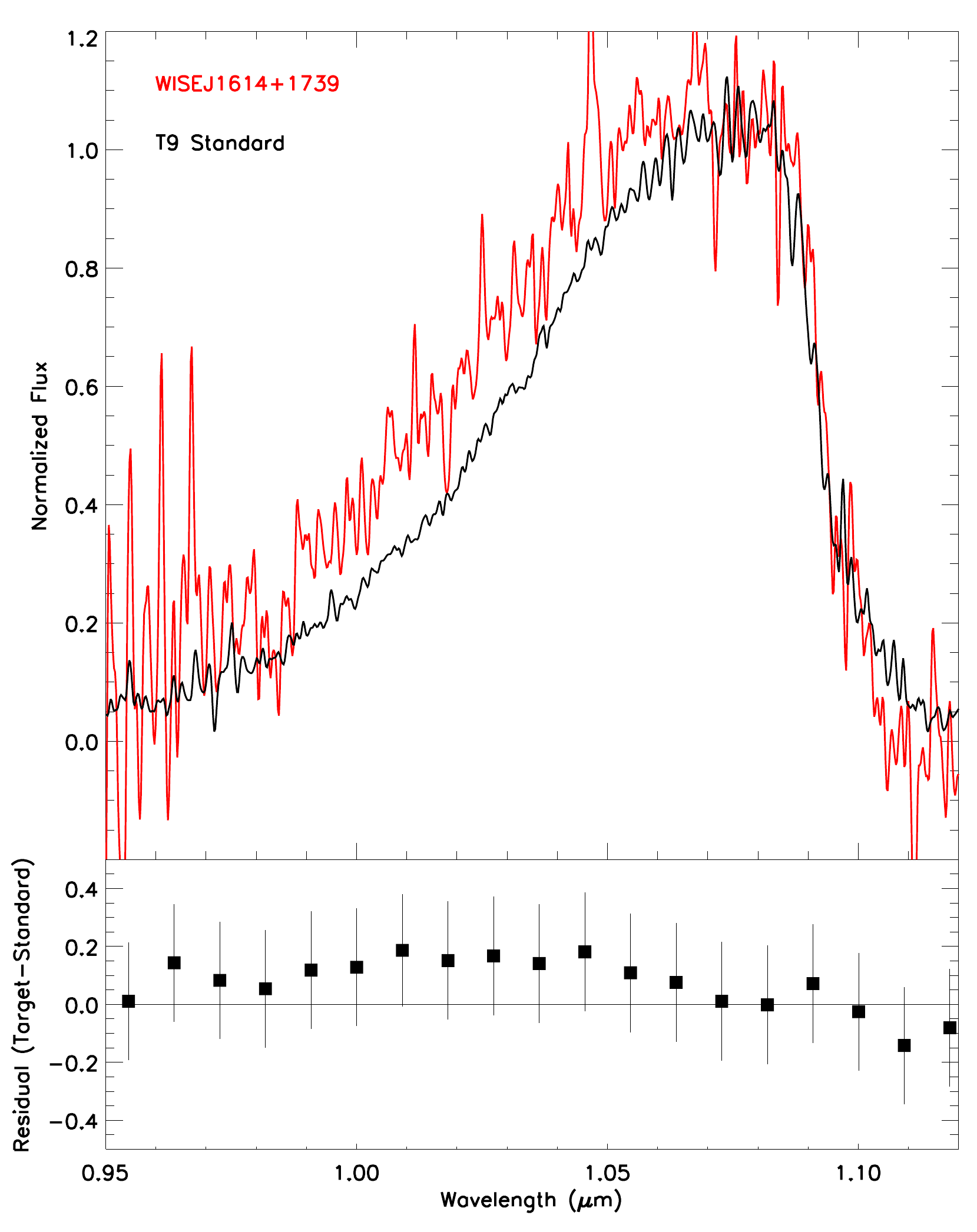}
\caption{WISE J1614+1739 (T9; \citealt{Kirkpatrick2011}) compared to the T9 spectral standard (black). Target $-$ standard residuals are plotted in black squares. We find that the \textit{Y}-band spectrum is overluminous in the blue wing when compared to the standard suggesting that this object may be metal-poor/high gravity.}
\label{fig:ulas1614+1739}
\end{figure}

\noindent
\textit{WISE J161441.46+173935.5:} \\ WISE J1614+1739 was identified as part of the follow-up of \textit{WISE} color-selected brown dwarf candidates (T9; Kirkpatrick et al. 2011\nocite{Kirkpatrick2011}). WISE J1614+1739 is not a known companion and Gelino et al. (2011)\nocite{Gelino2011} found no evidence of a close-in, substellar companion to WISE J1614+1739 in high-resolution imaging follow-up. Liu et al. (2012)\nocite{Liu2012} note that the \textit{$Y-J$} and \textit{$J-H$} band photometry of this object from \citet{Kirkpatrick2011} are unusual and they exclude it from their analysis. \citet{Leggett2013} also remark that WISE 1614+1739 stands out as red in \textit{$J-H$} color space and suggests that more photometric follow-up is required to ensure that the red color is physical, not instrumental. In spite of the large error bars on the residual plot due to the extremely faint nature (J \midtilde\ 19, see Table \ref{table:colors}) of this target, we find that WISE J1614+1739 stands out as overluminous in the blue wing of the \textit{Y} band when compared to the T9 spectral standard (see Figure \ref{fig:ulas1614+1739}), suggesting that it may be low-metallicity/high-gravity. Like ULAS J0950+0117, WISE J1614+1739 has discrepant \textit{H}-band photometry in the literature (see Table \ref{table:colors} and Section \ref{sec:discussion}). Both \textit{H}-band photometric and spectroscopic follow-up on this object would confirm whether this object has an enhanced \textit{Y} band and normal \textit{H} band like HIP 73786B or both an enhanced \textit{Y} and \textit{H} band like Wolf 1130C.

\subsection{Characteristics of other Objects in the Sample}\label{sec:characteristics_other}

Here we outline the known spectral properties of targets in our sample that do not deviate from their spectral standard. Targets are organized by spectral type. Individual standard comparison plots, including residuals, for the objects discussed in this section are available in Appendix \ref{sec:appendix}.

\noindent
\textit{CFBDS J092250+152741:} \\ CFBDS J0922+1524 was first discovered by Reyle et al. (2010)\nocite{Reyle2010}. They present an \textit{H}-band spectrum of CFBDS J0922+1524 and type it as a T7 dwarf based on the \textit{H}-band spectral indices from \citet{Burgasser2006b}. We present the first \textit{Y}-band spectrum of this object. We find that this blue object is well matched by the spectral standard in both the \textit{Y} band and the \textit{H} band. 

\clearpage
\noindent
\textit{ULAS J013939.77+004813.8:} \\ ULAS J0139+0048 (T7.5) was discovered by \citet{Chiu2008}. \citet{Chiu2008} present two sets of near-infrared photometry that classify the \textit{$J-H$} color of this object as red and blue, respectively (see Section \ref{sec:discussion} for a discussion of the photometry of the sample). Comparing ULAS 0139+0048's \textit{$H_{MKO}-ch2$} to evolutionary models based on \citet{Marley2002} and \citet{SaumonMarley2008}, \citet{Leggett2010} estimate a \teff\ $\approx 850 $ K, \logg\ $\approx 5$ dex, and [m/H] $\approx 0$ dex. We also do not find this object to be peculiar: neither spectral band shows significant variation from the spectral standards.

\noindent
\textit{WISE J075946.98-490454.0:} \\ WISE J0759-4904 was identified as part of the \textit{WISE} team follow-up of brown dwarf candidates (T8; Kirkpatrick et al. 2011\nocite{Kirkpatrick2011}). We find that this red object's \textit{Y}-band spectrum is well matched by the T8 spectral standard. 

\noindent
\textit{ULAS J101721.40+011817.9:} \\ Discovered by Burningham et al. (2008)\nocite{Burningham2008}, ULAS J1017+0118 is classified as T8p due to a reported dearth of methane absorption in its \textit{H}-band spectrum. \citet{Burningham2008} compare the spectrum to solar-metallicity BT-Settl model spectra and estimate its \teff\ to be between 750-850 K, and \logg\ between 5-5.5 dex. In contrast, \citet{Leggett2010} compare near-infrared and \textit{Spitzer} photometry of ULAS J1017+0118 to evolutionary models and suggest that the target may be low gravity (\logg\ = 4-4.5 dex) or high metallicity. Based on the \citetalias{Mace2013b} color criteria, ULAS J1017+0118 is classified as blue/young. The gravity results from this work's truncated \textit{H}-band model fits are highly model dependent and range from \logg\ = 3.5 dex (BT-Settl) to \logg\ = 5.5 dex (Saumon). Comparing to spectral standards, we find that the \textit{H}-band spectrum of ULAS J1017+0118 is slightly better matched to the T7 standard, though the T8 standard also provides a reasonable match to the target spectrum.

\noindent
\textit{WISE J161705.74+180714.1:} \\WISE J1617+1807 (T8) was identified by Burgasser et al. (2011)\nocite{Burgasser2011} as a potentially cloudy, cool (\teff\ = 600 $\pm$ 30 K), low-gravity (\logg\ = 4.0 $\pm$ 0.3 dex) T dwarf by comparison of a near-infrared spectrum from Magellan/FIRE \citep{Simcoe2010} to the Saumon \& Marley (2008)\nocite{SaumonMarley2008} atmospheric models. The blue \textit{$J-H$} color of this object supports the \citet{Burgasser2011} result. We find that the target is well matched by its spectral standard in both NIRSPEC \textit{Y} and \textit{H} bands. In general, the model fits to this object are in good agreement with the other T8/T8.5 targets in the sample, with the exception of the Morley fit in the \textit{Y}-band fixed \logg\ case, as discussed in Section \ref{sec:Y_fixedlogg}.

\noindent
\textit{WISE J054047.00+483232.4:} \\WISE J0540+4832 is a T8.5 dwarf discovered by Mace et al. (2013a)\nocite{Mace2013}. We present the first \textit{Y}- and \textit{H}-band spectra of this object. We find that this red object's spectrum is intermediate between the T8 and T9 standard in both the \textit{Y} and \textit{H} bands, as supported by its spectral type. 

\clearpage
\noindent
\textit{WISE J181210.85+272144.3:} \\ Like WISE J1617+1804, WISE J1812+2722 (T8.5:) was identified by Burgasser et al. (2011)\nocite{Burgasser2011} as a cool (\teff\ = 620 $\pm$ 30 K), low-gravity (\logg\ = 4.3 $\pm$ 0.3 dex), late-type T dwarf. Except for the Burrows fit, which prefers a \logg\ = 5.5 dex, our \textit{Y}-band model results are also consistent with low gravity. The truncated \textit{H}-band gravity fits are consistent with the other T8/T8.5 dwarfs in the sample. Compared to NIRSPEC BDSS spectral standards, the \textit{Y} band is perhaps slightly narrower than the T9 standard, but not significantly. The \textit{H} band is better fit by the T9 standard in the flux peak, but is equally well-fit by the T8 and T9 standards in the methane band.

\noindent
\textit{ULAS J214638.83-001038.7 (Wolf 940B):} \\ Discovered by \citet{Burningham2009}, Wolf 940B (T8.5) is the companion of an M4 dwarf. \citet{Burningham2009} derive a metallicity of  [Fe/H] = -0.06 $\pm$ 0.20 dex for the system. \citet{Leggett2010b} present low-resolution \textit{Spitzer} mid-infrared spectrum of Wolf 940B from 7.5 to 14.2 \microm\ and include a detailed analysis of the metallicity of Wolf 940B, concluding that the metallicity of Wolf 940B is within \midtilde 0.2 dex of solar. The blue \textit{$J-H$} color of this object supports the \citet{Leggett2010b} result. Both the \textit{Y}- and \textit{H}-band spectra of Wolf 940B are well matched by the T9 standard. 

\noindent
\textit{WISE J000517.48+373720.5:} \\ WISE J0005+3737 was identified as part of the \textit{WISE} team follow-up of brown dwarf candidates (T9; Kirkpatrick et al. 2012\nocite{Kirkpatrick2012}). Leggett et al. (2015)\nocite{Leggett2015} present new near-infrared photometry from Gemini/NIRI that is discrepant with the \textit{WISE} team photometry presented in Mace et al. (2013a)\nocite{Mace2013} (see Section \ref{sec:discussion}). We present the first \textit{Y}- and \textit{H}-band spectra of this object. We find that the T9 standard is an exceedingly good match to the \textit{H}-band spectrum of WISE J0005+3737. The \textit{Y}-band spectrum is marginally underluminous compared to the T9 standard in the \midtilde1.05-1.10 \microm\ region, but the rest of the \textit{Y}-band spectrum is an excellent match to the standard. There is no clear spectroscopic evidence of extreme metallicity/gravity in this object.

\section{Discussion}\label{sec:discussion}

In Section \ref{sec:analysis} we show that comparing the target spectra to atmospheric models by holding \teff\ or \logg\ fixed reveals similarities and differences among the four atmospheric model grids used in this analysis, but does not reveal any general physical trends in the sample that correlate with the \textit{$J-H$} color. Comparing each individual target in the sample to spectral standards from the BDSS further supports the overall homogeneity of the sample, though there are a few objects that stand out as unusual. In this section, we discuss the observational and physical mechanisms that could be driving the \textit{$J-H$} colors of the sample.

\clearpage
\subsection{Discrepant Photometry in the Literature}
\begin{deluxetable}{ccccccccccc}
\centering
\rotate
\tablecolumns{11}
\tablecaption{Near-infrared Photometry from the Literature}
\tabletypesize{\tiny}
\tablewidth{0in}
\tablehead{
\colhead{Short Name} &
\colhead{SpT} &
\colhead{\textit{J} (mag)}&
\colhead{\textit{J} err (mag)} &
\colhead{\textit{H} (mag)}&
\colhead{\textit{H} err (mag)}&
\colhead{\textit{$J-H$} (mag)}&
\colhead{\textit{$J-H$} err}&
\colhead{\citetalias{Mace2013b} Classification}&
\colhead{Instrument}&
\colhead{Reference\tablenotemark{b}}
}

\startdata
WISE J0005+3737	&	T9	&	18.33	&	0.12	&	18.10	&	0.08	&	0.23	&	0.14	&	red	&	SOAR/OSIRIS	&	M13a	\\
	&		&	17.59	&	0.02	&	17.98	&	0.02	&	-0.39	&	0.03	&	normal	&	Gemini/NIRI	&	L15	\\
ULAS J0139+0048	&	T7.5	&	18.43	&	0.04	&	19.12	&	0.05	&	-0.69	&	0.06	&	blue	&	UKIRT/UFTI	&	C08	\\
	&		&	18.69	&	0.09	&	18.61	&	0.17	&	0.08	&	0.19	&	red	&	UKIRT/WFCAM	&	C08	\\
CFBDS J0301-1614	&	T7p	&	18.34	&	0.07	&	18.99	&	0.10	&	-0.65	&	0.12	&	blue	&	CFHT/WIRCam	&	A11	\\
WISE J0540+4832	&	T8.5	&	18.49	&	0.02	&	18.62	&	0.05	&	-0.13	&	0.05	&	red	&	Palomar/WIRC	&	M13a	\\
WISE J0759-4904	&	T8	&	17.38	&	0.05	&	17.41	&	0.04	&	-0.03	&	0.06	&	red	&	Magellan/PANIC	&	M13a	\\
CFBDS J0922+1527	&	T7	&	18.28	&	0.04	&	18.81	&	0.10	&	-0.53	&	0.11	&	blue	&	CFHT/WIRCam	&	A11	\\
ULAS J0950+0117	&	T8p	&	18.09	&	0.07	&	18.07	&	0.08	&	-0.02	&	0.11	&	red	&	Palomar/WIRC	&	M13a	\\
	&		&	18.05	&	0.04	&	18.24	&	0.15	&	-0.19	&	0.16	&	normal	&	UKIRT/WFCAM	&	M13a	\\
	&		&	18.02	&	0.03	&	18.4	&	0.03	&	-0.38	&	0.04	&	normal	&	UKIRT/WFCAM\tablenotemark{a} 	&	B13	\\
ULAS J1017+0118	&	T8p	&	18.53	&	0.02	&	19.07	&	0.02	&	-0.54	&	0.03	&	blue	&	UKIRT/UFTI	&	B08	\\
HIP 73786B	&	T6p	&	16.59	&	0.02	&	17.05	&	0.04	&	-0.46	&	0.04	&	normal	&	UKIRT/WFCAM	&	S10; Mu11	\\
WISE J1614+1739	&	T9	&	19.084	&	0.059	&	18.471	&	0.216	&	0.613	&	0.224	&	red	&	SOAR/SpartanIRC	&	K11	\\
	&		&	18.9	&	0.02	&	19.31	&	0.04	&	-0.41	&	0.04	&	normal	&	Gemini/NIRI	&	L15	\\
WISE J1617+1807	&	T8	&	17.659	&	0.08	&	18.234	&	0.078	&	-0.575	&	0.112	&	blue	&	SOAR/SpartanIRC	&	K11	\\
WISE J1812+2721	&	T8.5	&	18.19	&	0.06	&	18.83	&	0.16	&	-0.64	&	0.17	&	blue	&	Palomar/WIRC	&	K11; Burg11	\\
Wolf 940B	&	T8.5	&	18.16	&	0.02	&	18.77	&	0.03	&	-0.61	&	0.04	&	blue	&	WHT/LIRIS; UKIRT/UFTI	&	B09	\\
\enddata
\label{table:colors}
\tablenotetext{a}{\citet{Burningham2013} obtained follow-up UKIRT/WFCAM photometry of ULAS J0950+0118 as part of their follow-up of UKIDSS brown dwarf candidates. All other UKIRT/WFCAM data included in this table were obtained from the UKIDSS database.}
\tablenotetext{b}{References: M13a \citep{Mace2013}; L15 \citep{Leggett2015};  C08 \citep{Chiu2008}; A11 \citep{Albert2011}; B13 \citep{Burningham2013}; B08 \citep{Burningham2008}; S10 \citep{Scholz2010}; Mu11 \citep{Murray2011}; K11 \citep{Kirkpatrick2011}; Burg11 \citep{Burgasser2011}; B09 \citep{Burningham2009}}

\end{deluxetable}

\citet{Leggett2015} report near-infrared MKO photometry for 18 known late-T and Y dwarfs. Fourteen of these objects had MKO or MKO-like photometry in the literature presented by the \textit{WISE} team (\citealt{Kirkpatrick2012, Kirkpatrick2013, Beichman2013, Beichman2014, Mace2013}). Of the 14 objects that overlap in the \textit{WISE} team and Leggett et al. samples, five objects have discrepant ($>$ 2 $\sigma$) \textit{J}- and/or \textit{H}-band magnitudes. Two of these objects are in our sample of color outliers: WISE J0005+3737 and WISE J1614+1739. Both objects are classified as unusually red based on \textit{WISE} team photometry, but would not be classified as color outliers based on the Leggett et al. (2015) photometry (see Table \ref{table:colors}).

\citet{Leggett2015} argue that atmospheric variability in the 14 overlapping T and Y dwarfs is likely too small to account for the discrepancies in the near-infrared photometry, which can differ by up to a magnitude between datasets. A large near-infrared study of late-T/Y dwarf variability is still needed to statistically characterize the impact of variability on the near-infrared photometry of these late-type objects (see also \citealt{Littlefair2017}). However, detections of near-infrared variability from a handful of mid- to late-T dwarfs with typical peak-to-peak amplitudes $<$ 10$\%$ (e.g. \citealt{Buenzli2012, Buenzli2014, Radigan2012, Radigan2014, Wilson2014, Rajan2015}) support this claim. Peak-to-peak variability amplitudes of $<$10$\%$ have also been detected for T and Y dwarfs using mid-infrared data from \textit{Spitzer} (\citealt{Metchev2015, Cushing2016, Leggett2016}). Leggett et al. (2015) suggest instead that the discrepant near-IR photometry of these objects may be due to previously ``unrecognized differences in photometric systems'' between the instruments at the Gemini Observatories used in Leggett et al. and the various instruments used by the \textit{WISE} team.

To further investigate whether instrumental differences could lead to the discrepant photometry in our sample, we examined the filter transmission curves for the instruments used to observe the targets in our sample with discrepant photometry. WISE J0005+3737 was observed by the \textit{WISE} team using the OSIRIS instrument \citep{DePoy1993} on the Southern Astrophysical Research (SOAR) Telescope and the results were presented in \citet{Mace2013}. While the wavelength coverage of the OSIRIS \textit{H}-band filter\footnote{\footnotesize SOAR/OSIRIS filter curves were obtained from \url{http://www.ctio.noao.edu/instruments/ir_instruments/osiris2soar/config/index.html}} is very similar to the wavelength coverage of an MKO \textit{H}-band filter, the OSIRIS \textit{J}-band wavelength coverage is broader than the MKO \textit{J} bandpass as defined in \citet{SimonsTokunaga2002}. \citet{Mace2013} list their OSIRIS photometry under the MKO heading in their Table 4, but the authors do note that the OSIRIS data is not on the MKO system. The WISE J0005+3737 photometry presented in \citet{Leggett2015} was observed with the Near InfraRed Imager (NIRI; \citealt{Hodapp2003}) on Gemini North. The NIRI filters\footnote{\footnotesize Gemini/NIRI filter curves were obtained from \url{http://www.gemini.edu/sciops/instruments/niri/imaging/filters}} are MKO and therefore we adopt the \citet{Leggett2015} measurements for WISE J0005+3737. With this new photometry, WISE J0005+3737 is no longer classified as a color outlier based on the \citetalias{Mace2013b} criteria. As discussed in Section 4.4 and presented in Appendix \ref{sec:appendix}, the \textit{Y}- and \textit{H}-band spectra of WISE J0005+3737 are well matched by the spectral standard. 

WISE J1614+1739 was observed by the \textit{WISE} team using the Spartan Infrared Camera (SpartanIRC; \citealt{Loh2012}) on the SOAR Telescope and by \citet{Leggett2015} using Gemini/NIRI. Like the Gemini/NIRI filters, the SpartanIRC/SOAR filters were designed as MKO filters, and the bandwidths of both the SpartanIRC and NIRI filter sets as listed on each instrument's website are nearly identical (see Table 5 in \citealt{Loh2012} or the SpartanIRC website at Michigan State University\footnote{\footnotesize \url{http://www.pa.msu.edu/~loh/SpartanIRCamera/}} for more information on the SpartanIRC filters). However, both the \textit{J}- and \textit{H}-band photometry differ by $\geq$ 3$\sigma$ between the two instruments ($\Delta$\textit{J} = 0.18 $\pm$ 0.06 mag; $\Delta$\textit{H} = 0.84 $\pm$ 0.22 mag). Either WISE J1614+1739 is variable, or there are other systematics in the data. Based on the SpartanIRC photometry, WISE J1614 is classified as ``red,'' but it is classified as ``normal'' using the NIRI photometry. Unlike WISE J0005+3737, WISE J1614+1739 does show enhanced \textit{Y}-band flux, suggesting low-metallicity/high-gravity, and further follow-up to resolve the photometric discrepancy would be valuable. 

Our sample also includes two targets, not included in \citet{Leggett2015}, that have discrepant MKO photometry in the literature: ULAS J0950+0117 and ULAS J0139+0048. As discussed in \citetalias{Mace2013b}, there are three sets of \textit{J}- and \textit{H}-band photometry for  ULAS J0950+0117 in the literature. \citet{Mace2013} present Palomar/WIRC \citep{Wilson2003} photometry and also report photometry from the UKIDSS archive \citep{Lawrence2007}. UKIDSS data were observed using the Wide Field CAMera (WFCAM) on UKIRT \citep{Casali2007}. \citet{Burningham2013} presented further UKIRT/WFCAM near-infrared photometry, as part of their follow-up of UKIDSS brown dwarf candidates. The \textit{J}-band photometry for all three data sets are consistent within 1$\sigma$, but the \textit{H}-band photometry ranges from \textit{$H_{\rm WIRC}$} = 18.07 $\pm$ 0.08 (\citealt{Mace2013}) to \textit{$H_{\rm WFCAM}$} = 18.40 $\pm$ 0.03 mag (\citealt{Burningham2013}). Based on the WIRC data, ULAS J0950+0048 is classified as ``red.'' Based on both sets of WFCAM data (from the UKIDSS database and as presented in \citealt{Burningham2013}), ULAS J0950+0117 is classified as ``normal.'' The WIRC\footnote{\footnotesize \url{http://www.astro.caltech.edu/palomar/observer/200inchResources/wircspecs.html\#filters}} and WFCAM\footnote{\footnotesize \url{http://www.ukirt.hawaii.edu/instruments/wfcam/user_guide/description.html}} filter sets were both developed to MKO specifications and the filter curves are nearly identical. Given the unusual nature of ULAS J0950+0117 and the range of \textit{H}-band photometry, further photometric follow-up on this object is recommended. 

\citet{Chiu2008} reported two sets of MKO \textit{J}- and \textit{H}-band measurements for the final object in our sample with discrepant photometry, ULAS J0139+0048. The first set of \textit{J}- and \textit{H}-band measurements came from the UKIDSS database and the second were observed with the UKIRT Fast-Track Imager\footnote{\footnotesize Filter transmission profiles were accessed here:\url{http://www.ukirt.hawaii.edu/instruments/ufti/PARAMETERS.html\#2}} (UFTI; \citealt{Roche2003}).  Both the \textit{J}- and \textit{H}-band UFTI photometry disagree at $\geq 2\sigma$ with the UKIDSS photometry. The UKIDSS photometry yields a ``red'' color designation, while the UFTI photometry yields a ``blue'' designation. While Chiu et al. (2008) note the inconsistent photometry, they do not suggest a cause for the discrepancy. A visual inspection of the UKIDSS data does not reveal any artifacts in the field. We have adopted the UFTI photometry for this analysis as it has smaller uncertainties. This maintains a ``blue'' color designation for ULAS J0139+0048, though the NIRSPEC \textit{Y}- and \textit{H}-band spectra of this object are not unusual. 

\begin{figure*}
\centering
\includegraphics[width=4.5in]{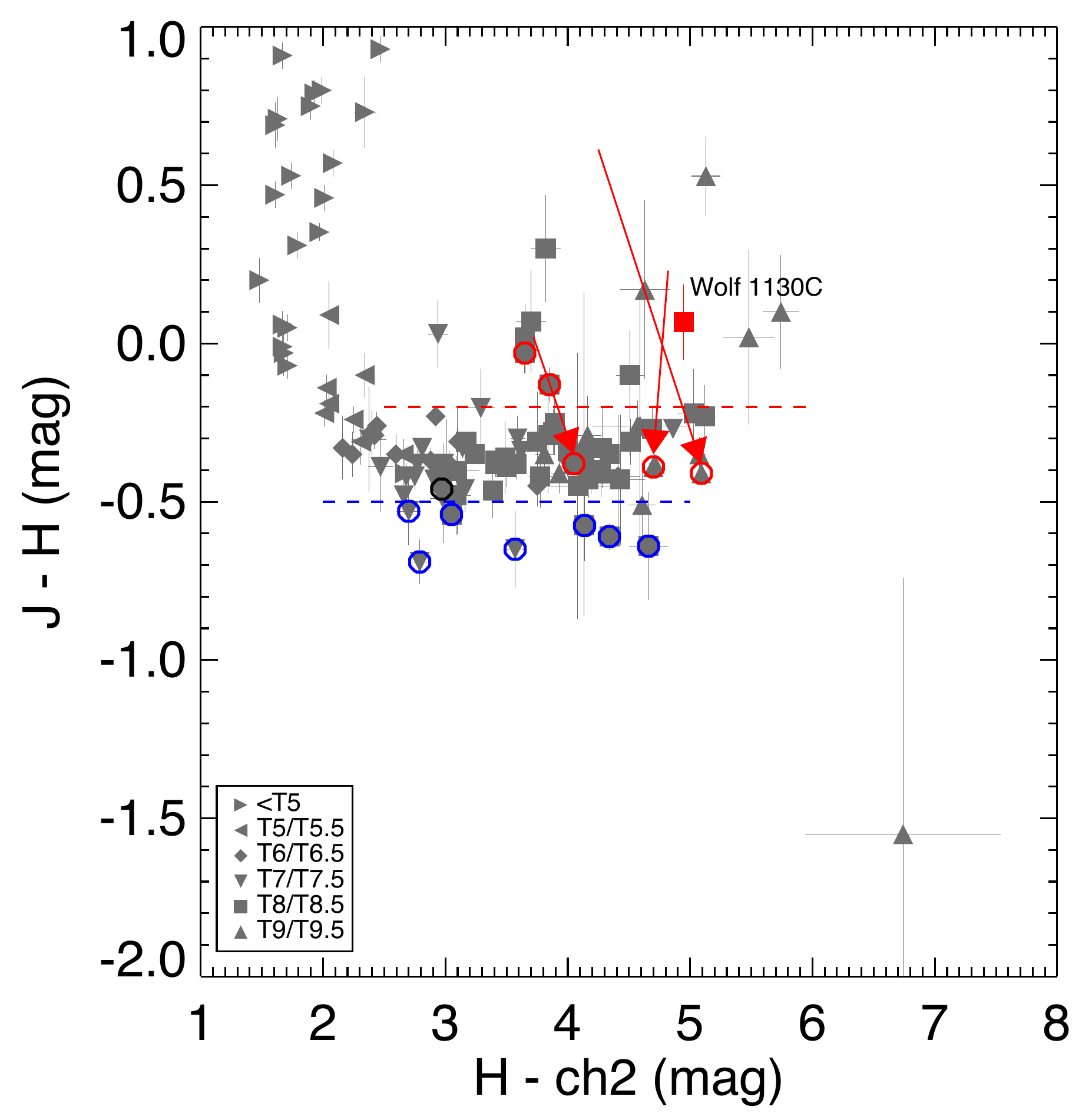}
\caption{Same as Figure \ref{fig:JH_vs_HCH2}, but with more recently published photometry for three of our four targets with discrepant photometry ($>$ 2 sigma) in the literature. The red arrows denote the movement of these three objects in color space, with the head of the arrow indicating photometry from \citet{Leggett2015} or \citet{Burningham2013} and the tail of the arrow indicating photometry from \citet{Kirkpatrick2011} or \citet{Mace2013}. The photometry from \citet{Kirkpatrick2011} or \citet{Mace2013} were used to select the \textit{$J-H$} outliers for this sample. All three objects would be reclassified from ``red'' to ``normal'' based on the Leggett et al. or Burningham et al. photometry. Of these three targets, two have NIRSPEC spectra with enhanced, peak-normalized \textit{Y}-band flux in comparison to their spectral standard.}
\label{fig:JH_vs_HCH2_leggett}
\end{figure*}

Of the four brown dwarfs in our sample with discrepant photometry in the literature, one object (WISE J0005+3737) does indeed appear to be discrepant due to significant differences in the filter transmission curves of the two instruments used to observe the target. This lends further support to the \citet{Leggett2015} assertion that instrumental differences can impact the photometry of these late-type T dwarfs and emphasizes the need for consistent photometry of the late-type T dwarf population on a well-calibrated instrument. However, the three remaining targets with discrepant photometry were observed with bonafide MKO filters and the differences in filter transmission between the instruments used is negligible. While it is possible that there are other instrumental systematics impacting the results, the unusual spectral morphologies of two of the three remaining dwarfs with discrepant photometry suggests that there may also be physical motivation for the photometric differences. Photometric monitoring of late-T dwarfs at \textit{YJH} bands may serve to separate the impact of atmospherically variability (e.g. clouds) from from morphological changes due to extreme physical parameters (e.g. [Fe/H], \logg). 

\subsection{Atmospheric Variability: Clouds and Temperature Variations}\label{sec:clouds}

If we adopt the more recent photometry for three of the four objects with discrepant photometry in the literature, all three of those objects move into ``normal'' \textit{$J-H$} color space (see Figure \ref{fig:JH_vs_HCH2_leggett}). Excluding those three objects, however, leaves ten objects that are classified as outliers based on the \citetalias{Mace2013b} color criteria, nine of which exhibit \textit{Y}- and \textit{H}-band spectral morphologies that closely match their respective spectral standards. If we assume robust photometry for these objects, the question then becomes, is it possible to induce an unusual \textit{$J-H$} color for these objects while simultaneously maintaining a uniform normalized flux shape?

\begin{figure}
\centering
\includegraphics[width=3.15in]{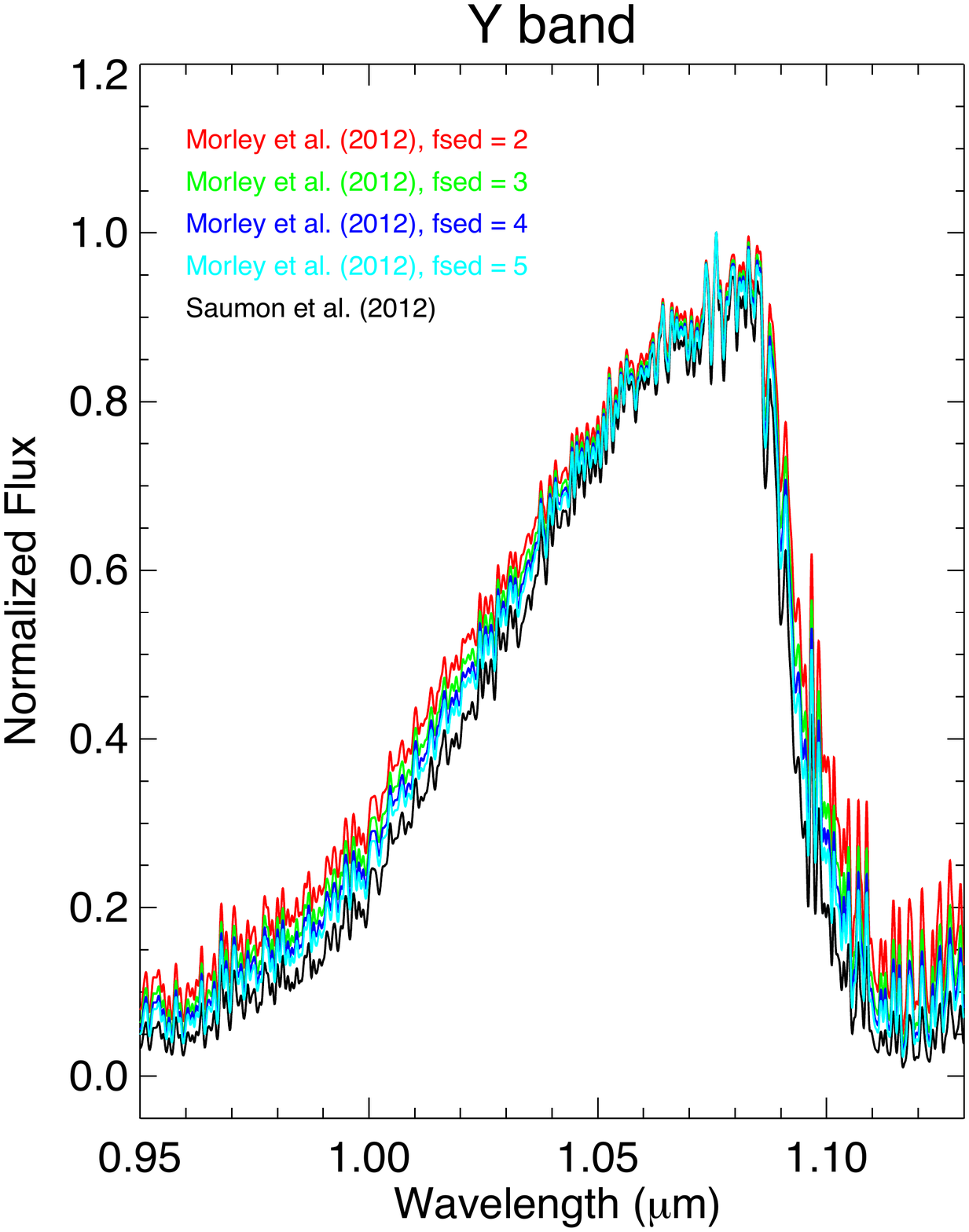}
\includegraphics[width=3.15in]{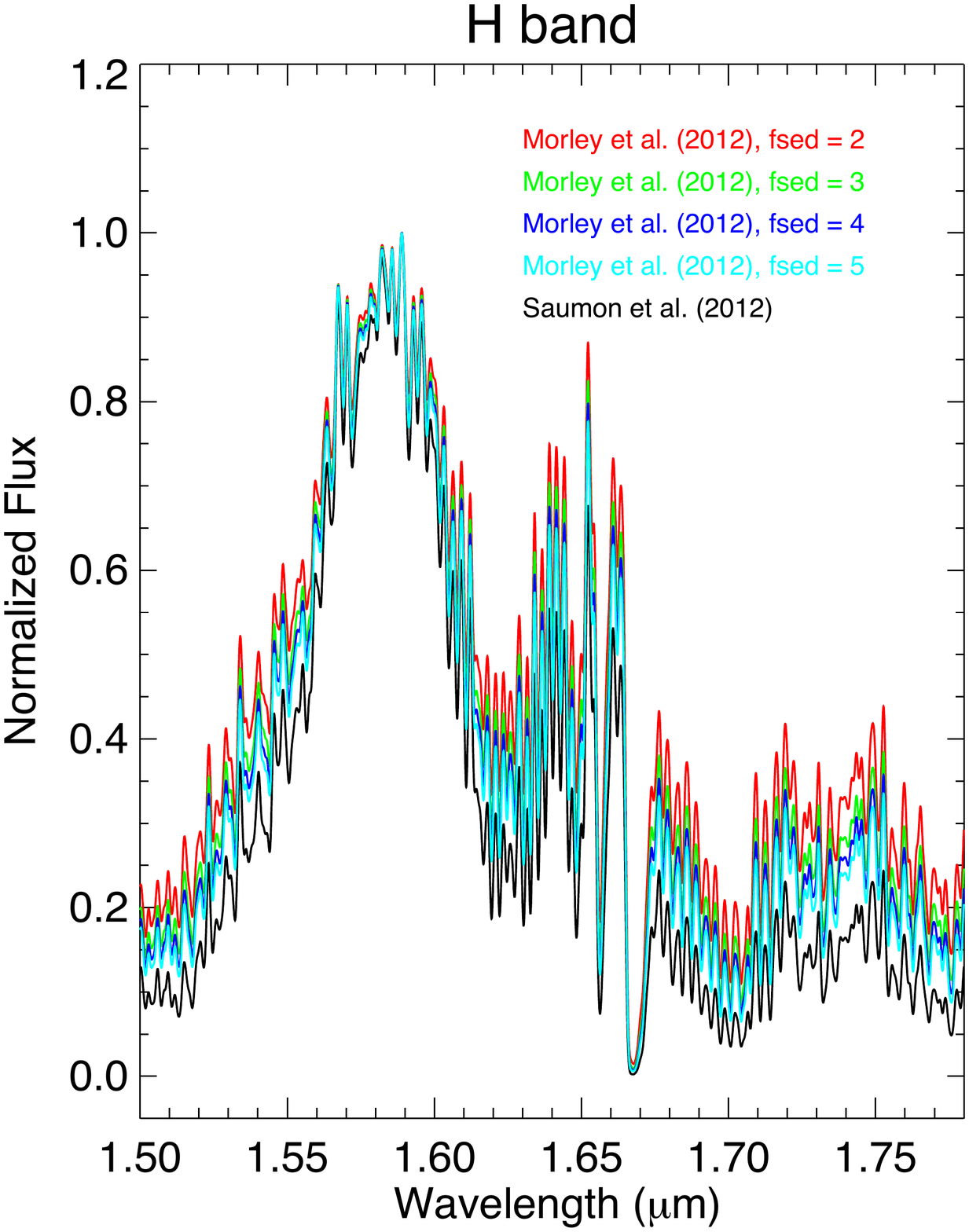}
\caption[The Impact of Clouds: Morley et al. 2012 models]{Peak-normalized \textit{Y}- (left) and \textit{H}-band (right) \citet{Morley2012} atmospheric model spectra at \teff\ = 700 K, \logg\ = 4.5 dex, solar metallicity, and varying \fsed. \citet{Saumon2012} cloudless models are also plotted for comparison. Holding everything else fixed, varying the sedimentation efficiency from \fsed\ = 5 (i.e. thin clouds) to \fsed\ = 2 (i.e. thick clouds), serves to broaden the wings of both the \textit{Y}- and \textit{H}-band spectra, but does not significantly impact the shape or breadth of the flux peak itself.}
\label{fig:morley_fsed}
\end{figure}

\citet{Morley2012} examine the impact of sulfide clouds on late-type T dwarfs. Analysis of the MKO photometry derived from the \citet{Morley2012} models\footnote{\footnotesize See \url{http://www.ucolick.org/~cmorley/cmorley/Models.html}} shows that, for a fixed \teff, \logg, and metallicity, \textit{$J-H$} colors move red-ward with decreasing \fsed\ (increasing clouds). In peak-normalized \textit{Y}- and \textit{H}-band model spectra, this translates to a broadening of the wings of each band, while the peaks remain relatively unchanged (see Figure \ref{fig:morley_fsed}), but the overall impact is small. Clouds \textit{do} alter the relative brightness of the \textit{YJHK} bands at these late-T dwarf temperatures, however (see, e.g. Figure 5 in \citealt{Morley2012}).  Thus, a change between the relative brightness of the individual bands would be detectable in flux-calibrated spectra, but we would not see such a change in this peak-normalized sample.

In \citet{Morley2014b}, the authors further examine the impact of patchy clouds and hot spots on the atmospheres of T and Y dwarfs. They find that increased cloud cover again drives the \textit{$J-H$} color of late-T dwarfs red-ward for effective temperatures $>$ 300 K, without significantly changing the mid-infrared colors or the peak morphology of the individual bands in the 1-2.5 \microm\ spectrum (see their Figures 1 and 2). Thin global clouds or minimal patchy clouds in these late-T dwarf spectra lead to bluer colors than thick global clouds or significant patchy cloud cover. Clouds can account for a spread in \textit{$J-H$} color consistent with the spread we see for all but a few of the reddest \textit{$J-H$} outliers in this sample, including Wolf 1130C. If clouds are prevalent in late-T dwarf atmospheres, then clouds make \textit{$J-H$} color an inconsistent selector for metallicity/gravity variation. However, \citet{Line2015,Line2017} do not generally find evidence for thick clouds in their atmospheric retrieval model fits to a sample of 11 late-T dwarfs.

Temperature fluctuations have also been proposed as a source of variability in brown dwarf atmospheres (e.g. \citealt{ShowmanKaspi2013, RobinsonMarley2014}). \citet{Morley2014b} investigate the impact of hot spots in a cloud-free atmospheric model and find that adding energy at different levels in the atmosphere (i.e. different pressures) does indeed impact \textit{$J-H$} color (see their Figure 5). However, like the patchy cloud models, the hot spot models do not account for the reddest \textit{$J-H$} outliers in our sample. Spectroscopically, the \citet{Morley2014b} hot spot models show the largest deviation in atmospheric absorption regions and are more prominent in the mid-infrared where cloud-induced variability is notably smaller. Thus, we would not expect to see the impact of hot spots in our ground-based, near-infrared spectra. 

While \citet{Morley2014b} show that variability can drive scatter in \textit{$J-H$} color, an object like Wolf 1130C, which shows broad deviation from the spectral standard in the near-infrared, cannot be explained by variability alone, making it an unambiguous outlier in metallicity/gravity. Thus, while variability may explain some of the measured spread in \textit{$J-H$} color, variability alone is not able to account for the near-infrared photometry of the late-T dwarf population.

\subsection{Gravity, Metallicity, Clouds, and Brown Dwarf Radii: A Complex, Interdependent Parameter Space}
Atmospheric composition can significantly impact the radius of a brown dwarf. \citet{Burrows2011} generated cloudy atmospheric models with [Fe/H] = 0.0 dex and 0.5 dex, and cloud-free models with [Fe/H] = -0.5, 0.0, and +0.5 dex to study the interplay between mass, radius, metallicity, clouds, and gravity. They find that, for a given mass, high metallicity and/or thick clouds, which increase atmospheric opacity, result in a larger radius (up to \midtilde10\% - 25\%) than a lower metallicity/cloud-free atmosphere. A cloud-induced spread in radii at a given mass and age also implies a spread in gravity for that mass and age. The impact on the emerging spectral morphology from differences in radii are thus complex and parameter interdependent. As discussed in Section \ref{sec:clouds}, \citet{Morley2012} show that sulfide clouds redden the near-infrared spectrum of late-T dwarfs at a fixed \logg. Increasing \logg\ strengthens the CIA of \hh\, which serves to  make the near-infrared spectrum bluer \citep{Saumon2012}. However, the precise interplay between clouds, gravity, and metallicity can be much more complicated, emphasizing the challenge of fully characterizing the atmospheres of late-type T dwarfs, particularly solitary T dwarfs in the field.

\subsection{Looking Forward}
Wolf 1130C illustrates the value of \textit{Y}-band spectroscopy for probing extreme metallicity and gravity in the late-T dwarf population. However, the overall homogeneity of this sample, and that of the  larger, late-T field population, suggests that the typical metallicity/gravity variation for late-T dwarfs is much smaller than the extremes exemplified by Wolf 1130C. This is perhaps unsurprising as this population is predominantly nearby and likely majority thin disk (e.g. see discussion in \citealt{Bensby2003}). To isolate the impact of metallicity/gravity on near-infrared, late-T dwarf spectra from the impact of spectral variability, high S/N spectra and models that are better calibrated at high- and low-metallicity will be required. In agreement with the literature (e.g. \citealt{Burgasser2006c, Leggett2007}), we find that \textit{Y}-band spectroscopy is more sensitive to age-dependent spectral features in late-T dwarfs than \textit{H}-band spectroscopy. \textit{K}-band spectroscopy is also a valuable probe of extreme gravity/metallicity (e.g. \citealt{Burgasser2006c, Leggett2007, Saumon2012}), but the faint nature of the \textit{K} band implies that high S/N, medium-resolution, \textit{K}-band spectra of late-T dwarfs will be easier to obtain with the next generation thirty-meter class telescopes or \textit{JWST}. For brighter targets, simultaneous, low- to medium-resolution JHK spectroscopy from the ground is also a valuable probe of the shape of the \textit{K} band and is a useful tool for identifying gravity outliers.

Is \textit{$J-H$} color a good predictor of age for late-type T dwarfs? The results from this study are inconclusive. Models indicate that \textit{$J-H$} color could be sensitive to atmospheric variability \citep{Morley2012, Morley2014b}. Observational variability studies indicate that the impact of variability on near-infrared photometry is typically $<10\%$ outside of the L/T transition (e.g. \citealt{Radigan2014}). Most near-infrared variability studies to date have focused on L/T transition objects and included only a few mid-to-late T dwarfs, and a dedicated late-T and Y dwarf multiwavelength variability study is needed. However, the intrinsically faint nature of these objects and the photometric errors associated with ground-based photometry make ground-based variability studies challenging (e.g. \citealt{Koen2013}). Moreover, the spread in \textit{J-H} color for late-type T dwarfs with more than one set of \textit{J}- and \textit{H}-band photometry in the literature can be larger than the expected change in magnitude driven by physical variability, emphasizing the need for consistent photometry to accurately make a definitive statement on the value of \textit{$J-H$} color as a probe of metallicity, gravity, clouds, and/or nonequilibrium chemistry.

\section{Summary}
In this work, we test the hypothesis set forth in \citet{Mace2013b} that T dwarfs with unusual \textit{$J-H$} colors may show evidence of extreme age compared to the typical field population. To do so, we used Keck/NIRSPEC \textit{Y}- and \textit{H}-band spectroscopy to look for spectroscopic tracers of age such as nonsolar metallicity or extreme gravity by comparison to spectral standards and four publicly available atmospheric model grids. The overall sample is surprisingly homogeneous, though four objects stand out as spectroscopic outliers HIP 73786B, ULAS J0950+0117, CFBDS J0301-1614, and WISE J1614+1739. 

HIP 73786B (T6p) and ULAS J0950+0117 (T8p) are previously known companions to stars with measured, subsolar metallicity. HIP 73786B was included in the study as a counter-example to the \citet{Mace2013b} hypothesis, with a ``normal'' \textit{J-H} color. \citet{Burningham2014} presented a low-resolution (\textit{R}\midtilde120) spectrum of HIP 73786B, which exhibited enhanced \textit{Y}-band flux  and a suppressed \textit{K} band. Our NIRPSEC \textit{Y} band of HIP 73786B is best matched by the \textit{Y}-band spectrum of the likely metal-poor/high-gravity dwarf, 2MASS J0937+2931, supporting the \citet{Burningham2014} result and consistent with the known low-metallicity nature of the primary. ULAS J0950+0117 was selected as a potentially low-metallicity/high-gravity follow-up object based on its red \textit{J-H} color. We find that the blue wing of the NIRSPEC \textit{Y}-band spectrum of ULAS J0950+0117 is slightly enhanced compared to the T8 spectral standard, consistent with the slightly enhanced \textit{Y} band in the low-S/N Magellan/FIRE spectrum presented in \citet{Mace2013}. These two enhanced \textit{Y}-band observations support the low metallicity inferred from the primary. However, additional \textit{H}-band photometry in the literature classifies ULAS J0950+0117 as normal based on our color criteria. Both HIP 73786B and ULAS J0950+0117 have \textit{H}-band spectra consistent with spectral standards, indicating that neither object has as extreme of a composition as Wolf 1130C. The other two stand-out objects in the sample are not known companions. CFBDS J0301-1614 was chosen for follow-up based on its blue \textit{J-H} color and has been previously identified as unusual and potentially young based on model comparisons to its near-IR photometry and the suggestion of an enhanced \textit{H}-band spectrum \citep{Albert2011}. We present a new \textit{Y}-band spectrum of CFBDs J0301-1614, which has a suppressed blue-wing flux compared to the spectral standard, consistent with high metallicity/low gravity. We also present a NIRSPEC \textit{H}-band spectrum that is slightly underluminous compared to the T7 standard, consistent with the \textit{H}-band spectrum in \citet{Albert2011} and potentially indicative of low gravity. The final object, WISE J1614+1739, is identified as spectroscopically unusual for the first time in this work. Like ULAS J0950+0117, discrepant photometry in the literature either classifies WISE J1614+1739 as ``red'' or ``normal.'' The new \textit{Y}-band spectrum of WISE J1614+1739 presented here shows an enhanced blue wing compared to the peak-normalized spectral standard and suggests that this object may be low metallicity/high gravity. While this work does not reveal an unambiguous trend between \textit{J-H} color and spectral tracers of age, HIP 73786B, ULAS J0950+0117, CFBDS J0301-1614, and WISE J1614+1739 all show evidence of unusual metallicity/gravity. Further follow-up of all four targets will be valuable as we begin to statistically probe the extremes of gravity and metallicity in the late-T dwarf population. 

Model comparisons illustrate that the current, publicly available atmospheric models are heterogeneous in the 1.05 \microm\ region of the \textit{Y} band and in the 1.65 \microm\ \meth\ absorption. Updated models with increased metallicity range, improved molecular line lists-- particularly of \meth-- and updated treatment of alkali line broadening, will ultimately be required to map the full parameter space probed by the late-T dwarf population. Also, discrepant photometry in the literature emphasizes the need for a near-infrared photometric survey carried out on a single, well-characterized instrument. Furthermore, time-series photometry will help disentangle the impact of variability from the impact of physical properties such as gravity and/or metallicity on the late-T dwarf population. 

\acknowledgements
Acknowledgements: We would like to thank the anonymous referee for their helpful comments that improved the clarity of the manuscript, Chris Gelino for a valuable discussion on near-infrared photometry, and Mike Cushing for a thoughtful discussion of SpeX and NIRSPEC spectral reduction. We would also like to thank the developers of the REDSPEC package, Sung-Soo Kim, Lisa Prato, and Ian McLean.

This work includes data obtained at the W.M. Keck Observatories on Maunakea, Hawaii. We wish to recognize and acknowledge the very significant cultural role and reverence that the summit of Maunakea has always had within the indigenous Hawaiian community.  We are most fortunate to have had the opportunity to conduct observations from this mountain. 

This research has further benefited from (1) the SpeX Prism Spectral Libraries, maintained by Adam Burgasser at http://pono.ucsd.edu/\midtilde adam/browndwarfs/spexprism, (2) the M, L, T, and Y dwarf compendium housed at DwarfArchives.org and curated by Chris Gelino, Davy Kirkpatrick, Mike Cushing, David Kinder, and Adam Burgasser, (3) the Brown Dwarf Spectroscopic Survey (BDSS) Archive housed at bdssarchive.org, (4) the NASA/ IPAC Infrared Science Archive, which is operated by the Jet Propulsion Laboratory, California Institute of Technology, under contract with the National Aeronautics and Space Administration, and (5) the SIMBAD database, developed and operated at CDS, Strasbourg, France.

Facility: \facility{Keck:II(NIRSPEC)}

\appendix
\section{Target Spectra versus Spectral Standards for the Objects in Section 4.4}\label{sec:appendix}

Here we present the \textit{Y}- and \textit{H}-band spectra of the targets described in Section \ref{sec:characteristics_other} that do not show spectral signatures of extreme gravity and/or metallicity. Each target is compared to the spectral standard of that type or of the two nearest spectral standards if the target has a half spectral type classification. The spectra are organized by spectral type and colored by \textit{$J-H$} color as described in Section \ref{sec:results}. Target $-$ standard residual plots are also provided, with the residual plots binned down for clarity. 

\begin{figure}[h]
\centering
\includegraphics[width=6in]{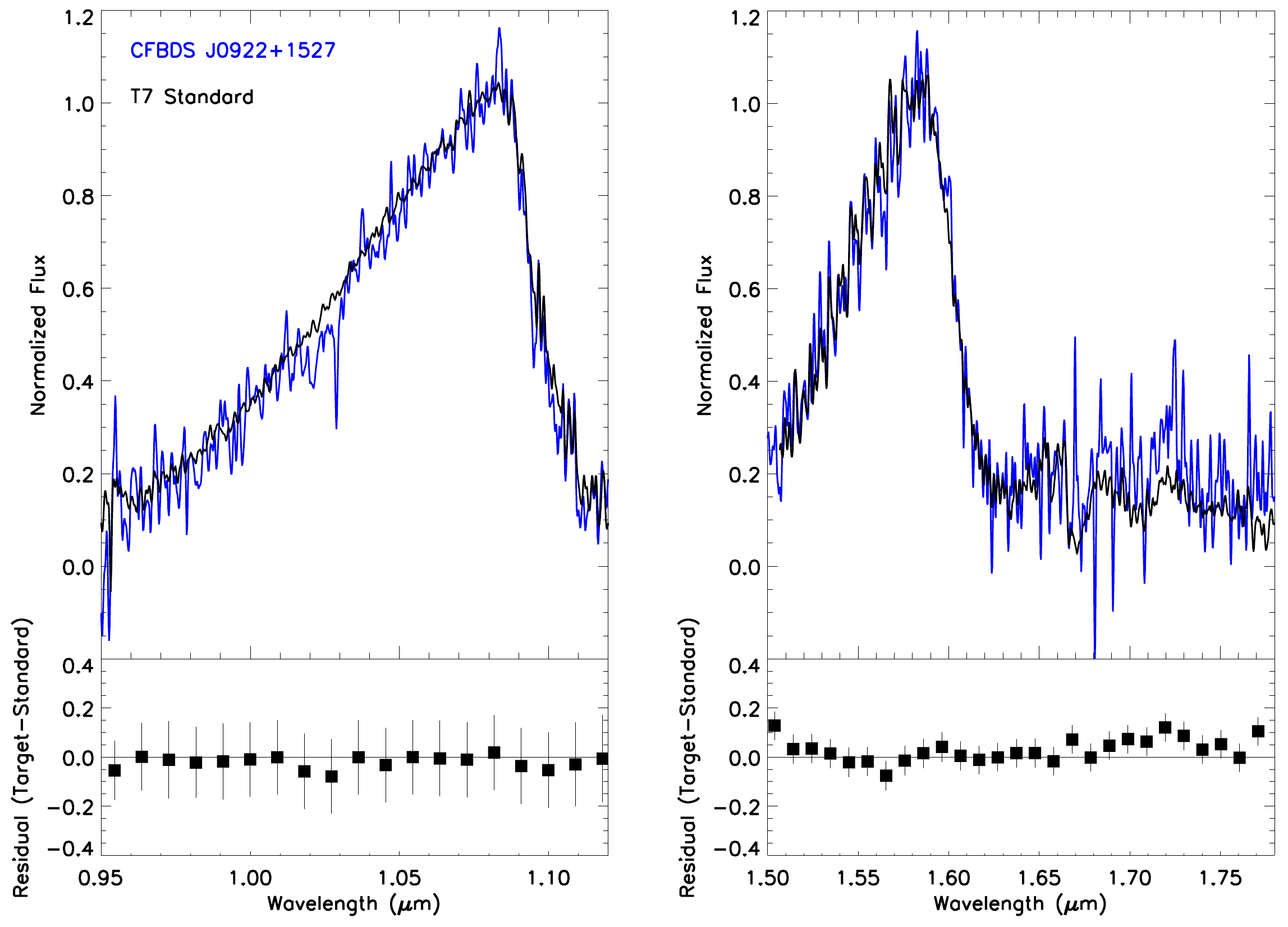}
\caption[Appendix: CFBDS J0922+1527]{CFBDS J0922+1527 (T7)}
\end{figure}

\begin{figure}[h]
\centering
\includegraphics[width=6in]{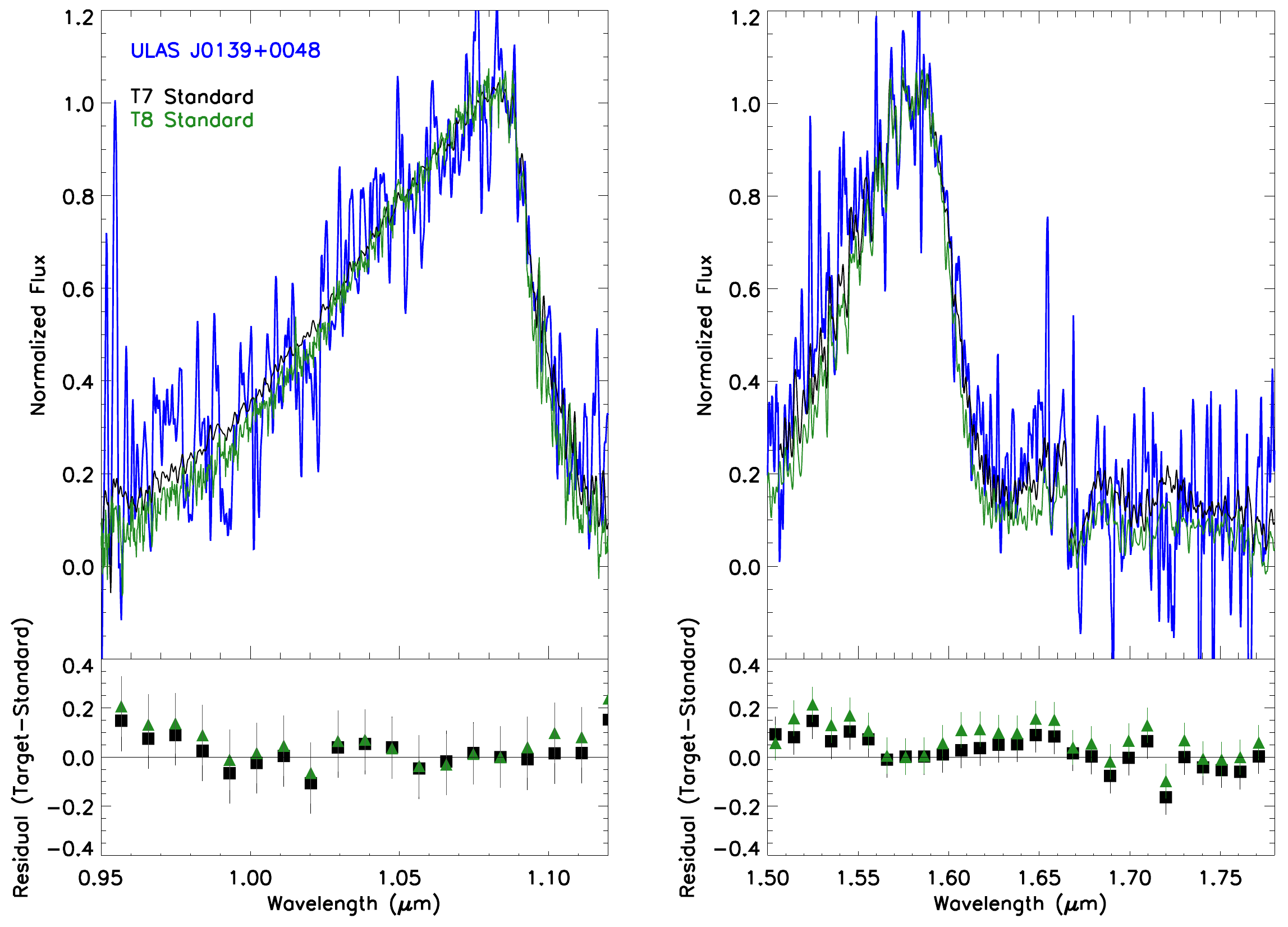}
\caption[Appendix: ULAS J0139+0048]{ULAS J0139+0048 (T7.5)}
\end{figure}

\begin{figure}[h]
\centering
\includegraphics[height=4in]{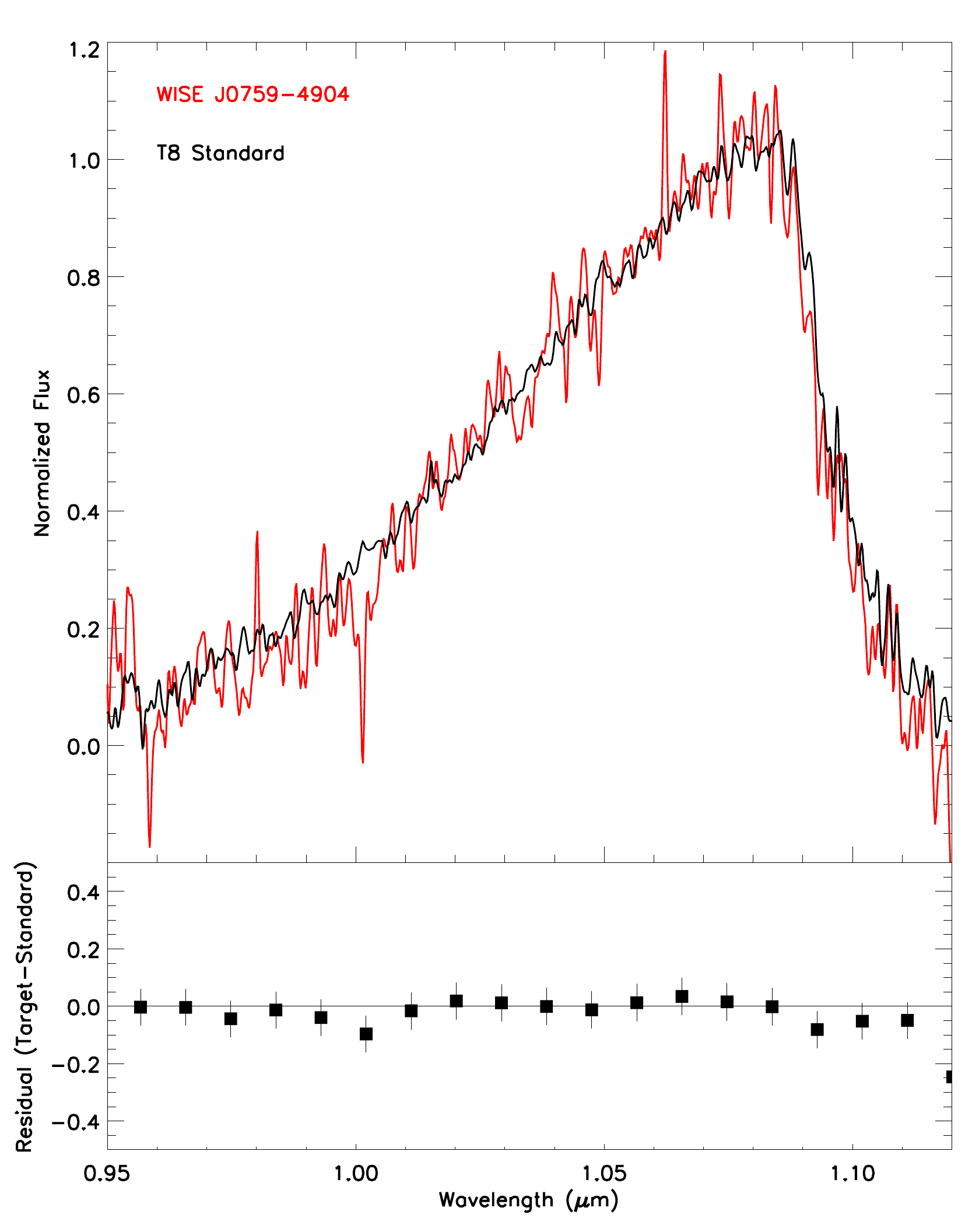}\hspace{1em}
\includegraphics[height=4in]{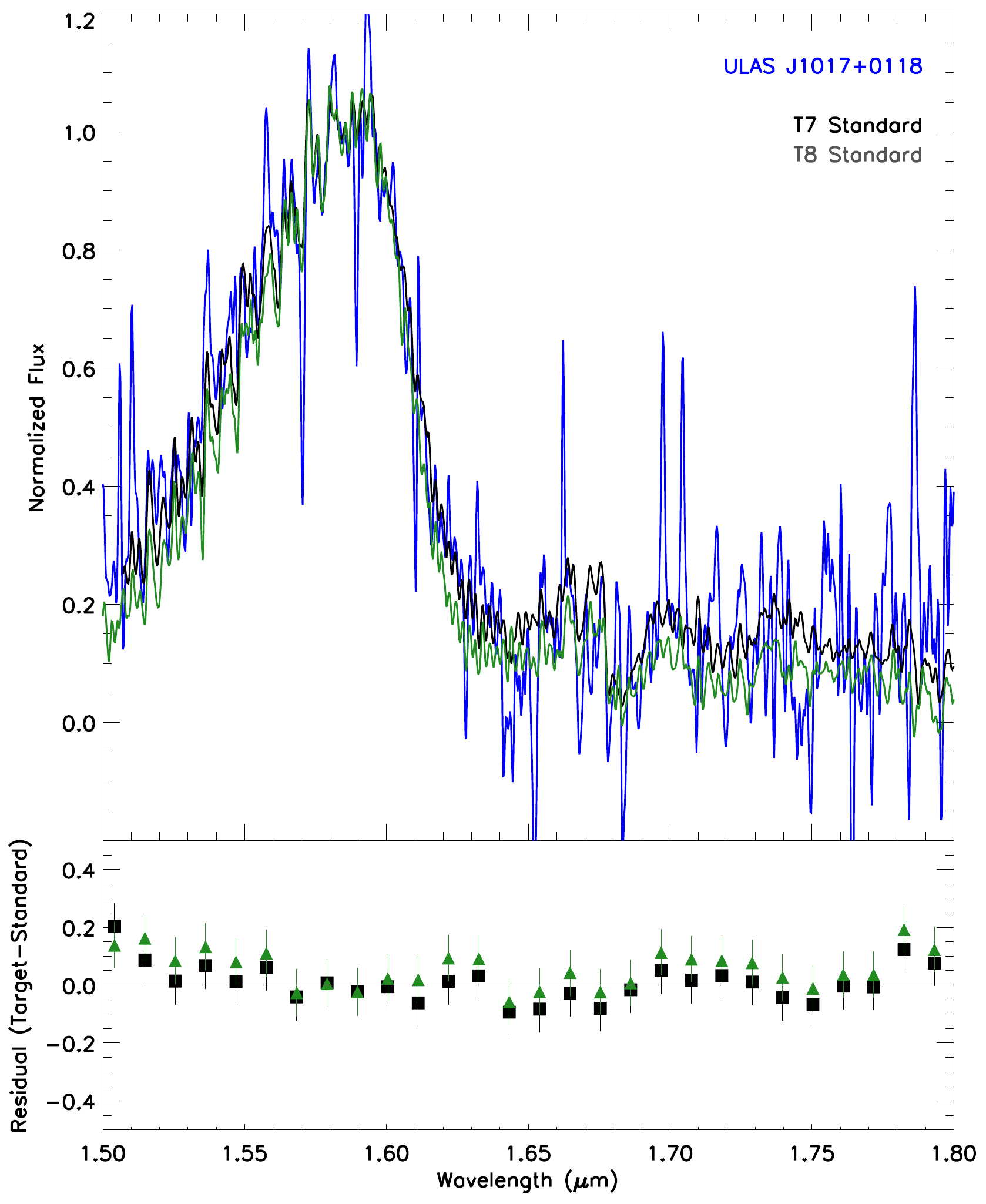}
\caption[Appendix: WISE J0759-4904 \& ULAS J1017+0118]{\textit{Y}-band spectrum of WISE J0759-4904 (T8; top) and \textit{H}-band spectrum of ULAS J1017+0118 (T8p).}
\end{figure}

\begin{figure}[h]
\centering
\includegraphics[width=5.5in]{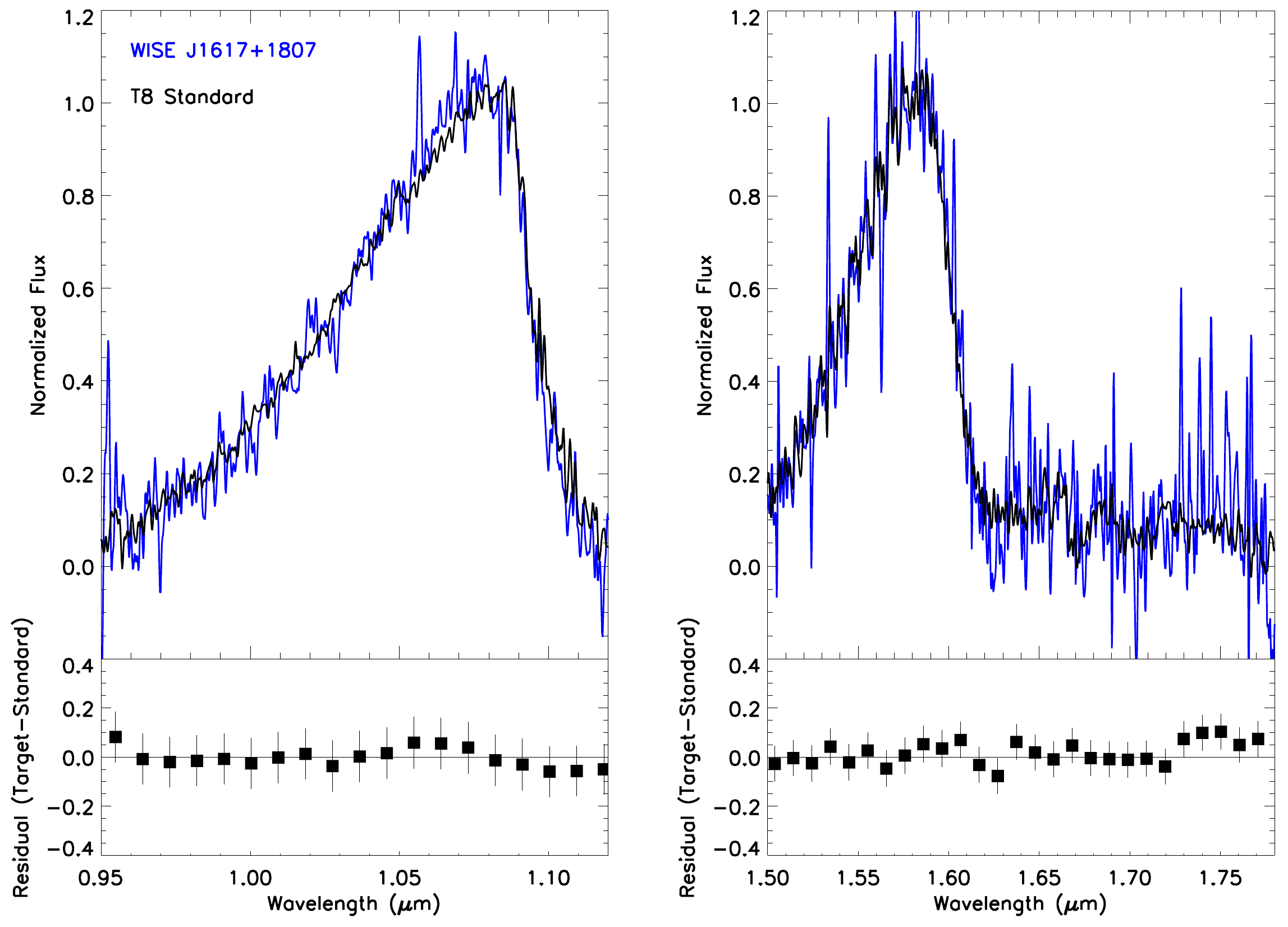}
\includegraphics[width=5.5in]{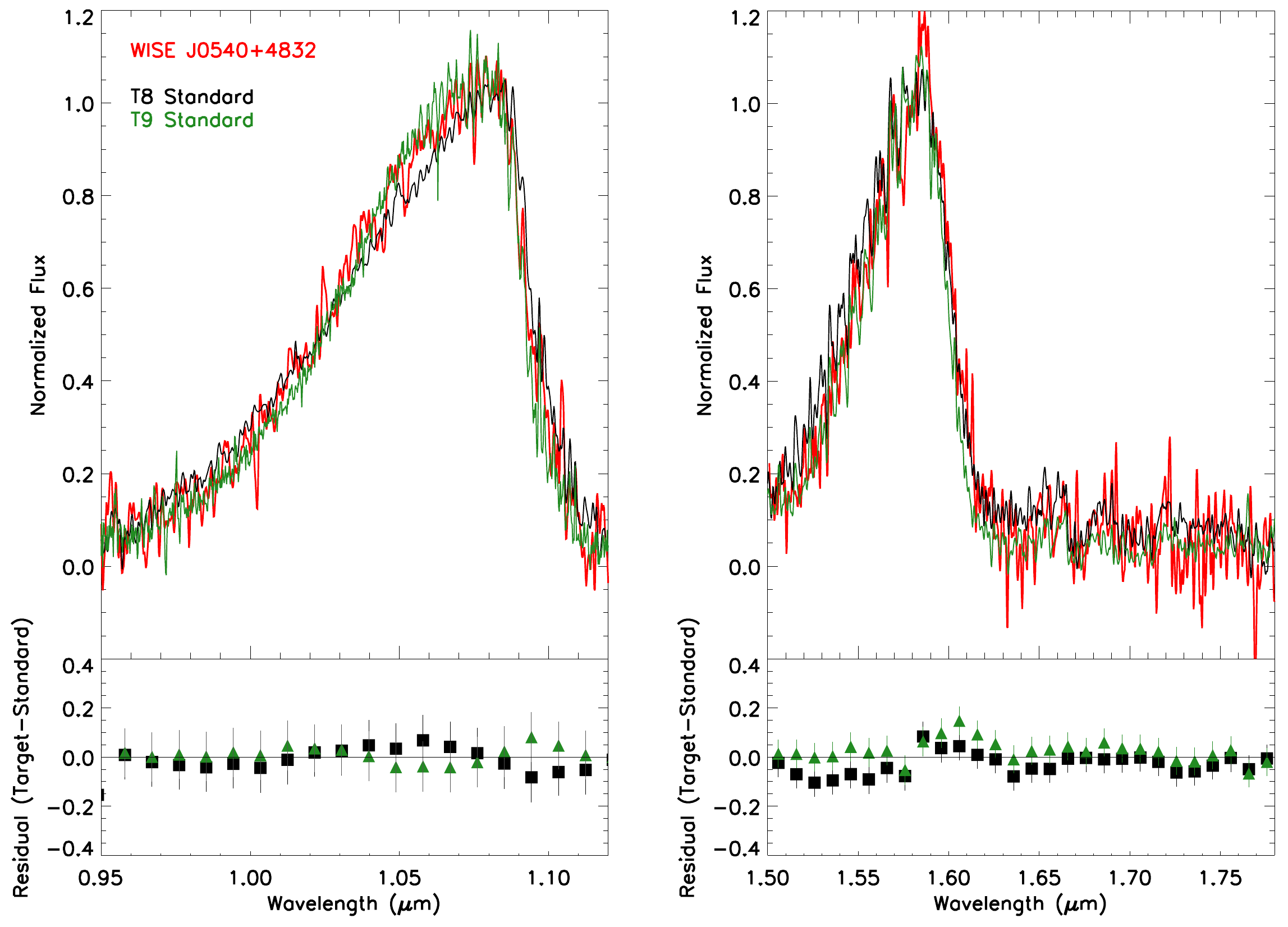}
\caption[Appendix: WISE J1617+1807 \& WISE J0540+4832]{WISE J1617+1807 (T8; top) and WISE J0540+4832 (T8.5)}
\end{figure}

\begin{figure}[h]
\centering
\includegraphics[width=5.5in]{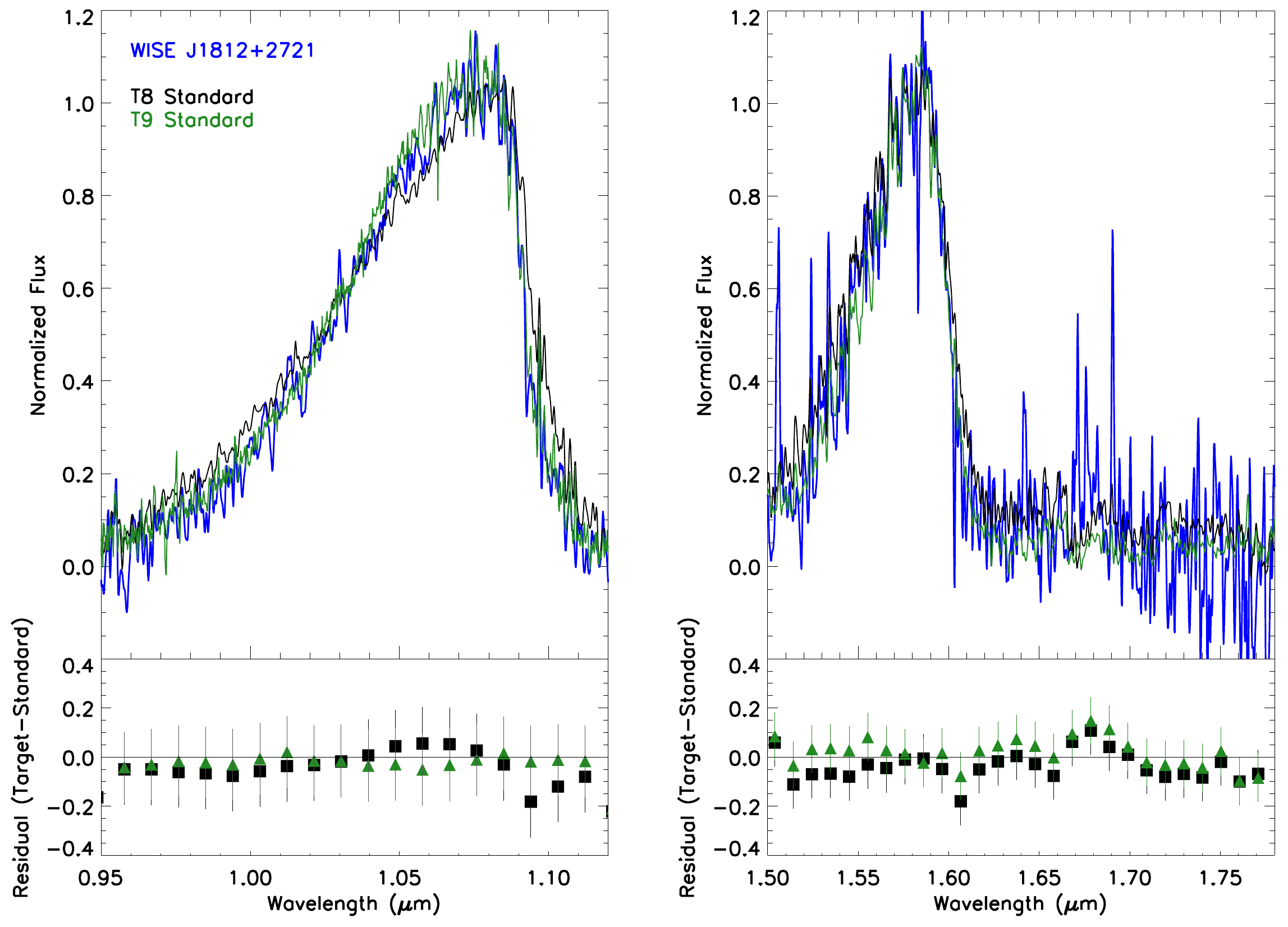}
\includegraphics[width=5.5in]{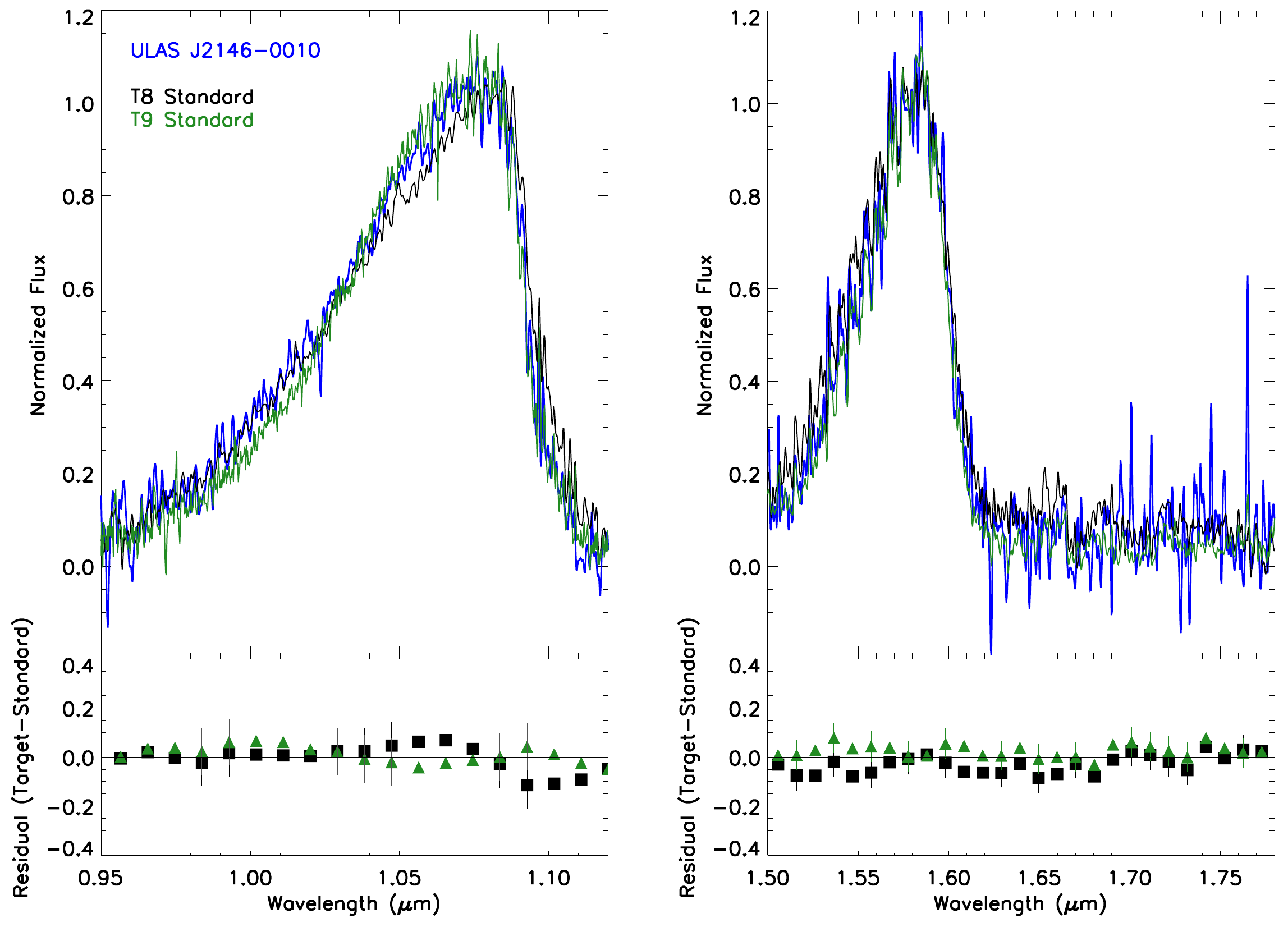}
\caption[Appendix: WISE J1812+2721 \& Wolf 940B]{WISE J1812+2721 (T8.5; top) and ULAS 2146-0010 (Wolf 940B; T8.5)}
\end{figure}

\begin{figure}[h]
\centering
\includegraphics[width=5.5in]{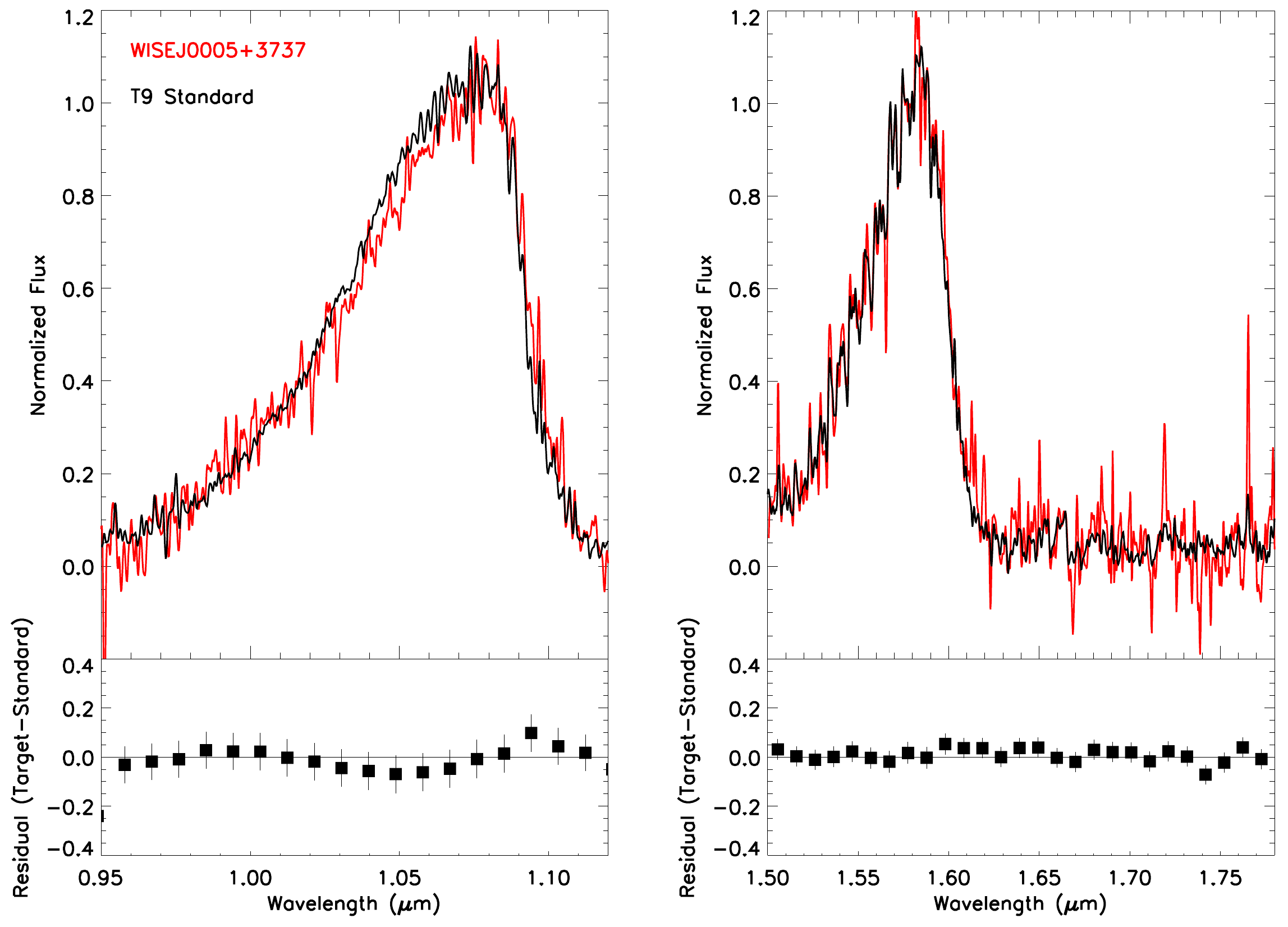}
\caption[Appendix: WISE J0005+3737]{WISE J0005+3737 (T9)}
\end{figure}

\clearpage

\bibliographystyle{apj}

\begin{thebibliography}{}
\expandafter\ifx\csname natexlab\endcsname\relax\def\natexlab#1{#1}\fi

\bibitem[{{Albert} {et~al.}(2011){Albert}, {Artigau}, {Delorme}, {Reyl{\'e}},
  {Forveille}, {Delfosse}, \& {Willott}}]{Albert2011}
{Albert}, L., {Artigau}, {\'E}., {Delorme}, P., {et~al.} 2011, \aj, 141, 203

\bibitem[{{Allard} {et~al.}(2011){Allard}, {Homeier}, \&
  {Freytag}}]{Allard2011}
{Allard}, F., {Homeier}, D., \& {Freytag}, B. 2011, in Astronomical Society of
  the Pacific Conference Series, Vol. 448, 16th Cambridge Workshop on Cool
  Stars, Stellar Systems, and the Sun, ed. C.~{Johns-Krull}, M.~K. {Browning},
  \& A.~A. {West}, 91

\bibitem[{{Allard} {et~al.}(2012){Allard}, {Homeier}, {Freytag}, \&
  {Sharp}}]{Allard2012}
{Allard}, F., {Homeier}, D., {Freytag}, B., \& {Sharp}, C.~M. 2012, in EAS
  Publications Series, Vol.~57, EAS Publications Series, ed. C.~{Reyl{\'e}},
  C.~{Charbonnel}, \& M.~{Schultheis}, 3--43

\bibitem[{{Allers} \& {Liu}(2013)}]{AllersLiu2013}
{Allers}, K.~N., \& {Liu}, M.~C. 2013, \apj, 772, 79

\bibitem[{{Beichman} {et~al.}(2013){Beichman}, {Gelino}, {Kirkpatrick},
  {Barman}, {Marsh}, {Cushing}, \& {Wright}}]{Beichman2013}
{Beichman}, C., {Gelino}, C.~R., {Kirkpatrick}, J.~D., {et~al.} 2013, \apj,
  764, 101

\bibitem[{{Beichman} {et~al.}(2014){Beichman}, {Gelino}, {Kirkpatrick},
  {Cushing}, {Dodson-Robinson}, {Marley}, {Morley}, \& {Wright}}]{Beichman2014}
---. 2014, \apj, 783, 68

\bibitem[{{Bensby} {et~al.}(2003){Bensby}, {Feltzing}, \&
  {Lundstr{\"o}m}}]{Bensby2003}
{Bensby}, T., {Feltzing}, S., \& {Lundstr{\"o}m}, I. 2003, \aap, 410, 527

\bibitem[{{Buenzli} {et~al.}(2014){Buenzli}, {Apai}, {Radigan}, {Reid}, \&
  {Flateau}}]{Buenzli2014}
{Buenzli}, E., {Apai}, D., {Radigan}, J., {Reid}, I.~N., \& {Flateau}, D. 2014,
  \apj, 782, 77

\bibitem[{{Buenzli} {et~al.}(2012){Buenzli}, {Apai}, {Morley}, {Flateau},
  {Showman}, {Burrows}, {Marley}, {Lewis}, \& {Reid}}]{Buenzli2012}
{Buenzli}, E., {Apai}, D., {Morley}, C.~V., {et~al.} 2012, \apjl, 760, L31

\bibitem[{{Burgasser} {et~al.}(2006{\natexlab{a}}){Burgasser}, {Burrows}, \&
  {Kirkpatrick}}]{Burgasser2006c}
{Burgasser}, A.~J., {Burrows}, A., \& {Kirkpatrick}, J.~D. 2006{\natexlab{a}},
  \apj, 639, 1095

\bibitem[{{Burgasser} {et~al.}(2006{\natexlab{b}}){Burgasser}, {Geballe},
  {Leggett}, {Kirkpatrick}, \& {Golimowski}}]{Burgasser2006b}
{Burgasser}, A.~J., {Geballe}, T.~R., {Leggett}, S.~K., {Kirkpatrick}, J.~D.,
  \& {Golimowski}, D.~A. 2006{\natexlab{b}}, \apj, 637, 1067

\bibitem[{{Burgasser} {et~al.}(2003{\natexlab{a}}){Burgasser}, {Kirkpatrick},
  {Liebert}, \& {Burrows}}]{Burgasser2003b}
{Burgasser}, A.~J., {Kirkpatrick}, J.~D., {Liebert}, J., \& {Burrows}, A.
  2003{\natexlab{a}}, \apj, 594, 510

\bibitem[{{Burgasser} {et~al.}(2000){Burgasser}, {Wilson}, {Kirkpatrick},
  {Skrutskie}, {Colonno}, {Enos}, {Smith}, {Henderson}, {Gizis}, {Brown}, \&
  {Houck}}]{Burgasser2000b}
{Burgasser}, A.~J., {Wilson}, J.~C., {Kirkpatrick}, J.~D., {et~al.} 2000, \aj,
  120, 1100

\bibitem[{{Burgasser} {et~al.}(2002){Burgasser}, {Kirkpatrick}, {Brown},
  {Reid}, {Burrows}, {Liebert}, {Matthews}, {Gizis}, {Dahn}, {Monet}, {Cutri},
  \& {Skrutskie}}]{Burgasser2002}
{Burgasser}, A.~J., {Kirkpatrick}, J.~D., {Brown}, M.~E., {et~al.} 2002, \apj,
  564, 421

\bibitem[{{Burgasser} {et~al.}(2003{\natexlab{b}}){Burgasser}, {Kirkpatrick},
  {Burrows}, {Liebert}, {Reid}, {Gizis}, {McGovern}, {Prato}, \&
  {McLean}}]{Burgasser2003}
{Burgasser}, A.~J., {Kirkpatrick}, J.~D., {Burrows}, A., {et~al.}
  2003{\natexlab{b}}, \apj, 592, 1186

\bibitem[{{Burgasser} {et~al.}(2010){Burgasser}, {Simcoe}, {Bochanski},
  {Saumon}, {Mamajek}, {Cushing}, {Marley}, {McMurtry}, {Pipher}, \&
  {Forrest}}]{Burgasser2010}
{Burgasser}, A.~J., {Simcoe}, R.~A., {Bochanski}, J.~J., {et~al.} 2010, \apj,
  725, 1405

\bibitem[{{Burgasser} {et~al.}(2011){Burgasser}, {Cushing}, {Kirkpatrick},
  {Gelino}, {Griffith}, {Looper}, {Tinney}, {Simcoe}, {Bochanski}, {Skrutskie},
  {Mainzer}, {Thompson}, {Marsh}, {Bauer}, \& {Wright}}]{Burgasser2011}
{Burgasser}, A.~J., {Cushing}, M.~C., {Kirkpatrick}, J.~D., {et~al.} 2011,
  \apj, 735, 116

\bibitem[{{Burningham} {et~al.}(2014){Burningham}, {Smith}, {Cardoso}, {Lucas},
  {Burgasser}, {Jones}, \& {Smart}}]{Burningham2014}
{Burningham}, B., {Smith}, L., {Cardoso}, C.~V., {et~al.} 2014, \mnras, 440,
  359

\bibitem[{{Burningham} {et~al.}(2008){Burningham}, {Pinfield}, {Leggett},
  {Tamura}, {Lucas}, {Homeier}, {Day-Jones}, {Jones}, {Clarke}, {Ishii},
  {Kuzuhara}, {Lodieu}, {Zapatero Osorio}, {Venemans}, {Mortlock}, {Barrado Y
  Navascu{\'e}s}, {Martin}, \& {Magazz{\`u}}}]{Burningham2008}
{Burningham}, B., {Pinfield}, D.~J., {Leggett}, S.~K., {et~al.} 2008, \mnras,
  391, 320

\bibitem[{{Burningham} {et~al.}(2009){Burningham}, {Pinfield}, {Leggett},
  {Tinney}, {Liu}, {Homeier}, {West}, {Day-Jones}, {Huelamo}, {Dupuy}, {Zhang},
  {Murray}, {Lodieu}, {Barrado Y Navascu{\'e}s}, {Folkes}, {Galvez-Ortiz},
  {Jones}, {Lucas}, {Calderon}, \& {Tamura}}]{Burningham2009}
---. 2009, \mnras, 395, 1237

\bibitem[{{Burningham} {et~al.}(2010){Burningham}, {Pinfield}, {Lucas},
  {Leggett}, {Deacon}, {Tamura}, {Tinney}, {Lodieu}, {Zhang}, {Huelamo},
  {Jones}, {Murray}, {Mortlock}, {Patel}, {Barrado Y Navascu{\'e}s}, {Zapatero
  Osorio}, {Ishii}, {Kuzuhara}, \& {Smart}}]{Burningham2010}
{Burningham}, B., {Pinfield}, D.~J., {Lucas}, P.~W., {et~al.} 2010, \mnras,
  406, 1885

\bibitem[{{Burningham} {et~al.}(2011){Burningham}, {Lucas}, {Leggett}, {Smart},
  {Baker}, {Pinfield}, {Tinney}, {Homeier}, {Allard}, {Zhang}, {Gomes},
  {Day-Jones}, {Jones}, {Kov{\'a}cs}, {Lodieu}, {Marocco}, {Murray}, \& {Sip{\H
  o}cz}}]{Burningham2011}
{Burningham}, B., {Lucas}, P.~W., {Leggett}, S.~K., {et~al.} 2011, \mnras, 414,
  L90

\bibitem[{{Burningham} {et~al.}(2013){Burningham}, {Cardoso}, {Smith},
  {Leggett}, {Smart}, {Mann}, {Dhital}, {Lucas}, {Tinney}, {Pinfield}, {Zhang},
  {Morley}, {Saumon}, {Aller}, {Littlefair}, {Homeier}, {Lodieu}, {Deacon},
  {Marley}, {van Spaandonk}, {Baker}, {Allard}, {Andrei}, {Canty}, {Clarke},
  {Day-Jones}, {Dupuy}, {Fortney}, {Gomes}, {Ishii}, {Jones}, {Liu},
  {Magazz{\'u}}, {Marocco}, {Murray}, {Rojas-Ayala}, \&
  {Tamura}}]{Burningham2013}
{Burningham}, B., {Cardoso}, C.~V., {Smith}, L., {et~al.} 2013, \mnras, 433,
  457

\bibitem[{{Burrows} {et~al.}(2002){Burrows}, {Burgasser}, {Kirkpatrick},
  {Liebert}, {Milsom}, {Sudarsky}, \& {Hubeny}}]{Burrows2002}
{Burrows}, A., {Burgasser}, A.~J., {Kirkpatrick}, J.~D., {et~al.} 2002, \apj,
  573, 394

\bibitem[{{Burrows} {et~al.}(2011){Burrows}, {Heng}, \&
  {Nampaisarn}}]{Burrows2011}
{Burrows}, A., {Heng}, K., \& {Nampaisarn}, T. 2011, \apj, 736, 47

\bibitem[{{Burrows} {et~al.}(2006){Burrows}, {Sudarsky}, \&
  {Hubeny}}]{Burrows2006}
{Burrows}, A., {Sudarsky}, D., \& {Hubeny}, I. 2006, \apj, 640, 1063

\bibitem[{{Burrows} {et~al.}(1997){Burrows}, {Marley}, {Hubbard}, {Lunine},
  {Guillot}, {Saumon}, {Freedman}, {Sudarsky}, \& {Sharp}}]{Burrows1997}
{Burrows}, A., {Marley}, M., {Hubbard}, W.~B., {et~al.} 1997, \apj, 491, 856

\bibitem[{{Casali} {et~al.}(2007){Casali}, {Adamson}, {Alves de Oliveira},
  {Almaini}, {Burch}, {Chuter}, {Elliot}, {Folger}, {Foucaud}, {Hambly},
  {Hastie}, {Henry}, {Hirst}, {Irwin}, {Ives}, {Lawrence}, {Laidlaw}, {Lee},
  {Lewis}, {Lunney}, {McLay}, {Montgomery}, {Pickup}, {Read}, {Rees}, {Robson},
  {Sekiguchi}, {Vick}, {Warren}, \& {Woodward}}]{Casali2007}
{Casali}, M., {Adamson}, A., {Alves de Oliveira}, C., {et~al.} 2007, \aap, 467,
  777

\bibitem[{{Cenarro} {et~al.}(2007){Cenarro}, {Peletier},
  {S{\'a}nchez-Bl{\'a}zquez}, {Selam}, {Toloba}, {Cardiel},
  {Falc{\'o}n-Barroso}, {Gorgas}, {Jim{\'e}nez-Vicente}, \&
  {Vazdekis}}]{Cenarro2007}
{Cenarro}, A.~J., {Peletier}, R.~F., {S{\'a}nchez-Bl{\'a}zquez}, P., {et~al.}
  2007, \mnras, 374, 664

\bibitem[{{Chiu} {et~al.}(2008){Chiu}, {Liu}, {Jiang}, {Allers}, {Stark},
  {Bunker}, {Fan}, {Glazebrook}, \& {Dupuy}}]{Chiu2008}
{Chiu}, K., {Liu}, M.~C., {Jiang}, L., {et~al.} 2008, \mnras, 385, L53

\bibitem[{{Crepp} {et~al.}(2012){Crepp}, {Johnson}, {Fischer}, {Howard},
  {Marcy}, {Wright}, {Isaacson}, {Boyajian}, {von Braun}, {Hillenbrand},
  {Hinkley}, {Carpenter}, \& {Brewer}}]{Crepp2012}
{Crepp}, J.~R., {Johnson}, J.~A., {Fischer}, D.~A., {et~al.} 2012, \apj, 751,
  97

\bibitem[{{Cushing} {et~al.}(2011){Cushing}, {Kirkpatrick}, {Gelino},
  {Griffith}, {Skrutskie}, {Mainzer}, {Marsh}, {Beichman}, {Burgasser},
  {Prato}, {Simcoe}, {Marley}, {Saumon}, {Freedman}, {Eisenhardt}, \&
  {Wright}}]{Cushing2011}
{Cushing}, M.~C., {Kirkpatrick}, J.~D., {Gelino}, C.~R., {et~al.} 2011, \apj,
  743, 50

\bibitem[{{Cushing} {et~al.}(2016){Cushing}, {Hardegree-Ullman}, {Trucks},
  {Morley}, {Gizis}, {Marley}, {Fortney}, {Kirkpatrick}, {Gelino}, {Mace}, \&
  {Carey}}]{Cushing2016}
{Cushing}, M.~C., {Hardegree-Ullman}, K.~K., {Trucks}, J.~L., {et~al.} 2016,
  \apj, 823, 152

\bibitem[{{Dahn} {et~al.}(2002){Dahn}, {Harris}, {Vrba}, {Guetter}, {Canzian},
  {Henden}, {Levine}, {Luginbuhl}, {Monet}, {Monet}, {Pier}, {Stone}, {Walker},
  {Burgasser}, {Gizis}, {Kirkpatrick}, {Liebert}, \& {Reid}}]{Dahn2002}
{Dahn}, C.~C., {Harris}, H.~C., {Vrba}, F.~J., {et~al.} 2002, \aj, 124, 1170

\bibitem[{{DePoy} {et~al.}(1993){DePoy}, {Atwood}, {Byard}, {Frogel}, \&
  {O'Brien}}]{DePoy1993}
{DePoy}, D.~L., {Atwood}, B., {Byard}, P.~L., {Frogel}, J., \& {O'Brien}, T.~P.
  1993, in \procspie, Vol. 1946, Infrared Detectors and Instrumentation, ed.
  A.~M. {Fowler}, 667--672

\bibitem[{{Dupuy} \& {Liu}(2012)}]{DupuyLiu2012}
{Dupuy}, T.~J., \& {Liu}, M.~C. 2012, \apjs, 201, 19

\bibitem[{{Dupuy} {et~al.}(2009){Dupuy}, {Liu}, \& {Ireland}}]{Dupuy2009}
{Dupuy}, T.~J., {Liu}, M.~C., \& {Ireland}, M.~J. 2009, \apj, 699, 168

\bibitem[{{Faherty} {et~al.}(2009){Faherty}, {Burgasser}, {Cruz}, {Shara},
  {Walter}, \& {Gelino}}]{Faherty2009}
{Faherty}, J.~K., {Burgasser}, A.~J., {Cruz}, K.~L., {et~al.} 2009, \aj, 137, 1

\bibitem[{Filippazzo {et~al.}(2015)Filippazzo, Rice, Faherty, Cruz, Gordon, \&
  Looper}]{Filipazzo2015}
Filippazzo, J.~C., Rice, E.~L., Faherty, J., {et~al.} 2015, The Astrophysical
  Journal, 810, 158

\bibitem[{{Gelino} {et~al.}(2011){Gelino}, {Kirkpatrick}, {Cushing},
  {Eisenhardt}, {Griffith}, {Mainzer}, {Marsh}, {Skrutskie}, \&
  {Wright}}]{Gelino2011}
{Gelino}, C.~R., {Kirkpatrick}, J.~D., {Cushing}, M.~C., {et~al.} 2011, \aj,
  142, 57

\bibitem[{Gray {et~al.}(2003)Gray, Corbally, Garrison, McFadden, \&
  Robinson}]{Gray2003}
Gray, R.~O., Corbally, C.~J., Garrison, R.~F., McFadden, M.~T., \& Robinson,
  P.~E. 2003, The Astronomical Journal, 126, 2048

\bibitem[{{Hodapp} {et~al.}(2003){Hodapp}, {Jensen}, {Irwin}, {Yamada},
  {Chung}, {Fletcher}, {Robertson}, {Hora}, {Simons}, {Mays}, {Nolan}, {Bec},
  {Merrill}, \& {Fowler}}]{Hodapp2003}
{Hodapp}, K.~W., {Jensen}, J.~B., {Irwin}, E.~M., {et~al.} 2003, \pasp, 115,
  1388

\bibitem[{{Kirkpatrick}(2005)}]{Kirkpatrick2005}
{Kirkpatrick}, J.~D. 2005, \araa, 43, 195

\bibitem[{{Kirkpatrick} {et~al.}(2013){Kirkpatrick}, {Cushing}, {Gelino},
  {Beichman}, {Tinney}, {Faherty}, {Schneider}, \& {Mace}}]{Kirkpatrick2013}
{Kirkpatrick}, J.~D., {Cushing}, M.~C., {Gelino}, C.~R., {et~al.} 2013, \apj,
  776, 128

\bibitem[{{Kirkpatrick} {et~al.}(2011){Kirkpatrick}, {Cushing}, {Gelino},
  {Griffith}, {Skrutskie}, {Marsh}, {Wright}, {Mainzer}, {Eisenhardt},
  {McLean}, {Thompson}, {Bauer}, {Benford}, {Bridge}, {Lake}, {Petty},
  {Stanford}, {Tsai}, {Bailey}, {Beichman}, {Bloom}, {Bochanski}, {Burgasser},
  {Capak}, {Cruz}, {Hinz}, {Kartaltepe}, {Knox}, {Manohar}, {Masters},
  {Morales-Calder{\'o}n}, {Prato}, {Rodigas}, {Salvato}, {Schurr}, {Scoville},
  {Simcoe}, {Stapelfeldt}, {Stern}, {Stock}, \& {Vacca}}]{Kirkpatrick2011}
---. 2011, \apjs, 197, 19

\bibitem[{{Kirkpatrick} {et~al.}(2012){Kirkpatrick}, {Gelino}, {Cushing},
  {Mace}, {Griffith}, {Skrutskie}, {Marsh}, {Wright}, {Eisenhardt}, {McLean},
  {Mainzer}, {Burgasser}, {Tinney}, {Parker}, \& {Salter}}]{Kirkpatrick2012}
{Kirkpatrick}, J.~D., {Gelino}, C.~R., {Cushing}, M.~C., {et~al.} 2012, \apj,
  753, 156

\bibitem[{{Kirkpatrick} {et~al.}(2014){Kirkpatrick}, {Schneider},
  {Fajardo-Acosta}, {Gelino}, {Mace}, {Wright}, {Logsdon}, {McLean}, {Cushing},
  {Skrutskie}, {Eisenhardt}, {Stern}, {Balokovi{\'c}}, {Burgasser}, {Faherty},
  {Lansbury}, {Rich}, {Skrzypek}, {Fowler}, {Cutri}, {Masci}, {Conrow},
  {Grillmair}, {McCallon}, {Beichman}, \& {Marsh}}]{Kirkpatrick2014}
{Kirkpatrick}, J.~D., {Schneider}, A., {Fajardo-Acosta}, S., {et~al.} 2014,
  \apj, 783, 122

\bibitem[{{Knapp} {et~al.}(2004){Knapp}, {Leggett}, {Fan}, {Marley}, {Geballe},
  {Golimowski}, {Finkbeiner}, {Gunn}, {Hennawi}, {Ivezi{\'c}}, {Lupton},
  {Schlegel}, {Strauss}, {Tsvetanov}, {Chiu}, {Hoversten}, {Glazebrook},
  {Zheng}, {Hendrickson}, {Williams}, {Uomoto}, {Vrba}, {Henden}, {Luginbuhl},
  {Guetter}, {Munn}, {Canzian}, {Schneider}, \& {Brinkmann}}]{Knapp2004}
{Knapp}, G.~R., {Leggett}, S.~K., {Fan}, X., {et~al.} 2004, \aj, 127, 3553

\bibitem[{Koen(2013)}]{Koen2013}
Koen, C. 2013, Monthly Notices of the Royal Astronomical Society, 428, 2824

\bibitem[{{Lawrence} {et~al.}(2007){Lawrence}, {Warren}, {Almaini}, {Edge},
  {Hambly}, {Jameson}, {Lucas}, {Casali}, {Adamson}, {Dye}, {Emerson},
  {Foucaud}, {Hewett}, {Hirst}, {Hodgkin}, {Irwin}, {Lodieu}, {McMahon},
  {Simpson}, {Smail}, {Mortlock}, \& {Folger}}]{Lawrence2007}
{Lawrence}, A., {Warren}, S.~J., {Almaini}, O., {et~al.} 2007, \mnras, 379,
  1599

\bibitem[{{Leggett} {et~al.}(2007){Leggett}, {Marley}, {Freedman}, {Saumon},
  {Liu}, {Geballe}, {Golimowski}, \& {Stephens}}]{Leggett2007}
{Leggett}, S.~K., {Marley}, M.~S., {Freedman}, R., {et~al.} 2007, \apj, 667,
  537

\bibitem[{Leggett {et~al.}(2015)Leggett, Morley, Marley, \&
  Saumon}]{Leggett2015}
Leggett, S.~K., Morley, C.~V., Marley, M.~S., \& Saumon, D. 2015, The
  Astrophysical Journal, 799, 37

\bibitem[{{Leggett} {et~al.}(2013){Leggett}, {Morley}, {Marley}, {Saumon},
  {Fortney}, \& {Visscher}}]{Leggett2013}
{Leggett}, S.~K., {Morley}, C.~V., {Marley}, M.~S., {et~al.} 2013, \apj, 763,
  130

\bibitem[{{Leggett} {et~al.}(2010{\natexlab{a}}){Leggett}, {Burningham},
  {Saumon}, {Marley}, {Warren}, {Smart}, {Jones}, {Lucas}, {Pinfield}, \&
  {Tamura}}]{Leggett2010}
{Leggett}, S.~K., {Burningham}, B., {Saumon}, D., {et~al.} 2010{\natexlab{a}},
  \apj, 710, 1627
  
  \bibitem[{{Leggett} {et~al.}(2010{\natexlab{b}}){Leggett}, {Saumon},
  {Burningham}, {Cushing}, {Marley}, \& {Pinfield}}]{Leggett2010b}
{Leggett}, S.~K., {Saumon}, D., {Burningham}, B., {et~al.} 2010{\natexlab{b}},
  \apj, 720, 252

\bibitem[{{Leggett} {et~al.}(2016){Leggett}, {Cushing}, {Hardegree-Ullman},
  {Trucks}, {Marley}, {Morley}, {Saumon}, {Carey}, {Fortney}, {Gelino},
  {Gizis}, {Kirkpatrick}, \& {Mace}}]{Leggett2016}
{Leggett}, S.~K., {Cushing}, M.~C., {Hardegree-Ullman}, K.~K., {et~al.} 2016,
  \apj, 830, 141

\bibitem[{{Leggett} {et~al.}(2017){Leggett},{Tremblin},{Esplin},{Luhman}\&{Morley}}]{Leggett2017}
{Leggett}, S.~K., {Tremblin}, P., {et~al.} 2017, \apj, 842, 118

\bibitem[{{Line} {et~al.}(2017){Line}, {Marley}, {Liu}, {Morley}, {Burningham},
  {Hinkel}, {Teske}, \& {Fortney}}]{Line2017}
{Line}, M.~R., {Marley}, M.~S., {Liu}, M.~C., {et~al.} 2017, \apj,
  848, 83

\bibitem[{{Line} {et~al.}(2015){Line}, {Teske}, {Burningham}, {Fortney}, \&
  {Marley}}]{Line2015}
{Line}, M.~R., {Teske}, J., {Burningham}, B., {Fortney}, J.~J., \& {Marley},
  M.~S. 2015, \apj, 807, 183

\bibitem[{{Littlefair} {et~al.}(2017){Littlefair}, {Burningham}, \&
  {Helling}}]{Littlefair2017}
{Littlefair}, S.~P., {Burningham}, B., \& {Helling}, C. 2017, \mnras, 466, 4250

\bibitem[{{Liu} {et~al.}(2012){Liu}, {Dupuy}, {Bowler}, {Leggett}, \&
  {Best}}]{Liu2012}
{Liu}, M.~C., {Dupuy}, T.~J., {Bowler}, B.~P., {Leggett}, S.~K., \& {Best},
  W.~M.~J. 2012, \apj, 758, 57

\bibitem[{{Liu} {et~al.}(2008){Liu}, {Dupuy}, \& {Ireland}}]{Liu2008}
{Liu}, M.~C., {Dupuy}, T.~J., \& {Ireland}, M.~J. 2008, \apj, 689, 436

\bibitem[{{Loh} {et~al.}(2012){Loh}, {Biel}, {Davis}, {Laporte}, {Loh}, \&
  {Verhanovitz}}]{Loh2012}
{Loh}, E.~D., {Biel}, J.~D., {Davis}, M.~W., {et~al.} 2012, \pasp, 124, 343

\bibitem[{{Lucas} {et~al.}(2010){Lucas}, {Tinney}, {Burningham}, {Leggett},
  {Pinfield}, {Smart}, {Jones}, {Marocco}, {Barber}, {Yurchenko}, {Tennyson},
  {Ishii}, {Tamura}, {Day-Jones}, {Adamson}, {Allard}, \&
  {Homeier}}]{Lucas2010}
{Lucas}, P.~W., {Tinney}, C.~G., {Burningham}, B., {et~al.} 2010, \mnras, 408,
  L56

\bibitem[{{Luhman} {et~al.}(2012){Luhman}, {Loutrel}, {McCurdy}, {Mace},
  {Melso}, {Star}, {Young}, {Terrien}, {McLean}, {Kirkpatrick}, \&
  {Rhode}}]{Luhman2012}
{Luhman}, K.~L., {Loutrel}, N.~P., {McCurdy}, N.~S., {et~al.} 2012, \apj, 760,
  152

\bibitem[{{Mace}(2014)}]{MacePHD2014}
{Mace}, G.~N. 2014, PhD thesis, University of California, Los Angeles

\bibitem[{{Mace} {et~al.}(2013{\natexlab{a}}){Mace}, {Kirkpatrick}, {Cushing},
  {Gelino}, {Griffith}, {Skrutskie}, {Marsh}, {Wright}, {Eisenhardt}, {McLean},
  {Thompson}, {Mix}, {Bailey}, {Beichman}, {Bloom}, {Burgasser}, {Fortney},
  {Hinz}, {Knox}, {Lowrance}, {Marley}, {Morley}, {Rodigas}, {Saumon},
  {Sheppard}, \& {Stock}}]{Mace2013}
{Mace}, G.~N., {Kirkpatrick}, J.~D., {Cushing}, M.~C., {et~al.}
  2013{\natexlab{a}}, \apjs, 205, 6

\bibitem[{{Mace} {et~al.}(2013{\natexlab{b}}){Mace}, {Kirkpatrick}, {Cushing},
  {Gelino}, {McLean}, {Logsdon}, {Wright}, {Skrutskie}, {Beichman},
  {Eisenhardt}, \& {Kulas}}]{Mace2013b}
---. 2013{\natexlab{b}}, \apj, 777, 36

\bibitem[{{Mace} {et~al.}(2018){Mace}, {Mann}, {Skiff}, {Sneden},
  {Kirkpatrick}, {Schneider}, {Kidder}, {Gosnell}, {Kim}, {Mulligan}, {Prato},
  \& {Jaffe}}]{Mace2018}
{Mace}, G.~N., {Mann}, A.~W., {Skiff}, B.~A., {et~al.} 2018, \apj, 854, 145

\bibitem[{{Marley} \& {Robinson}(2015)}]{MarleyRobinson2015}
{Marley}, M.~S., \& {Robinson}, T.~D. 2015, \araa, 53, 279

\bibitem[{{Marley} {et~al.}(2010){Marley}, {Saumon}, \&
  {Goldblatt}}]{Marley2010}
{Marley}, M.~S., {Saumon}, D., \& {Goldblatt}, C. 2010, \apjl, 723, L117

\bibitem[{Marley {et~al.}(2002)Marley, Seager, Saumon, Lodders, Ackerman,
  Freedman, \& Fan}]{Marley2002}
Marley, M.~S., Seager, S., Saumon, D., {et~al.} 2002, The Astrophysical
  Journal, 568, 335

\bibitem[{{Marocco} {et~al.}(2017){Marocco}, {Pinfield}, {Cook}, {Zapatero
  Osorio}, {Montes}, {Caballero}, {G{\'a}lvez-Ortiz}, {Gromadzki}, {Jones},
  {Kurtev}, {Smart}, {Zhang}, {Cabrera Lavers}, {Garc{\'{\i}}a {\'A}lvarez},
  {Qi}, {Rickard}, \& {Dover}}]{Marocco2017}
{Marocco}, F., {Pinfield}, D.~J., {Cook}, N.~J., {et~al.} 2017, \mnras, 470,
  4885

\bibitem[{Martin {et~al.}(2017)Martin, Mace, McLean, Logsdon, Rice,
  Kirkpatrick, Burgasser, McGovern, \& Prato}]{Martin2017}
Martin, E.~C., Mace, G.~N., McLean, I.~S., {et~al.} 2017, The Astrophysical
  Journal, 838, 73

\bibitem[{{McGovern} {et~al.}(2004){McGovern}, {Kirkpatrick}, {McLean},
  {Burgasser}, {Prato}, \& {Lowrance}}]{McGovern2004}
{McGovern}, M.~R., {Kirkpatrick}, J.~D., {McLean}, I.~S., {et~al.} 2004, \apj,
  600, 1020

\bibitem[{{McLean} {et~al.}(2003){McLean}, {McGovern}, {Burgasser},
  {Kirkpatrick}, {Prato}, \& {Kim}}]{McLean2003}
{McLean}, I.~S., {McGovern}, M.~R., {Burgasser}, A.~J., {et~al.} 2003, \apj,
  596, 561

\bibitem[{{McLean} {et~al.}(1998){McLean}, {Becklin}, {Bendiksen}, {Brims},
  {Canfield}, {Figer}, {Graham}, {Hare}, {Lacayanga}, {Larkin}, {Larson},
  {Levenson}, {Magnone}, {Teplitz}, \& {Wong}}]{McLean1998}
{McLean}, I.~S., {Becklin}, E.~E., {Bendiksen}, O., {et~al.} 1998, in
  \procspie, Vol. 3354, Infrared Astronomical Instrumentation, ed. A.~M.
  {Fowler}, 566--578

\bibitem[{{Metchev} {et~al.}(2015){Metchev}, {Heinze}, {Apai}, {Flateau},
  {Radigan}, {Burgasser}, {Marley}, {Artigau}, {Plavchan}, \&
  {Goldman}}]{Metchev2015}
{Metchev}, S.~A., {Heinze}, A., {Apai}, D., {et~al.} 2015, \apj, 799, 154

\bibitem[{{Morley} {et~al.}(2012){Morley}, {Fortney}, {Marley}, {Visscher},
  {Saumon}, \& {Leggett}}]{Morley2012}
{Morley}, C.~V., {Fortney}, J.~J., {Marley}, M.~S., {et~al.} 2012, \apj, 756,
  172

\bibitem[{{Morley} {et~al.}(2014){Morley}, {Marley}, {Fortney}, \&
  {Lupu}}]{Morley2014b}
{Morley}, C.~V., {Marley}, M.~S., {Fortney}, J.~J., \& {Lupu}, R. 2014, \apjl,
  789, L14

\bibitem[{{Murray} {et~al.}(2011){Murray}, {Burningham}, {Jones}, {Pinfield},
  {Lucas}, {Leggett}, {Tinney}, {Day-Jones}, {Weights}, {Lodieu}, {P{\'e}rez
  Prieto}, {Nickson}, {Zhang}, {Clarke}, {Jenkins}, \& {Tamura}}]{Murray2011}
{Murray}, D.~N., {Burningham}, B., {Jones}, H.~R.~A., {et~al.} 2011, \mnras,
  414, 575

\bibitem[{{Pinfield} {et~al.}(2006){Pinfield}, {Jones}, {Lucas}, {Kendall},
  {Folkes}, {Day-Jones}, {Chappelle}, \& {Steele}}]{Pinfield2006}
{Pinfield}, D.~J., {Jones}, H.~R.~A., {Lucas}, P.~W., {et~al.} 2006, \mnras,
  368, 1281

\bibitem[{{Pinfield} {et~al.}(2012){Pinfield}, {Burningham}, {Lodieu},
  {Leggett}, {Tinney}, {van Spaandonk}, {Marocco}, {Smart}, {Gomes}, {Smith},
  {Lucas}, {Day-Jones}, {Murray}, {Katsiyannis}, {Catalan}, {Cardoso},
  {Clarke}, {Folkes}, {G{\'a}lvez-Ortiz}, {Homeier}, {Jenkins}, {Jones}, \&
  {Zhang}}]{Pinfield2012}
{Pinfield}, D.~J., {Burningham}, B., {Lodieu}, N., {et~al.} 2012, \mnras, 422,
  1922

\bibitem[{{Pinfield} {et~al.}(2014){Pinfield}, {Gromadzki}, {Leggett}, {Gomes},
  {Lodieu}, {Kurtev}, {Day-Jones}, {Ruiz}, {Cook}, {Morley}, {Marley},
  {Marocco}, {Smart}, {Jones}, {Lucas}, {Beletsky}, {Ivanov}, {Burningham},
  {Jenkins}, {Cardoso}, {Frith}, {Clarke}, {G{\'a}lvez-Ortiz}, \&
  {Zhang}}]{Pinfield2014}
{Pinfield}, D.~J., {Gromadzki}, M., {Leggett}, S.~K., {et~al.} 2014, \mnras,
  444, 1931

\bibitem[{{Radigan} {et~al.}(2012){Radigan}, {Jayawardhana}, {Lafreni{\`e}re},
  {Artigau}, {Marley}, \& {Saumon}}]{Radigan2012}
{Radigan}, J., {Jayawardhana}, R., {Lafreni{\`e}re}, D., {et~al.} 2012, \apj,
  750, 105

\bibitem[{{Radigan} {et~al.}(2014){Radigan}, {Lafreni{\`e}re}, {Jayawardhana},
  \& {Artigau}}]{Radigan2014}
{Radigan}, J., {Lafreni{\`e}re}, D., {Jayawardhana}, R., \& {Artigau}, E. 2014,
  \apj, 793, 75

\bibitem[{{Rajan} {et~al.}(2015){Rajan}, {Patience}, {Wilson}, {Bulger}, {De
  Rosa}, {Ward-Duong}, {Morley}, {Pont}, \& {Windhorst}}]{Rajan2015}
{Rajan}, A., {Patience}, J., {Wilson}, P.~A., {et~al.} 2015, \mnras, 448, 3775

\bibitem[{{Rayner} {et~al.}(2003){Rayner}, {Toomey}, {Onaka}, {Denault},
  {Stahlberger}, {Vacca}, {Cushing}, \& {Wang}}]{Rayner2003}
{Rayner}, J.~T., {Toomey}, D.~W., {Onaka}, P.~M., {et~al.} 2003, \pasp, 115,
  362

\bibitem[{{Reyl{\'e}} {et~al.}(2010){Reyl{\'e}}, {Delorme}, {Willott},
  {Albert}, {Delfosse}, {Forveille}, {Artigau}, {Malo}, {Hill}, \&
  {Doyon}}]{Reyle2010}
{Reyl{\'e}}, C., {Delorme}, P., {Willott}, C.~J., {et~al.} 2010, \aap, 522,
  A112

\bibitem[{{Robinson} \& {Marley}(2014)}]{RobinsonMarley2014}
{Robinson}, T.~D., \& {Marley}, M.~S. 2014, \apj, 785, 158

\bibitem[{{Roche} {et~al.}(2003){Roche}, {Lucas}, {Mackay}, {Ettedgui-Atad},
  {Hastings}, {Bridger}, {Rees}, {Leggett}, {Davis}, {Holmes}, \&
  {Handford}}]{Roche2003}
{Roche}, P.~F., {Lucas}, P.~W., {Mackay}, C.~D., {et~al.} 2003, in \procspie,
  Vol. 4841, Instrument Design and Performance for Optical/Infrared
  Ground-based Telescopes, ed. M.~{Iye} \& A.~F.~M. {Moorwood}, 901--912

\bibitem[{Saumon \& Marley(2008)}]{SaumonMarley2008}
Saumon, D., \& Marley, M.~S. 2008, The Astrophysical Journal, 689, 1327

\bibitem[{{Saumon} {et~al.}(2012){Saumon}, {Marley}, {Abel}, {Frommhold}, \&
  {Freedman}}]{Saumon2012}
{Saumon}, D., {Marley}, M.~S., {Abel}, M., {Frommhold}, L., \& {Freedman},
  R.~S. 2012, \apj, 750, 74

\bibitem[{{Schmidt} {et~al.}(2007){Schmidt}, {Cruz}, {Bongiorno}, {Liebert}, \&
  {Reid}}]{Schmidt2007}
{Schmidt}, S.~J., {Cruz}, K.~L., {Bongiorno}, B.~J., {Liebert}, J., \& {Reid},
  I.~N. 2007, \aj, 133, 2258

\bibitem[{{Scholz}(2010)}]{Scholz2010}
{Scholz}, R.-D. 2010, \aap, 515, A92

\bibitem[{{Showman} \& {Kaspi}(2013)}]{ShowmanKaspi2013}
{Showman}, A.~P., \& {Kaspi}, Y. 2013, \apj, 776, 85

\bibitem[{{Simcoe} {et~al.}(2010){Simcoe}, {Burgasser}, {Bochanski},
  {Schechter}, {Bernstein}, {Bigelow}, {Pipher}, {Forrest}, {McMurtry},
  {Smith}, \& {Fishner}}]{Simcoe2010}
{Simcoe}, R.~A., {Burgasser}, A.~J., {Bochanski}, J.~J., {et~al.} 2010, in
  \procspie, Vol. 7735, Ground-based and Airborne Instrumentation for Astronomy
  III, 773514

\bibitem[{{Simons} \& {Tokunaga}(2002)}]{SimonsTokunaga2002}
{Simons}, D.~A., \& {Tokunaga}, A. 2002, \pasp, 114, 169

\bibitem[{{Thompson} {et~al.}(2013){Thompson}, {Kirkpatrick}, {Mace},
  {Cushing}, {Gelino}, {Griffith}, {Skrutskie}, {Eisenhardt}, {Wright},
  {Marsh}, {Mix}, {Beichman}, {Faherty}, {Toloza}, {Ferrara}, {Apodaca},
  {McLean}, \& {Bloom}}]{Thompson2013}
{Thompson}, M.~A., {Kirkpatrick}, J.~D., {Mace}, G.~N., {et~al.} 2013, \pasp,
  125, 809

\bibitem[{{van Leeuwen}(2007)}]{vanLeeuwen2007}
{van Leeuwen}, F. 2007, \aap, 474, 653

\bibitem[{{Vrba} {et~al.}(2004){Vrba}, {Henden}, {Luginbuhl}, {Guetter},
  {Munn}, {Canzian}, {Burgasser}, {Kirkpatrick}, {Fan}, {Geballe},
  {Golimowski}, {Knapp}, {Leggett}, {Schneider}, \& {Brinkmann}}]{Vrba2004}
{Vrba}, F.~J., {Henden}, A.~A., {Luginbuhl}, C.~B., {et~al.} 2004, \aj, 127,
  2948

\bibitem[{{Wilson} {et~al.}(2003){Wilson}, {Eikenberry}, {Henderson},
  {Hayward}, {Carson}, {Pirger}, {Barry}, {Brandl}, {Houck}, {Fitzgerald}, \&
  {Stolberg}}]{Wilson2003}
{Wilson}, J.~C., {Eikenberry}, S.~S., {Henderson}, C.~P., {et~al.} 2003, in
  \procspie, Vol. 4841, Instrument Design and Performance for Optical/Infrared
  Ground-based Telescopes, ed. M.~{Iye} \& A.~F.~M. {Moorwood}, 451--458

\bibitem[{{Wilson} {et~al.}(2014){Wilson}, {Rajan}, \& {Patience}}]{Wilson2014}
{Wilson}, P.~A., {Rajan}, A., \& {Patience}, J. 2014, \aap, 566, A111

\bibitem[{{Wright} {et~al.}(2010){Wright}, {Eisenhardt}, {Mainzer}, {Ressler},
  {Cutri}, {Jarrett}, {Kirkpatrick}, {Padgett}, {McMillan}, {Skrutskie},
  {Stanford}, {Cohen}, {Walker}, {Mather}, {Leisawitz}, {Gautier}, {McLean},
  {Benford}, {Lonsdale}, {Blain}, {Mendez}, {Irace}, {Duval}, {Liu}, {Royer},
  {Heinrichsen}, {Howard}, {Shannon}, {Kendall}, {Walsh}, {Larsen}, {Cardon},
  {Schick}, {Schwalm}, {Abid}, {Fabinsky}, {Naes}, \& {Tsai}}]{Wright2010}
{Wright}, E.~L., {Eisenhardt}, P.~R.~M., {Mainzer}, A.~K., {et~al.} 2010, \aj,
  140, 1868

\bibitem[{{Zhang} {et~al.}(2017){Zhang}, {Pinfield}, {G{\'a}lvez-Ortiz},
  {Burningham}, {Lodieu}, {Marocco}, {Burgasser}, {Day-Jones}, {Allard},
  {Jones}, {Homeier}, {Gomes}, \& {Smart}}]{Zhang2017}
{Zhang}, Z.~H., {Pinfield}, D.~J., {G{\'a}lvez-Ortiz}, M.~C., {et~al.} 2017,
  \mnras, 464, 3040

\end{thebibliography}

\end{document}